\documentstyle[epsfig,graphicx,12pt]{article}
\begin{document}

\title{Describing the Nucleon and its Form Factors\\ at High  $q^2$}
\author{
B. DESPLANQUES$^1$\thanks{{\it E-mail address:}  desplanq@isn.in2p3.fr}, 
B. SILVESTRE-BRAC$^1$\thanks{{\it E-mail address:} 
   silvestre@isn.in2p3.fr},   
F. CANO$^2$\thanks{{\it E-mail address:}    cano@xaloc.ific.uv.es},\\ 
P. GONZALEZ$^{3,4}$\thanks{{\it E-mail address:} 
   gonzalep@evalvx.ific.uv.es} 
and S. NOGUERA$^3$\thanks{{\it E-mail address:}   Santiago.Noguera@uv.es}  \\
$^1$Institut des Sciences Nucl\'eaires,      UMR CNRS-UJF, \\
     F-38026 Grenoble Cedex, France \\
$^2$Dipartamento di Fisica, Universit\'a degli Studi di Trento, \\
     I-38050 Povo (Trento), Italy\\
$^3$Departamento de F\'{\i}sica Te\'orica, Universidad de Valencia, \\
     E-46100 Burjassot (Valencia), Spain \\
$^4$IFIC, Centro Mixto Universidad de Valencia-CSIC, \\ 
     E-46100 Burjassot    (Valencia), Spain \\
}

\sloppy

\maketitle
\begin{abstract}
\small{The nucleon form factors are calculated using a non-relativistic 
description in terms of constituent quarks. The emphasis is put on 
the reliability of present numerical methods used to solve the three-body 
problem in order to correctly reproduce the expected asymptotic behavior 
of form factors. Nucleon wave functions obtained in the hyperspherical 
formalism or employing Faddeev equations have been considered. While 
a $q^{-8}$ behavior is expected at high $q$ for a quark-quark force 
behaving like $\frac{1}{r}$ at short distances, it is found that 
the hypercentral approximation in the hyperspherical formalism 
($K=0$) leads to a $q^{-7}$ behavior. An infinite set of waves is 
required to get the correct behavior. Solutions of the Faddeev 
equations lead to the $q^{-8}$ behavior. The amplitude of the 
corresponding term however depends on the number of partial waves 
retained in the Faddeev amplitude. The convergence to the asymptotic 
behavior has also been studied. Sizeable departures are observed 
in some cases at squared momentum transfers as high 
as $50 \, ({\rm GeV/c})^2$. It is not clear whether these departures 
are of the order $\frac{1}{q}$ or $\frac{1}{q^2}\, {\rm log} \, q$ 
relatively to the dominant contribution and whether the bad convergence 
results from truncations in the calculations. From a comparison with the 
most complete Faddeev results, a  $q^{2}$ validity range is obtained for 
the calculation made in the hyperspherical formalism or in the Faddeev 
approach with the minimum number of amplitudes.}
\end{abstract}

\vspace{1cm} 
\vspace{1cm} 

%%%%%%%%%%%%%%%%%1111111111111111111111%%%%%%%%%%%%%%%%%%%%%%%%%%%%%%%%%%%%%%%%

\section{Introduction}
A great interest is currently devoted to the nucleon form factor and, 
in particular, to its behavior at high momentum transfers. Indeed, 
from $QCD$ in the perturbative regime \cite{BROD}, one expects the nucleon 
form factor, $G_M(q^2)$ (or $F_1(q^2)$), to scale like $q^{-4}$, 
up to ${\rm log}$ terms. Experimentally, this behavior seems to be reached 
quite rapidly, around $10 \, ({\rm GeV/c})^2$ \cite{STOL}.

Many theoretical works tend to make a bridge between the low and high 
momentum domains \cite{STRO,IVAN}. They deal with the non-perturbative 
regime of $QCD$, where predictions are more difficult, and rely on 
non-relativistic calculations for the lower part of the momentum range. 
Among the questions that may be raised, there is the sensitivity of the 
predictions to the theoretical framework, including particular techniques, 
or the rapidity of the asymptotic behavior onset. Obviously, in view 
of the large momentum transfers that the asymptotic regime supposes, 
a definite statement would require a relativistic treatment. Some works 
along these lines are in progress \cite{STRO,IVAN,SALM,KARM}. We 
nevertheless believe that a non-relativistic calculation may be of 
some help to provide qualitative, if not quantitative answers to the 
above questions. We will concentrate on them in the following.

Quite simple descriptions of nucleons in terms of constituent quarks 
have relied on the harmonic oscillator wave function \cite{ISGU}. 
In such models , the calculated form factors drop exponentially to 
zero beyond $q^2=3 \, ({\rm GeV/c})^2$ \cite{PFEI}. Curiously, it has been 
sometimes deduced from this result that a constituent quark model could not 
give rise to a power law behavior for form factors at high $q^2$. 
Better non-relativistic descriptions of the nucleon involve solving 
the Schr\"odinger equation, using for instance the hyperspherical 
harmonic formalism or the Faddeev equations. These approaches 
are often considered as exact ones, or almost. Calculations performed 
with the same quark-quark force both in Valencia (hyperspherical 
formalism) \cite{VALE} and Grenoble (Faddeev equations) \cite{SILV} 
have evidenced discrepancies in the binding energy of the low lying baryons 
of the order of a few MeV to be compared to a total contribution of 
the order of $1 \, {\rm GeV}$ for the kinetic energy and the non-constant 
part of the potential \cite{DESP1}. The small difference may be due 
to approximations made in either approach: the restriction to the 
lowest values of the grand orbital, $K$, in the hyperspherical 
formalism and the number of amplitudes in the Faddeev approach. 
In such conditions, it is tempting to go further and calculate 
the charge and magnetic form factors of both the proton and the 
neutron and see whether the discrepancy remains at the same level 
as for the binding energy. It is what we did, with the idea to 
check the sensitivity of the results to the approach and, within 
each of them, to the truncations that are currently made in the 
calculations. We did it also with the aim to compare the results 
to the asymptotic power law in $q^{-8}$, which is expected in 
non-relativistic approaches with Coulomb or Yukawa type 
potentials \cite{ALAB}. While doing so, we have been led to elaborate 
simple models to understand our results. In view of their possible 
usefulness, some of them will be presented. A few remarks as for 
deducing the force between quarks from the nucleon electromagnetic 
form factor \cite{GAVI}, or about higher order $QCD$ corrections 
will be made. 

The plan of the paper will be as follows. The second 
section is devoted to reminding the argument for a $q^{-8}$ power 
law asymptotic behavior of the nucleon form factors in a non-relativistic 
approach. The origin of the difference with the $q^{-4}$ $QCD$ behavior 
is briefly explained. The third section shows the importance of 
the description of the wave function at short distances for an 
accurate prediction of the form factor. This is done on an hydrogenic type 
two-body system. In the fourth section, we give a few details as to 
the calculation of the nucleon wave function in the hyperspherical 
formalism or using Faddeev equations. It includes general features 
concerning these approaches as well as a few numerical results 
pertinent to the nucleon. Results for the form factors calculated with the 
quark-quark force of Bhaduri et al. \cite{BHAD} and from different 
approaches are presented in Sect. 5. The onset of their asymptotic behavior is 
discussed.  A detailed discussion about understanding some of the previous 
results is made in Sects. 6 and 7. In Sect. 8, we consider a few corrections 
that should be accounted for to make a realistic comparison with experiment. 
These include intrinsic quark form factor for the lower $q^2$ domain of 
the nucleon form factors, the consideration of interaction models with some 
three-body forces, an improved description of the spin-spin force and 
relativistic effects for the higher $q^2$ domain.   As most of the work 
presented here deals with a non-relativistic picture, there is no need to 
introduce the extra variable $Q^2$, which is often introduced in relativistic 
approaches to remedy the inconvenience of a possibly negative squared 4-momentum 
transfer. The following notation $q^2=\vec{q}^{2}$, where $q^2$ is 
also equal to the quantity $Q^2$ in the Breit frame, is therefore adopted. 
%%%%%%%%%%%%%%%%%%%%%%%222222222222222222222222%%%%%%%%%%%%%%%%%%%%%%%%%%%%%%%%
\section{ Power law expectations for the nucleon  form factor}

Predictions for the form factor of two-and three-body systems at 
high $q$ have been made long ago by Alabiso and Schierholz \cite{ALAB} 
in the case of spinless constituents, assuming a non-relativistic 
as well as a relativistic treatment. For our purpose, we 
remind some of their results that may be useful for the following.

\begin{figure}[htb!]
\begin{center}
\mbox{ \epsfig{ file=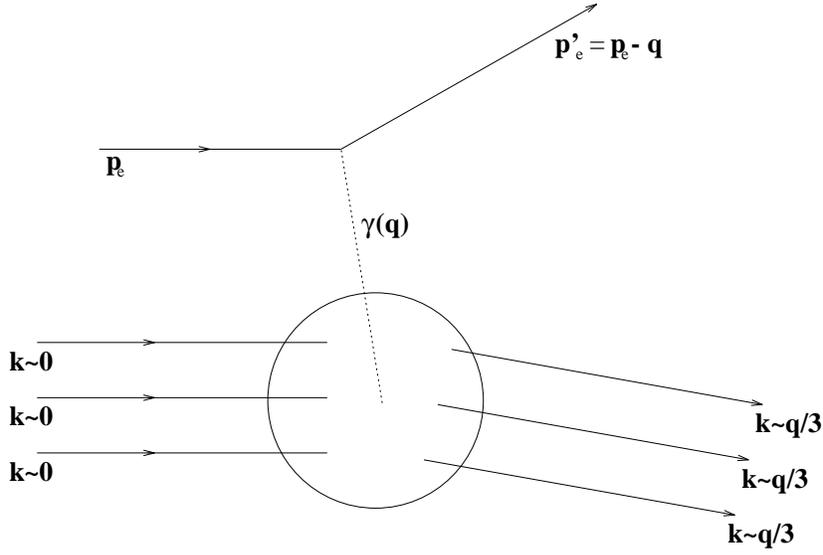, angle=270, width=10cm}}
\vspace*{-3mm}  
\end{center}
\caption{\small{Representation of electron scattering on a nucleon at rest 
(laboratory 
system). The kinematics relative to quarks 
indicated in the figure is that of a high momentum transfer where the internal 
momentum of quarks within the nucleon can be neglected.}}
\end{figure}

Electron scattering on the nucleon is represented in Fig. 1. The 
kinematics is pertinent to a high momentum transfer process. 
In the initial nucleon at rest, quarks have a small momentum, 
essentially zero, while in the final state the three-quarks 
share equally the momentum transferred to the nucleon, $\vec{q}$, 
and therefore carry the momentum $\frac{\vec{q}}{3}$. Electron 
scattering on the nucleon involves many diagrams that differ 
by their time ordering and include the exchange of two gluons 
at least. Two of them are represented in Fig. 2. The first 
one, (a), is most often shown. The virtual photon transfers 
to a quark, essentially at rest, a momentum $\vec{q}$ that is 
shared with the other two quarks by the successive exchange 
of two gluons. This diagram can be considered as representing a 
final state interaction. The second diagram, (b), is completely 
symmetric of the first one in time. It corresponds to an interaction 
effect in the initial state and has the advantage to 
show that the form factor at high $\vec{q}$ is sensitive to the 
high momentum components of the nucleon wave function, a feature that 
is not so transparent on the first diagram. We insist on this 
point because interaction effects in the initial and final 
states have often been considered on a different footing in the past 
\cite{DESP2}, with the obvious idea to simplify some calculations.

\begin{figure}[htb!]
\begin{center}
\mbox{ \epsfig{ file=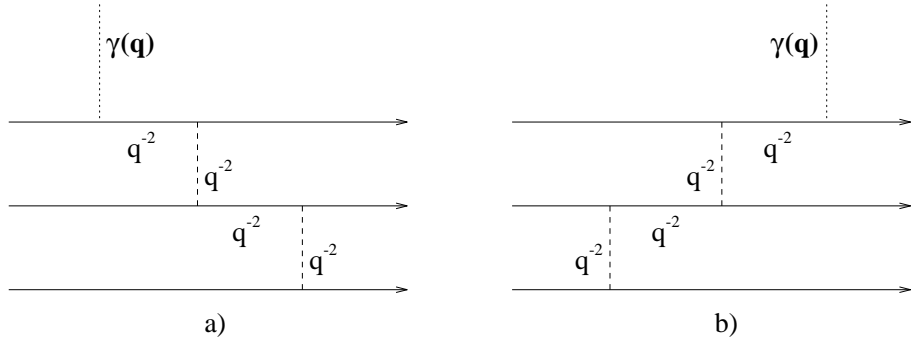, angle=270, width=12cm}}
\vspace*{-3mm}  
\end{center}
\caption{\small{Representation of two processes with gluon exchange 
contributions 
relevant to the nucleon form factor at high momentum transfer. 
The first one (a) corresponds to an interaction in the final state and the 
second (b) to an interaction in the initial state. 
The various $\frac{1}{q^2}$ factors that contribute to the asymptotic 
$\frac{1}{q^8}$ form factor in the non-relativistic 
approach and arise from the gluon or quark propagators are indicated in the 
figure. Many other diagrams with different orderings of 
the gluon exchanges contribute. Some of them are shown in Fig. 9.}}
\end{figure}  

As expected from the previous observation, the form factor at 
high $\vec{q}$ is directly proportional to the high momentum 
component of the wave function, and not to its square. Let's also 
mention that the contribution of the two diagrams, (a) 
and (b), and similar ones not shown in Fig. 2, tend to cancel 
each other in an inelastic charge scattering process where some 
energy, but no momentum, is transferred to the system (this 
simply stems from the orthogonality of the states under 
consideration).

Examination of Fig. 2  provides a quick estimate of the behavior 
of the form factor (or the $\gamma^* N \rightarrow N$ 
amplitude) at high $q^2$. Each gluon propagator introduces a 
factor $\frac{1}{q^2}$. Furthermore, in the non-relativistic 
limit, each intermediate quark also introduces a factor $\frac{1}{q^2}$. 
Hence the form factor is expected to have the following behavior:
\begin {equation}
F^{n.r.}_{N} (q^{2})_{q^{2} \rightarrow \infty} \propto q^{-8}
\end{equation}
The above behavior may be invalidated if the interaction contains terms 
which do not scale like the square inverse power of the momentum at high 
momenta, such as a gaussian type force, or also if higher order 
effects are not small enough, in which case larger $q^{2}$ 
should be considered. This is an important issue, which is aimed to be 
considered in Sects. 6, 7, 8.2 .

\begin{figure}[htb!]
\begin{center}
\mbox{ \epsfig{ file=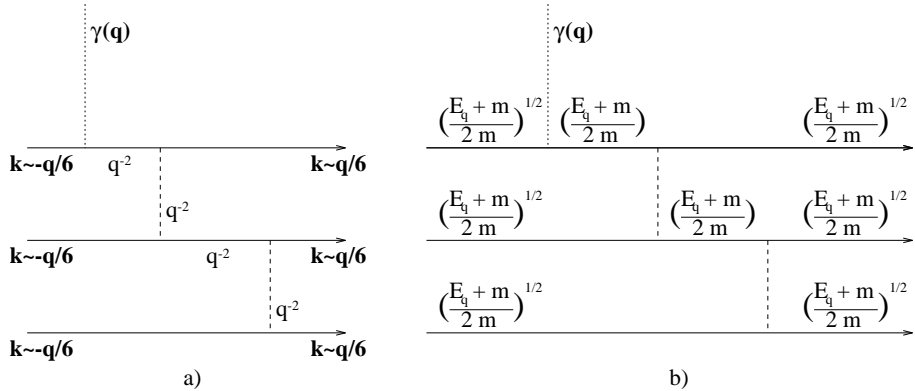, angle=270, width=12cm}} 
\vspace*{-3mm} 
\end{center}
\caption{\small{Representation in the Breit frame of a process contributing to 
the asymptotic form factor for a system of spin $1/2$ particles 
exchanging spin 1 bosons in both a non-relativistic approach 
(nucleon case: $\frac{1}{q^8}$ asymptotic behavior, diagram a) and
a relativistic one ($\frac{1}{q^4}$ asymptotic behavior, diagram b). The 
diagram b) only shows the extra factors responsible for a difference with the 
first case. For simplicity, the same q factor appears everywhere, but it should 
be understood that the appropriate fraction of q is to be used, depending on 
the diagram and on the position in that diagram. The representation in the 
lab. frame is also possible, but some caution is required in determining the 
asymptotic behavior as the time component of the four-momentum transfer,
$q_0$, varies like $q^2$. Results for the non-relativistic case also hold 
for spinless particles exchanging spinless bosons.}}
\end{figure}  

The difference with the $QCD$ power law expectation, $q^{-4}$, deserves some 
explanation. First of all, we notice that the form factor of a system of 
three spinless particles, interacting via usual scalar boson exchanges 
and treated relativistically, does have a behavior given by Eq. (1), as can be 
checked from Fig. 2 with the quark propagators substituted by scalar particle 
propagators and gluon exchanges substituted by scalar boson exchanges.
However, if one considers spin 1 boson exchange, at each particle-boson-particle 
vertex, a factor q is introduced as a consequence of the vector coupling and a 
$q^{-4}$ behavior is obtained. In the non-relativistic limit, Eq. (1) holds for 
both types of boson exchanges. For quarks, spin $\frac{1}{2}$, the exchanges of 
scalar or vector bosons are also equivalent in the non-relativistic limit, 
giving rise to a  $q^{-8}$ power law (Fig. 3a). In the relativistic treatment, 
the spin 1 case evidences  five extra factors, $q$, related to the definition 
of the quark spinors (see Fig. 3b), of which one has to be absorbed 
in the definition of the initial and final nucleon spinors. These 
factors, which arise from boosting the nucleon at rest, have a well 
defined origin and, therefore, can be accounted for, at least 
approximately, when a more realistic estimate of the form factor or 
a comparison with experiment is intended to be performed. They should 
not be confused with factors originating from the 
dynamics. The case of spin zero bosons coupling to spin $\frac{1}{2}$ particles 
may also be considered. The above analysis needs to be refined but a 
$q^{-4}$ behavior is expected too.

Summarizing this section, a non-relativistic calculation of the nucleon form 
factor does provide a power law $q^{-8}$ at high $q^{2}$ (provided that the 
interaction between quarks is mediated by the usual exchange of bosons). The 
difference with the prediction, $q^{-4}$, is due to both relativity and the 
nature of the $QCD$ interaction, which involves the coupling of spin 1 bosons 
to spin $\frac{1}{2}$ quarks.  
%%%%%%%%%%%%%%%333333333333333333333%%%%%%%%%%%%%%%%%%%%%%%%%%%%%%%%%%%%%%%%%%%

\section{Asymptotic form factor of a two-body system}

It is well known that the form factor of a two-body system at 
high momentum transfer is very sensitive to its short-range 
description. In a few cases, a close relation can be established. 
Some results that may be found in the literature 
are reminded here for an hydrogenic type atom (see for instance \cite{MILL}). 
We emphasize points that could be relevant for the understanding of 
results concerning the three-quark system. 

Discarding the center of mass motion, the wave function of the first 
s-state of the hydrogenic atom obeys the following Schr\"odinger equation:
\begin{equation}
\left(  \frac{\vec{p}^2}{2m_r} -\frac{\alpha}{r}-E \right) \; \psi(\vec{r}) = 0,
\end{equation}
where $m_r$ is the reduced mass, $m_r=\frac{m_1 m_2}{m_1+m_2}$, - 
$\frac{\alpha}{r}$ the Coulomb interaction 
and $ r = |\vec{r}_1-\vec{r}_2 |$. The normalized solution is given by:
\begin{equation}
\psi(\vec{r}) = \frac{2}{\sqrt{4 \pi}}  (\alpha m_r)^{3/2} e^{-\alpha m_r r},
\end{equation}
while the corresponding binding energy is $E = - \frac{\alpha^2m_r}{2}$.
Assuming now that only the particle 1 carries a unit charge, the form 
factor of this state is easily calculated:
\begin{equation}
F(\vec{q}^{2})= \int d\vec{r} \; \psi^2(r) \; e^{i \vec{q}.\vec{r} 
\frac{m_2}{m_1+m_2} } \\
= \frac{1}{ ( 1 + \frac{\vec{q}^{2}}{4\alpha^2m^2_1} )^2 }.
\end{equation}
This expression shows that the form factor scales like $q^{-4}$ at high momentum 
transfers. A direct relation of this behavior to a perturbative calculation is 
obtained in momentum space. There, the form factor $F(\vec{q}^{2})$ gets the 
expression:
\begin{equation}
F(\vec{q}^{2}) = \int \frac{d\vec{k}}{(2 \pi)^3} \; \varphi(\vec{k}) 
\; \varphi( \vec{k}+\vec{q} \; \frac{m_2}{m_1+m_2} ),
\end{equation}
where $\varphi(\vec{k})$ is the Fourier transform of $ \psi(r)$:
\begin{equation}
\varphi(\vec{k}) = \int d\vec{r} \; e^{-i \vec{k}.\vec{r} } \; \psi(\vec{r}) \\
=\sqrt{4 \pi} \frac{4(\alpha m_r)^{5/2}}{ (k^2+\alpha^2m^2_r)^2}
=\varphi(-\vec{k}).
\end{equation}
As $\varphi(\vec{k})$ is concentrated at small values of  $\vec{k}$, two 
domains contribute equally to the integral in Eq. (5), around $\vec{k} =0$, 
and $\vec{k} = -\vec{q}\;\frac{m_2}{m_1+m_2}$, which allows one to write:
\begin{equation}
F(\vec{q}^{2}) \simeq 2 \, \varphi(\vec{q} \; \frac{m_2}{m_1+m_2}) \int 
\frac{d\vec{k}}{(2\pi)^3} \; \varphi(\vec{k}).
\end{equation}
Now, the integral in (7) is nothing but the wave function in r-space at the 
origin, $\psi(0)$, while $\varphi(\vec{k})$ obeys the Schr\"odinger equation, 
which in momentum space reads:
\begin{equation}
( \frac{\vec{k}^2-2m_rE}{2m_r} ) \; \varphi(\vec{k})= \int 
\frac{d\vec{k'}}{(2\pi)^3} 
\frac{4\pi \alpha }{ (\vec{k}-\vec{k'})^2 } \; \varphi(\vec{k}').
\end{equation}
Again using the property that  $\varphi(\vec{k}) $ is concentrated at small 
values of $k$ allows one to write:
\begin{equation}
\varphi(\vec{k})_{k \rightarrow \infty} = \frac{8\pi \alpha m_r }{k^4} \psi(0).
\end{equation}
Gathering results given by Eqs. (7) and (9), one gets:
\begin{equation}
F(\vec{q}^{2})_{\vec{q}^{\,2} \rightarrow \infty} = \frac{16\pi \alpha m_r 
}{q^4} \; (\frac{m_1+m_2}{m_2})^4 \; \psi^2(0),
\end{equation}
which is in agreement with Eq. (4) in the same limit and, at the same time, 
shows the sensitivity of the form factor to the radial wave function at the 
origin. This one contains non-perturbative effects. 

A more precise statement can be made by looking back at the expression of the 
form factor given by Eq. (4) and making an expansion of $\psi(r)$ around the 
origin:
\begin{equation}
\psi^2(\vec{r}) = \psi^2(0) + 2 \, r \; \psi(0) \; \psi'(0)+...
\end{equation}
Inserting this expression in Eq. (4), one obtains from the first term a 
$\delta(\vec{q})$ function, which obviously vanishes at high $q$. From terms 
beyond the first derivative, dots in Eq. (11), one obtains contributions that 
tend to zero faster than $q^{-4}$ (dimensional argument). As a dominant 
contribution at high $q$, one is therefore left with:
\begin{equation}
F(\vec{q}^{2})_{q^2 \rightarrow \infty}=-\frac{16\pi}{q^4}  
(\frac{m_1+m_2}{m_2})^4 \; \psi(0) \psi'(0).
\end{equation}
In order to get this result, we employed the relation:
\begin{equation}
[\int dr \, r^3 \; j_0(qr)]_{q \neq 0} 
  = {\rm lim}_{\epsilon \rightarrow 0} [\int dr 
r^3 j_0(qr) e^{-\epsilon r}]_{q \neq 0} = -\frac{2}{q^4}.
\end{equation}

Equation (12) shows the sensitivity of the form factor at 
high $q$ to the radial wave function at the origin, but also to 
its derivative at the same point. The calculation of the form factor 
at high $q$ therefore supposes to correctly determine the 
slope of the wave function at the origin, which, itself, is determined 
by the Coulomb potential, independently of 
the energy of the state under consideration. This can be checked 
on the first radial, $l =0$, excitation, whose 
wave function and form factor are respectively given by:
\begin{equation}
\psi^{*}(\vec{r}) = \frac{1}{ \sqrt{4\pi} }   \frac{ (\alpha m_r)^{3/2} }{ 
\sqrt{2} } 
e^{-\frac{1}{2}\alpha m_r r} \; (1-\frac{\alpha m_r r}{2})
\end{equation}
\begin{equation}
F^{*}(\vec{q}^{2}) = \frac{   (1-\frac{q^2}{(\alpha m_1)^2} ) (1-2 
\frac{q^2}{(\alpha m_1)^2}) }{ ( 1+\frac{q^2}{(\alpha m_1)^2} )^4   }.
\end{equation}

In the limit $q \rightarrow \infty$, the above form factor exhibits for the 
$1/q^4$ term a coefficient different from that obtained for the ground state 
form factor, $F(\vec{q}^2)$, given by Eq. (4), but this only reflects the 
difference in 
the value of the square of the wave functions at the origin 
for the states under consideration, $ \frac{(\alpha m_r)^3}{8\pi} $ 
and $ \frac{(\alpha m_r)^{3}}{\pi} $. For a general potential, the $q^{-4}$  
behavior of the form factor holds provided that the interaction 
inserted in the Schr\"odinger equation is as singular at small 
distances as a Coulomb or Yukawa potential ($\propto \frac{1}{r}$). 
It has to do with the fact that the first derivative of the wave 
function at the origin is determined by this piece of the 
interaction. It is lost when the singularity is 
weaker. It can be checked for instance that the form factor corresponding 
to the wave function, $e^{-br}(1+br)$, cooked up in such a way 
to behave like $1+0(r^2)$ at small distances has no $q^{-4}$ 
component at large $q$, the first non-zero component being $q^{-6}$. 
An other interesting example is the gaussian wave 
function, $\psi(r) = \pi^{-3/4} e^{-b^2 r^2/2}$, whose form factor: 
\begin{equation}
F^{G}(q^2) = e^{-q^2/4b'^2},\;\;\;{\rm with}\; b'=b \; \frac{m_1+m_2}{m_2},
\end{equation}
has obviously no $q^{-4}$ component at large $q$, as expected from Eq. (12).

While the first derivative of the wave function at the origin 
is determined by the $1/r$ term of the potential, the second one 
furthermore depends on the ``binding energy'', i.e. the total mass 
of the system minus the masses of the constituents and other constant 
terms (see Eq. (96) for the corresponding situation in the three-body 
case). It is therefore state dependent. With this respect, we should 
mention that hadronic systems to be discussed here are sensitive to a confining 
potential. This one does not directly influence the calculation of the second 
derivative at $r=0$, but it does indirectly through its contribution to the 
total energy of the system. As will be seen in Sects. 4.3 and 6.1, this one is 
making the second derivative smaller than for a pure Coulomb problem, with 
possibly an opposite sign in some cases. This feature may have consequences for 
the rapidity of the onset of the asymptotic power law behavior of the form 
factor of hadronic systems and for the manner how it occurs.
%%%%%%%%%%%%%%%%%%%44444444444444444444444444%%%%%%%%%%%%%%%%%%%%%%%%%%%%%%%%%%
\section{Description of the nucleon wave function  in terms of quarks}
Determining a nucleon wave function in terms of quarks has been 
done in many papers, see refs. \cite{BHAD2, ALVA} for general presentations and 
refs. \cite{ISGU,SILV,CARL,CAPS,STAS,BASD,GIAN,METS,SALA,BOFF,GRAZ} for more 
specific works. We nevertheless remember a few details relative to our own 
calculations \cite{DESP1} and, especially, to some of the approximations that 
have been made. The quark-quark force employed in our 
calculations is that of Bhaduri et al. \cite{BHAD}:
\begin{eqnarray}
V_{BHA} = \frac{1}{2} \sum_{i<j} \left( - \frac{\kappa}{r_{ij}} + 
\frac{r_{ij}}{a^{2}} - D
+ \frac{ \kappa_{\sigma} }{ m_{i} m_{j} } 
\frac{ {\rm exp}(-r_{ij}/r_{0}) }{ r^2_0 r_{ij} } 
\vec{\sigma}_{i}.\vec{\sigma}_{j} \right) \nonumber \\
\equiv \;\; \sum_{i<j} V_{ij} \;\; \equiv \;\;\sum_{i<j} 
V(\vec{r_i} - \vec{r_j}), \hspace*{3.9cm}
\end{eqnarray}
where $\kappa=\kappa_{\sigma}=102.67 \, {\rm MeV}\, {\rm fm}, \; a=0.0326 \, 
{\rm MeV}^{-1/2} \, {\rm fm}^{1/2}, \; r_0=0.4545 \, {\rm fm}$, 
$m_i=m_j=m_q=337 \, {\rm MeV}, \; D=913.5 \, {\rm MeV}$.
With these definitions, distances are expressed in units of fermi and the 
potential in MeV. For characterizing the strength of the 
spin-spin force, we will most often refer to the following quantity:
\begin{equation}
\kappa '_{\sigma}= \frac{ \kappa_{\sigma} }{ (r_0 m_q)^2}=1.66 \, 
\kappa_{\sigma} .
\end{equation}
Fitted on the meson spectrum, the above interaction provides a 
reasonably good account of the baryon spectrum \cite{SILV}, although it 
misses the Roper resonance. It is reminded however that the choice of 
the model is not essential here and that references to other quark 
interaction models will be made in any case, see Sects. 4.3 and 8.2. 

Having neither spin-orbit nor tensor component, 
the force given by Eq. (17) has the particular feature to conserve the 
spin. The baryon wave function can thus be factorized into a spin and an 
orbital part. The simplification may be a drawback with some respects, 
for a discussion of the helicity conservation in $QCD$ for 
instance. Here, on the contrary, it represents an advantage as it 
avoids unnecessary admixture of the effects we are looking at 
with other ones that are irrelevant for our purpose.

Four spin-isospin quark wave functions with the spin and isospin 
of the nucleon, $S=1/2, T=1/2$, are available. In notations of 
ref. \cite{SILV}, they are:
\begin{eqnarray}
 | S > \;\; = \frac{1}{\sqrt{2}}  | \chi^0 \eta^0 + \chi^1 \eta^1 >,\nonumber \\
 | A >\;\;  = \frac{1}{\sqrt{2}}  | \chi^0 \eta^1 - \chi^1 \eta^0 >,\nonumber \\
 |MS > = \frac{1}{\sqrt{2}}  | \chi^0 \eta^0 - \chi^1 \eta^1 >,\nonumber \\
 |MA > = \frac{1}{\sqrt{2}}  | \chi^0 \eta^1 + \chi^1 \eta^0 >,
\end{eqnarray}
where $\chi^0(\eta^0)$ and $\chi^1(\eta^1)$ correspond to quarks 1 and 2 
coupled with an intermediate spin (isospin) equal respectively to  0 
and 1. All of them have a definite character under the exchange 
of quarks 1 and 2 (symmetric : $  | S> $ and $ | MS>$, 
antisymmetric: $ |A > $ and  $ |MA >$ ). Under the exchange 
of quarks 1, 2 and 3, the spin-isospin wave functions $  |S > $ 
and $ |A >$ are respectively symmetric and antisymmetric. The two 
other ones, $ | MS > $ and $ | MA > $, have a mixed character 
and transform into a combination of each other under the exchange 
of quarks 1 and 3, or 2 and 3, see Eqs. (127, 128). These wave functions 
have to be combined with spatial wave functions that have appropriate 
transformation properties under the exchange of quarks 1, 2 
and 3 : complete symmetry for  $ |S > $, complete antisymmetry 
for $ |A > $ and mixed symmetry for $ |MS > $ and $ |MA > $. 
The antisymmetry is ensured, as well known, by the color wave 
function, which we omit to write down as its effect 
factorizes out. The total wave function of momentum $\vec{P}$ may thus be 
written:
\begin{eqnarray}
\Psi _{\vec{P}} (\vec{r}_1,\vec{r}_2,\vec{r}_3)= 
\psi_{S}(\vec{r}_1,\vec{r}_2,\vec{r}_3) \, | S > + 
\psi_{A} ( \vec{r}_1,\vec{r}_2,\vec{r}_3 ) \, | A > \nonumber \\ + 
\frac{1}{\sqrt{2}} \; \left( \psi_{MS} ( \vec{r}_1,\vec{r}_2,\vec{r}_3) \, |M S> 
 + \psi_{MA} (  \vec{r}_1,\vec{r}_2,\vec{r}_3 )  \, |MA > \right) .
\end{eqnarray}
It obeys the Schr\"odinger equation:
\begin{eqnarray}
\left( \frac{\vec{p}_1^2}{2m_q} + \frac{\vec{p}_2^2}{2m_q} 
+ \frac{\vec{p}_3^2}{2m_q} + 
V(\vec{r}_1-\vec{r}_2) + V(\vec{r}_2-\vec{r}_3) + 
V(\vec{r}_3-\vec{r}_1) - E \right) \nonumber \\ \Psi_{\vec{P}} 
(\vec{r}_1,\vec{r}_2,\vec{r}_3)=0.
\end{eqnarray}

For the considered potential, the center of mass motion factorizes out. Being 
irrelevant for the description of the nucleon, we omit from now on the 
corres\-ponding phase factor as well as any reference to the total momentum, 
implicitly assuming that the system is at rest, with $\vec{P}=0$. As to the 
internal wave functions 
$\psi_{S,A,MS,MA}(\vec{r}_1,\vec{r}_2,\vec{r}_3)$, they may be expressed 
in terms of Jacobi variables, $\vec{r}_1 -\vec{r}_2$ and $ \vec{r}_3- 
\frac{ \vec{r}_1+\vec{r}_2 }{2} $, or any combination of them, 
see end of App. C. 
The normalization of the nucleon wave function is chosen to be:
\begin{eqnarray}
\int d(\vec{r}_1 -\vec{r}_2) \;d( \vec{r}_3 -  \frac{\vec{r}_1+\vec{r}_2}{2} )
\left( \psi^{2}_{S}(\vec{r}_1,\vec{r}_2,\vec{r}_3) + 
\psi_{A}^{2}(\vec{r}_1,\vec{r}_2,\vec{r}_3) \;\;\;\;\;\;\;\;\; \nonumber 
\right. 
\\  \left.+ 
\frac{1}{2} ( \psi^{2}_{MS}(\vec{r}_1,\vec{r}_2,\vec{r}_3) + 
\psi^2_{MA}(\vec{r}_1,\vec{r}_2,\vec{r}_3) ) \right) = 1.
\end{eqnarray}
The symmetry properties of $\psi_{MS}(\vec{r}_1,\vec{r}_2,\vec{r}_3)$ and
$\psi_{MA}(\vec{r}_1,\vec{r}_2,\vec{r}_3)$ under the exchange of 
particles 1 and 3, or 2 and 3:
\begin{eqnarray}
\psi_{MS}(\vec{r}_3,\vec{r}_2,\vec{r}_1)= - \frac{1}{2} 
\psi_{MS}(\vec{r}_1,\vec{r}_2,\vec{r}_3)+ \frac{\sqrt{3}}{2} \; 
\psi_{MA}(\vec{r}_1,\vec{r}_2,\vec{r}_3),\nonumber \\
\psi_{MA}(\vec{r}_3,\vec{r}_2,\vec{r}_1)= + \frac{1}{2} 
\psi_{MA}(\vec{r}_1,\vec{r}_2,\vec{r}_3)+ \frac{\sqrt{3}}{2} \; 
\psi_{MS}(\vec{r}_1,\vec{r}_2,\vec{r}_3),
\end{eqnarray}
also imply the relation:
\begin{equation}
\int d(\vec{r}_1 -\vec{r}_2) d(\vec{r}_3 - \frac{\vec{r}_1+\vec{r}_2}{2}) 
\left( 
\psi^{2}_{MS}(\vec{r}_1,\vec{r}_2,\vec{r}_3) - \psi^{2}_{MA} 
(\vec{r}_1,\vec{r}_2,\vec{r}_3) \right) = 0.
\end{equation}
%%%%%%%%%%%%%%%%%%%%%%%%%%%%%
\subsection{Faddeev approach}
In the Faddeev approach, the wave function $\Psi(\vec{r}_1,\vec{r}_2, 
\vec{r}_3) $ is written as a sum of 3 terms:
\begin{equation}
\Psi(\vec{r}_1 ,\vec{r}_2 ,\vec{r}_3)  = \Psi_{12,3} + 
\Psi_{13,2}+\Psi_{23,1},
\end{equation}
where $\Psi_{12,3}$ is symmetric in the exchange of particles 1 
and 2 while $\Psi_{13,2}$ and $\Psi_{23,1}$ are obtained 
from $\Psi_{12,3}$ by performing the corresponding permutations, 
so that $\Psi(\vec{r}_1 ,\vec{r}_2,\vec{r}_3)$ is 
symmetrical in the exchange of particles 1, 2, and 3. The amplitude 
$\Psi_{12,3}$ obeys the equation:
\begin{equation}
(\frac{\vec{p}_1^2}{2m_q} + \frac{\vec{p}_2^2}{2m_q} +\frac{\vec{p}_3^2}{2m_q} 
+ V(\vec{r}_1 -\vec{r}_2) - E ) \Psi_{12,3} 
= - V(\vec{r}_1 -\vec{r}_2) ( \Psi_{13,2}+\Psi_{32,1} ).
\end{equation}

For calculations, $\Psi_{12,3} $ is developped on the spin-isospin basis (19), 
or an equivalent one, and the spatial part, which in the present 
case has a total orbital angular momentum $L=0$, is decomposed in terms of the 
spherical harmonics relative to the Jacobi variables  
$\vec{x} = \vec{r}_1 -\vec{r}_2 $ and $\vec{y} = \frac{2}{\sqrt{3}} ( 
\vec{r}_3-\frac{\vec{r}_1+\vec{r}_2}{2} )$:
\begin{equation}
\psi_{12,3} (\vec{r}_1 -\vec{r}_2,\vec{r}_3-\frac{\vec{r}_1+\vec{r}_2}{2})
= \sum^{\infty}_{\lambda=0} \sum^{\lambda}_{m=-\lambda} 
Y_{\lambda}^{m*}(\hat{x}) \, Y^m_{\lambda}(\hat{y}) \, \varphi_{\lambda}(x,y).
\end{equation}

The sum includes even and odd values of $\lambda$. Through determining  
combinations with well defined symmetry character (see equation below), these 
values will be associated with spin-isospin wave functions that have the same 
parity  under the exchange of particles 1 and 2, respectively $(  |S >, |MS>)$ 
and $ ( |A >,| MA > )$. In practice, only a finite number is retained. The 
minimum is to retain the lo\-west value, $\lambda = 0$, which gives 
two amplitudes in $\Psi_{12,3}$ corresponding to the two spin-isospin 
components, $ | \chi^0 \eta^0 >$ and $ | \chi^1 \eta^1>$, or 
equivalently $  |S >$ and $ | MS>$. The restriction to these two 
components, $\varphi_0^0(x,y)$ and $\varphi_0^1(x,y)$, respectively 
associated with the spin-isospin wave functions $ | \chi^0 \eta^0>$ 
and $ | \chi^1 \eta^1>$, is motivated by the expected dominance of the 
interaction in these channels, which should be somewhat corrected 
however for the long range of the gluon exchange. As it may be useful, 
we give here the full expression of the wave function:
\begin{eqnarray}
\Psi_{{\rm F}}(\vec{r}_1,\vec{r}_2,\vec{r}_3)
     = \;\;\;\; \left[ \frac{\psi^0_{12,3}+\psi^1_{12,3}}{\sqrt{2}}+ (1 
\leftrightarrow 
3) + (2 \leftrightarrow 3) \right]  |S >\nonumber \;\;\;\;\;\;\;\;\; \\
     +  \left[ \frac{\psi^0_{12,3}-\psi^1_{12,3}}{\sqrt{2}}-\frac{1}{2} (1 
\leftrightarrow 3) - \frac{1}{2} (2 \leftrightarrow 3) \right] | MS >
\nonumber \\
     + \; \frac{\sqrt{3}}{2} \left[ 
\frac{\psi^0_{32,1}-\psi^1_{32,1}}{\sqrt{2}}-(1 \leftrightarrow 2) \right]  |MA 
> , \;\;\;\;\;\;\;\;\;\;\;\;\;
\end{eqnarray}
where, in the two component case:
\begin{eqnarray}
\psi^{0,1}_{12,3} & = & \varphi^{0,1}_{0}(x,y),\nonumber 
\;\;\;\;\;\;\;\;\;\;\;\;\;\;\;\;\;\;\;\;\;\;\;\;\;\;\;\;\;\;\;\;\;\;\;\;\;\;\;
\; \\
\psi^{0,1}_{32,1} &  = & \varphi^{0,1}_{0}( \; |\frac{1}{2} \vec{x}+ 
\frac{\sqrt{3}}{2} \vec{y}|\; , |-\frac{1}{2} \vec{y} +\frac{\sqrt{3}}{2} 
\vec{x}|\;  ),
 \nonumber \\
\psi^{0,1}_{13,2} & =  & \varphi^{0,1}_{0}( \; |\frac{1}{2} \vec{x}- 
\frac{\sqrt{3}}{2} \vec{y}|\; , |-\frac{1}{2} \vec{y} -\frac{\sqrt{3}}{2} 
\vec{x}|\; ).
\end{eqnarray}
While values of $\lambda$ higher than 0 are neglected in the above minimal 
calculation of the Faddeev amplitude $\Psi_{12,3}$, their contribution to the 
total wave function 
are not totally neglected. They appear through the exchange terms $\Psi_{32,1}$ 
and $\Psi_{13,2}$, whose expansion in terms of spherical 
harmonics $ \sum Y_{\lambda '}^{m'*}(\hat{x}) \; Y_{\lambda '}^{m'} (\hat{y})$ 
and spin-isospin wave functions, in principle infinite, 
has been limited to 10 terms in calculating the observables 
(form factors) presented in this paper. A calculation consisting 
in retaining 8 amplitudes in the expansion of the Faddeev 
amplitude $\Psi_{12,3}$ (not to be confused with the wave function) 
has also been performed. The comparison with the 2 amplitude 
calculation will provide an interesting test. Indeed an important 
issue is to know whether some truncation in the Faddeev 
amplitude affects the power law behavior of the form factor 
at high momentum transfers.
%%%%%%%%%%%%%%%%%%%%%%%
\subsection{Hyperspherical formalism approach}
In the hyperspherical formalism, a different choice of the variables 
is made (see App. A). The modulus of the Jacobi variables are expressed 
in terms of an hyperradius, $\rho$, and an extra angle, $\phi \,$:
\begin{eqnarray}
\frac{x}{\sqrt{2}} = \rho_{12} = \rho \; {\rm sin} \phi, \nonumber \\
\frac{y}{\sqrt{2}} =  \; \rho_{3} \; = \rho \; {\rm cos} \phi.
\end{eqnarray}
In terms of these variables, the total wave function reads:
\begin{equation}
\Psi_{ {\rm HH} } \equiv |\psi(\rho,\Omega)> = \sum_{K,L,sym} 
\psi_{K,L}(\rho) 
\; \left[ Y_{[K,sym]}^{(L,M_{L})}(\Omega) \; |sym > \right]_{J^P},
\end{equation}
where $  | sym >$ represents the spin-isospin wave functions given by (19) 
and $\Omega$ the set of the various angles relative to the unit vectors, 
$\hat{\rho}_{12}$ and $\hat{\rho}_3$ together with the hyperspherical angle 
$\phi$. $Y_{[K,sym]}^{(L,M_{L})}(\Omega)$ 
stands for the hyperspherical harmonic (HH) of definite symmetry, orbital 
angular momentum and quantum numbers specified by K. The hyperradial wave 
functions, $\psi_{K,L}(\rho)$, satisfy a (infinite) set of coupled equations
written in detail in App. A. 

In practice, as for solving the Faddeev equation, a truncation 
is required. Two amplitudes have been retained. They involve 
the totally symmetric spin-isospin wave function $ |S > $ and 
a combination of the mixed ones, $ | MS >$ and  $ |MA>$. They respectively have 
the lowest allowed $K$ value, 0 and 2. By identifying 
\begin{eqnarray}
\psi_{0,L=0}(\rho) \equiv \psi_1 (\rho),\\
\psi_{2,L=0}(\rho) \equiv \psi_3 (\rho), \; 
\end{eqnarray}
an explicit expression of the wave function is:
\begin{eqnarray}
\Psi_{ {\rm HH} }( \vec{r}_1,\vec{r}_2,\vec{r}_3 ) \simeq 
\frac{1}{\pi\sqrt{\pi}} 
\left[  \psi_1 (\rho) | S > \;\;\;\;\;\;\;\;\;\;\;\;\;\;
\;\;\;\;\;\;\;\;\;\;\;\;\;\;\;\;\;\;\;\;\;\;\;\;\;\;\;\;\;\;\;\;\; \nonumber 
\right. \\
\left. + \sqrt{2} \; \psi_3(\rho) 
\left( ( {\rm cos}^2 \phi - {\rm sin}^2 \phi )| MS >- 2 \; 
{\rm sin} \phi \;  {\rm cos} \phi \; \hat{\rho}_{12}.\hat{\rho}_3 |MA >
\right) \right].
\end{eqnarray}
The front factor ensures the normalization over the various angles, 
\begin{equation}
\int \frac{d\Omega}{\pi^3}=1, 
\end{equation}
and the radial wave functions are normalized as:
\begin{equation}
\int d\rho \,\rho^5 \; (\psi_1^2(\rho) + \psi_3^2(\rho))=1.
\end{equation}

As for the Faddeev approach, one may wonder whether the truncation 
of the wave function has some consequence for the prediction of 
the form factor at high momentum transfers. From now on, it is 
remarked that the description of the system with the totally symmetric 
radial wave function $\psi_1(\rho)$ is more economical but
poorer than the corresponding one in the Faddeev approach, first bracket 
on the r.h.s. of Eq. (28). The latter contains extra terms that, 
in the hyperspherical formalism, suppose to introduce components with 
$K=4,\, 6....$. This can be seen by realizing that the completely 
symmetric hyperspherical harmonic for $K=0$ only contains 
$Y_{\lambda=0}(\hat{x}) \, Y_{\lambda=0}(\hat{y})$, whereas the Faddeev 
amplitudes $\Psi_{32,1}$ and $\Psi_{13,2}$ in the simplest case (2 amplitude 
calculation) contain terms with $\lambda \neq 0$. Let's also notice that 
the radially symmetric component, $\psi_1(\rho)$, is the dominant one, while the 
mixed symmetry component, $\psi_3(\rho)$, which is smaller, is directly 
determined by the spin-spin interaction.
%%%%%%%%%%%%%%%%%%%%%%%%%%%%%%%%%%%4.3
\subsection{Static properties and wave functions}
\begin{table}[htb!]
\caption{\small{Static properties pertinent to the description of the nucleon, 
as 
calculated from different approaches with the same force: Faddeev equations 
with 2 and 8 amplitudes (first and second columns, F) and hyperspherical 
harmonic formalism (HH) with a restriction to the $K$ values, 0 and 2 (third 
column). Results calculated in the hyperspherical harmonic formalism with $K=0$ 
are given separately in the fourth column for a different force. This one only 
includes the Coulomb and confining parts of the Bhaduri et al.'s potential (no 
spin-spin component). The listed quantities are successively the mass, the 
square matter radius, the mixed symmetry probability, the proton and neutron 
square ``charge'' radius, and the density of the nucleon at the origin.}}
\begin{center}
\begin{tabular}{lcccccc}
\hline
      &  (F)    &  (F)    &             (HH)                & &  (HH) \\ [1.ex]
      & $2A  $ & $8A $ & $\kappa_{\sigma'}=1.66 \,\kappa $ & \hspace*{1cm}
& $\kappa_{\sigma'}=0 $ 
\\ [2.ex]
\hline
                      &          &          &        & &       \\
 $M_N({\rm MeV}) $          &  1031    &   1020   & 1039   & &  1201  \\ [2.ex]
 $<r^2_m>({\rm fm}^2)$      &  0.219   &  0.219   & 0.218  & & 0.255  \\ [2.ex] 
 $P_{S'} $            & 1.0\%    &   2.1\%  &  1.5\% & &    0   \\ [2.ex]
 $<r^2_p>({\rm fm}^2)$      &  0.234   &  0.243   & 0.238  & &  0.255 \\ [2.ex] 
 $<r^2_n>({\rm fm}^2)$      & -0.015   &  -0.024  & -0.020 & &    0   \\ [2.ex]
 $\psi^2(0)({\rm fm}^{-6})$ &   386    &  386     & 398  & &  136   \\ [2.ex]
 
\hline
\end{tabular}
\end{center}
\end{table}

We give in Table 1 a few results relative to the nucleon wave function 
calculated with the Bhaduri et al.'s force in the different 
approaches : Faddeev with 2 and 8 amplitudes and hyperspherical 
formalism. They concern the mass of the nucleon and its square 
matter radius, the proton and neutron charge squared radii and 
the mixed symmetry probability. Results where the spin-spin force 
in the hyperspherical formalism is neglected are also given.
The comparison of the results incorporating the spin-spin interaction 
does not evidence much difference for the mass or matter radius. The 
sensitivity to the approximation is more important for the neutron 
charge squared radius and the mixed symmetry probability, which 
both involve the spin-spin interaction at the first and second order 
respectively.

\begin{figure}[htb!]
\begin{center}
\mbox{ \epsfig{file=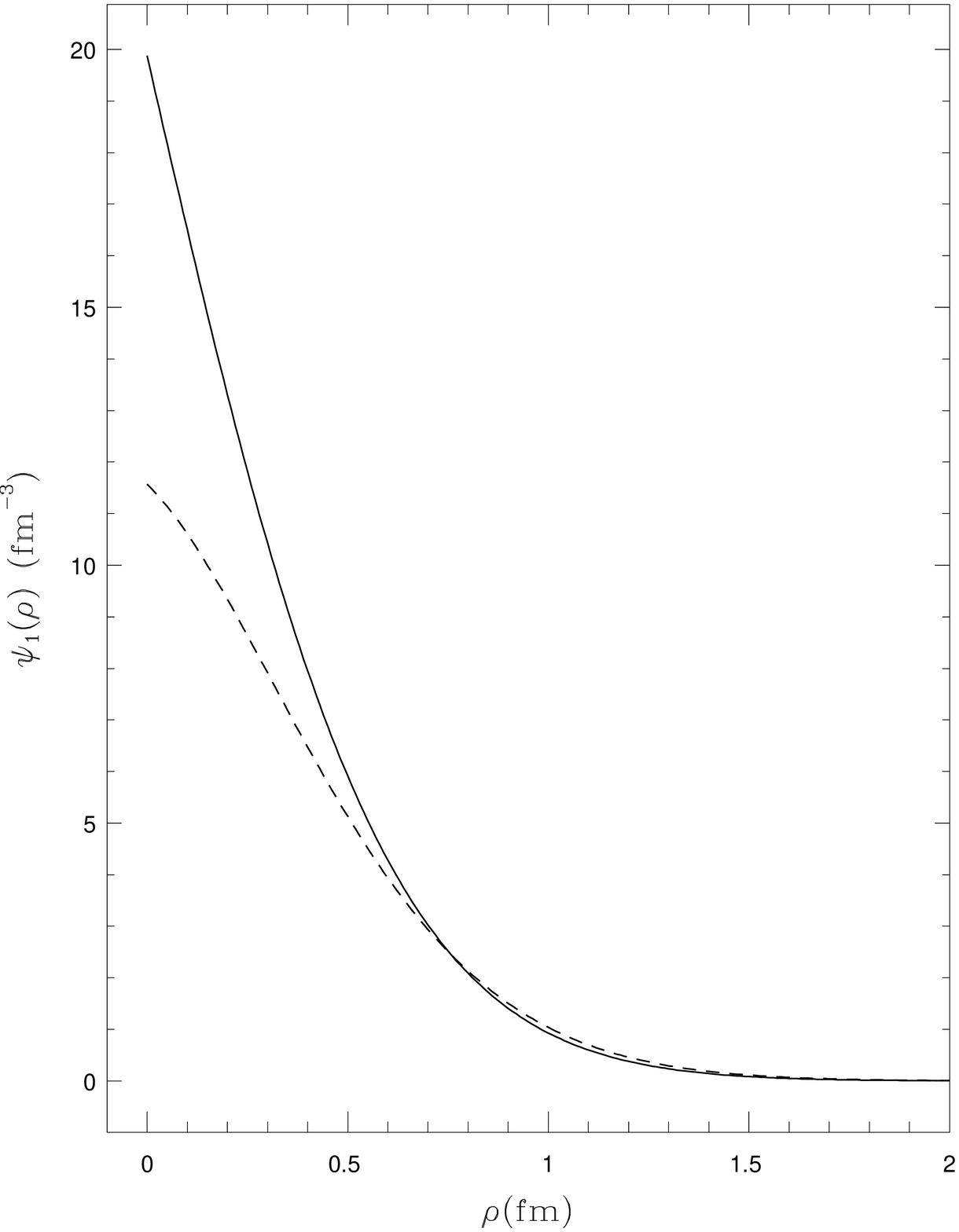, width=6cm} \hspace*{1cm} 
  \epsfig{file=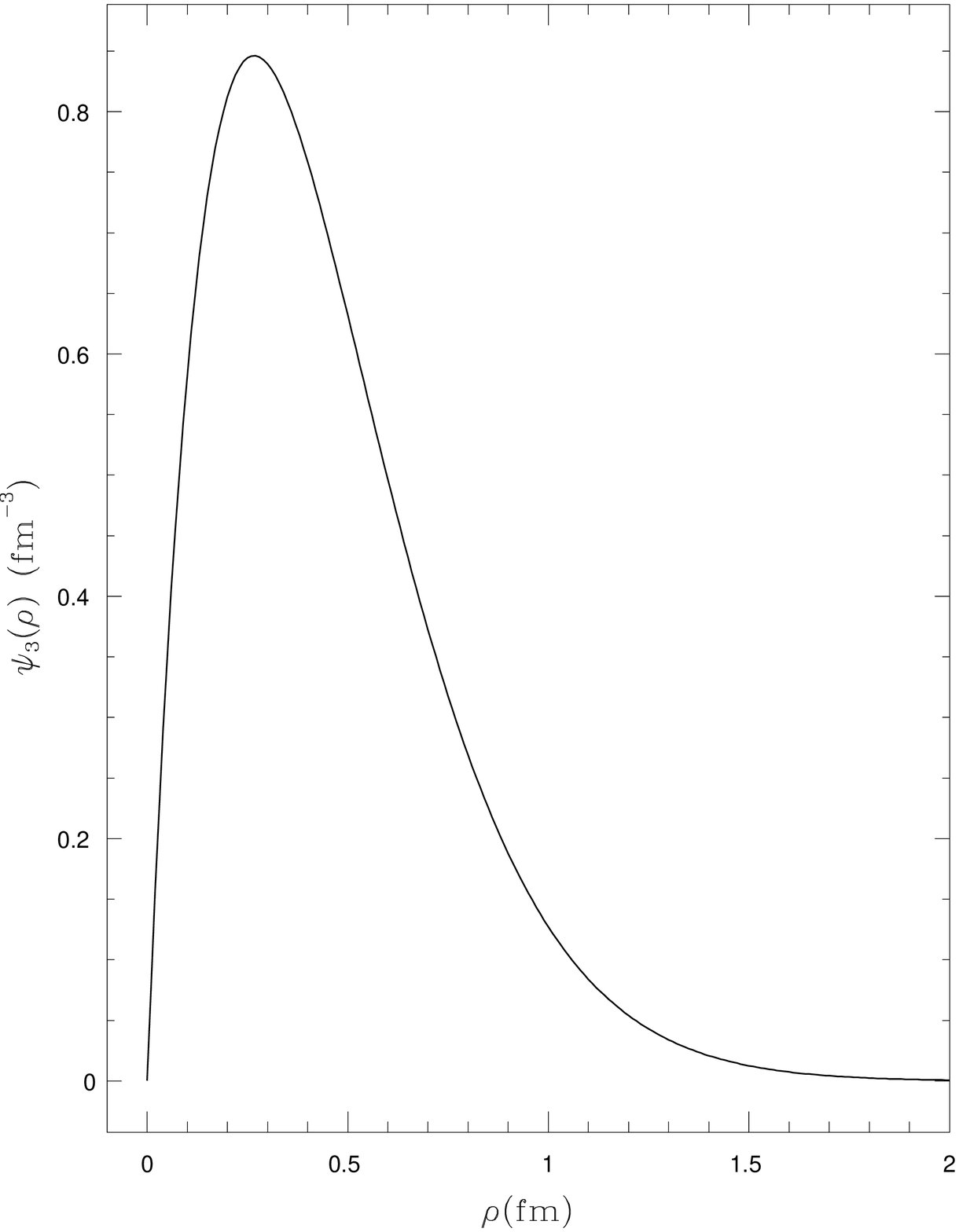, width=6cm} }
\end{center}
\vspace{-3mm}
\caption{\small{Radial wave functions $\psi_1(\rho)$  and $\psi_3(\rho)$ 
obtained 
in the hyperspherical formalism with the Bhaduri et al.'s force, Eq. (17) 
(continuous line). The wave function $\psi_1(\rho)$ in the case where 
the spin-spin force is turned off is also shown (dashed line).}}
\end{figure}

As we are interested in the form factor at high $q$, which depends 
on the description of the system at short distances, we also 
show some wave functions. This is done only for the hyperspherical 
formalism where the graphical presentation of the results is much 
simpler than for the Faddeev approach since it reduces to the wave 
functions $\psi_1(\rho)$ and $ \psi_3(\rho)$. They are given in Fig. 4.
Examination of $ \psi_1(\rho)$ at short distances evidences at first 
sight an exponential behavior. This one is approximately given by:
\begin{equation}
\psi_1(\rho) = \alpha_1 e^{-\beta_1\rho},
\end{equation}
with $ \alpha_1 =19.95\, {\rm fm}^{-3}$ and $ \beta_1 =1.715 \, {\rm fm}^{-1}$.

The slope, $\beta_1$, which is entirely determined by the Coulomb like
part of the potential (including the spin-spin part which also 
behaves as $\frac{1}{r}$ at small $r$), is close to what is theoretically 
expected (see App. A):
\begin{equation}
\frac{8\sqrt{2}}{5\pi} (\kappa +  \kappa'_{\sigma} ) \; m_q = 1.702 
\, {\rm fm}^{-1}.
\end{equation}

Some departure from the exponential form expected in a pure Coulombian problem 
is however observed, which is seen in particular on the wave function, 
$\psi_1(\rho)$, calculated when the spin-spin part of the interaction is turned 
off (see Fig. 4). For this case, a sizeable change occurs  in the slope of the 
wave function at short distances when going from $\rho=0$ to $\rho=0.25 \, 
{\rm fm}$. It can lead to specific features in the form factor at high $q$, 
which, as reminded in Sect. 3 for the two-body case, is sensitive to that part 
of the wave function. Numerically, this one up to $\rho^2$ terms is given by:
\begin{equation}
\psi_1(\rho) = \alpha'_1 (1-\beta'_1\rho + \gamma'_1 \rho^2),
\end{equation}
with $\alpha'_1=11.65 \, {\rm fm}^{-3}, \beta'_1=0.645 \, {\rm fm}^{-1} $ 
and $ \gamma'_1 = - 2.0 \, {\rm fm}^{-2}$. \\
The value of $\beta'_1$ is close to the theoretical expectation, 
$\frac{8\sqrt{2}}{5\pi }  \kappa m_q =0.640 \, {\rm fm}^{-1}$. 
The value of $\alpha'_1$ is smaller than the value of $\alpha_1$ in (37). 
This is in direct relation with the absence of the short range attraction that 
the spin-spin interaction provides on the average. As to the $\gamma'_1$ 
coefficient, it evidences a striking departure from its value in the pure 
Coulombian problem, $\gamma = \frac{\beta^2}{2} = 0.20 \, {\rm fm}^{-2}$ 
(according to the notation employed in App. A), since it has 
an opposite sign and is one order of magnitude larger. It depends on the 
constant terms appearing in Eq. (88) and should be compared to the 
theoretical expectation given by Eq. (96):
\begin{equation}
 -\frac{m_q}{6} ( E - 3m_q + \frac{3}{2}D ) 
 +\frac{5}{12} (\frac{8\sqrt{2}}{5\pi } \kappa m_q )^2 = 
- 2.1 \, {\rm fm}^{-2}.
\end{equation}
Through $E$, it mainly involves the non-perturbative part of the potential 
(17), arising from the confinement. In comparison, the role played by the 
perturbative part due to one-gluon exchange, represented by the last term in 
the l.h.s. of (40), is negligible.

The behavior of the component, $\psi_3(\rho)$, at short distances 
also deserves some comments. An approximate expression is:
\begin{equation}
\psi_3(\rho) = \alpha_3 \, \rho \, e^{-\beta_3 \rho}
\end{equation}
with $\alpha_3 = 8.4 \, {\rm fm}^{-4}, \; \beta_3 = 3.76 \, {\rm fm}^{-1}$.
\noindent
The value of $\alpha_3$ may be compared to the theoretical 
expectation given by Eq. (101):
\begin{equation}
 \frac{32}{35\pi} \kappa'_{\sigma} m_q \alpha_1 = 8.23 \,{\rm 
fm}^{-4}.
\end{equation} 
As to the $\beta_3$ parameter, it has no well determined interpretation. While 
the exponential factor in Eq. (41) turns out to be a (very!) good approximation 
to the numerical solution at small $\rho$, a more complete theoretical analysis 
indicates that the second term of the expansion of the solution should 
contain a well defined term, $\rho^2 \; {\rm log} \, \rho $, besides a $\rho^2$ 
term. While the former is determined by the knowledge of $\alpha_3$, $\alpha_1$ 
and $\beta_1$,  the latter is related to the solution of the homogenous 
Eq. (100) and cannot be similarly determined. The insertion of the $\rho^2$ 
term in the l.h.s. of this equation indeed leads to an indertermination for its 
coefficient.

The comparison of the phenomenological parameters $\beta_1$, $\beta'_1$, 
$\gamma'_1$ and $\alpha_3$ with the theoretical expectations evidences some 
slight discrepancy. We believe that this one has its origin in the numerical 
methods used to solve Eqs. (99, 100). Notwithstanding it, we think that 
the results show the adequacy of the Numerov algorithm that we employed to 
solve the equations of the HH approach and determine the short range 
part of the hyperradial wave function. 
%%%%%%%%%%%%%%%%%%%55555555555555555555555%%%%%%%%%%%%%%%%%%%%%%%%%%%%%%%%%%%%%
\section{Expression and calculation of the nucleon form factors}
Having determined the nucleon wave functions, we now consider the form factors 
that can be calculated from them. The charge and magnetic 
form factors of the proton and the neutron, which represent four quantities, 
are the object of a current interest. They can be obtained 
from the charge and magnetization densities. In the approximation used for 
describing the wave function, where the total orbital momentum 
of the nucleon is $L=0$, in agreement with the quark-quark force we used, the 
starting point is given by the following formulas:
\begin{eqnarray}
 < N |\frac{1+\tau^z}{2} G_E^p (\vec{q}^{2}) + \frac{1-\tau^z}{2} G_E^n 
(\vec{q}^{2})| N> (2\pi)^3 \delta(\vec{P}^f-\vec{P}^{i}-\vec{q}) = 
 \;\;\;\;\;\;\;\;\;\; \nonumber \\
< N(\vec{P}^f)  | \sum_i ( \frac{1}{2} ( \frac{1}{3} + \tau_i^z ) 
e^{i\vec{q}.\vec{r}_i}) | N(\vec{P}^{i})>,\\
 < N | \vec{\sigma} ( \frac{1+\tau^z}{2} \frac{G_M^p(\vec{q}^{2})}{2m_N} + 
\frac{1-\tau^z}{2} \frac{G_M^n(\vec{q}^{2})}{2m_N} ) |N  > 
(2\pi)^3 \delta(\vec{P}^f- \vec{P}^{i}-\vec{q}) = \;\;\; \nonumber \\
 < N (\vec{P}^f) | \sum_i  \frac{1}{2m_q} \frac{\vec{\sigma}_i}{2} 
( \frac{1}{3} + \tau_i^z ) e^{i \vec{q}.\vec{r}_i} | N(\vec{P}^{i}) >.
\end{eqnarray}

Using the symmetry of the nucleon wave function, the sum over the three-quarks 
appearing in (43, 44) can be restricted to one of them, 
the matrix element being multiplied by 3. The third quark is here chosen as the 
expression of our wave functions for mixed symmetry 
states have a definite symmetry property under the exchange of particles 1 and 
2. After performing the algebra relative to spin and 
isospin operators, one is left with the following expressions:
\begin{eqnarray}
G_E^p(\vec{q}^{2}) = (< S | O | S > + \frac{1}{2} <MS | O | MS>) + \sqrt{2} 
< S | O | MS > \;\;\;\;\;\;\;\;\;\;\;\;\;\; \nonumber  \\ 
+ (<A | O |A > + \frac{1}{2} < MA| O |MA>)  - \sqrt{2} < A | O | MA>,  \\
G_E^n(\vec{q}^{2}) = - \sqrt{2} <  S | O | MS > + \sqrt{2} < A | O | MA > 
,\;\;\;\;\;\;\;\;\;\;\;\;\;\;\;\;\;\;\;\;\;\;\;\;\;\;\;\;
\;\;\;\;\;\;\;\; \\
G_M^p(\vec{q}^{2}) \; \frac{m_q}{m_N} =  (< S | O | S + \frac{1}{2}< MS | O | 
MS >) + \sqrt{2} < S | O | MS > \;\;\;\;\;\;\;\;\; \nonumber \\ 
- \frac{1}{3} (< A | O | A > + \frac{1}{2} < MA | O | MA >) + 
\frac{\sqrt{2}}{3} 
< A | O |MA >,  \\
G_M^n(\vec{q}^{2}) \; \frac{m_q}{m_N} = - \frac{2}{3} (<S| O |S> + 
\frac{1}{2} 
< MS | O | MS >) - \frac{\sqrt{2}}{3} < S | O | MS >  \nonumber \\ 
+\frac{2}{3} (<A | O | A > + \frac{1}{2} < MA| O | MA >) + \frac{\sqrt{2}}{3} 
< A | O | MA > .
\end{eqnarray}
The matrix elements are defined as:
\begin{eqnarray}
< X | O | Y > = \int d(\frac{\vec{r}_1-\vec{r}_2}{2}) \; 
d(\vec{r}_3 - \frac{\vec{r}_1+\vec{r}_2}{2}) \;\;\;\;\;\;\;\;\;\;\;\;\;\; 
\nonumber \\
\psi_X(\vec{r}_1,\vec{r}_2,\vec{r}_3) \; e^{ 
i\vec{q}.(\frac{2}{3}\vec{r}_3-\frac{\vec{r}_1+\vec{r}_2}{3} ) } \; 
\psi_Y(\vec{r}_1,\vec{r}_2,\vec{r}_3),
\end{eqnarray}
where $X,Y$ stand for the different symmetry character of the components of the 
wave function, $S, A, MS$ and $MA$.
The expressions (45-48) have been written in such a way to emphasize their 
dependence on four independent quantities. As one of them involves 
the completely antisymmetric component $\psi_A(\vec{r}_1,\vec{r}_2,\vec{r}_3)$ 
which is expected to be small (when it is 
not simply put to zero as a result of approximating the wave function as in the 
hyperspherical formalism or the two amplitude Faddeev calculations),
one should approximately obtain the following relation:
\begin{equation}
\frac{1}{3} ( \, G_E^p(\vec{q}^{2}) - G_E^n(\vec{q}^{2}) \, ) - 
G_M^p(\vec{q}^{2}) 
\; \frac{m_q}{m_N}-G_M^n(\vec{q}^{2}) \; \frac{m_q}{m_N} = 0.
\end{equation}
This is trivially satisfied for:
\begin{eqnarray}
G_E^p(\vec{q}^{2}) = f(\vec{q}^{2}),\;\;\;\;\; G_E^n(\vec{q}^{2})=0, 
\hspace*{20mm} \nonumber \\
G_M^p(\vec{q}^{2}) \frac{m_q}{m_N} = f(\vec{q}^{2}),\;\;\; 
G_M^n(\vec{q}^{2}) \frac{m_q}{m_N} = - \frac{2}{3} f(\vec{q}^{2}),
\end{eqnarray}
a choice which is sometimes made in the litterature, $f(\vec{q}^{2})$ being 
given most often a dipole expression. The validity of Eq. (50) has been 
tested in ref. \cite{BSB5} for both the Faddeev wave functions used in the 
present work and the measurements. At $\vec{q}^2=0$, expressions different from 
Eqs. (45-48)  may be found in the 
literature  \cite{VALE}. They correspond to another type of component $|A>$
with a non-zero orbital angular momentum (1 instead of 0). In the harmonic 
oscillator model, the two kinds of states represent at least a 2$\hbar \omega$ 
($L=1$) and a 6$\hbar \omega$ ($L=0$) excitations respectively.

\begin{figure}[htb!]
\vspace{4mm}
\begin{center}
\mbox{ \epsfig{file=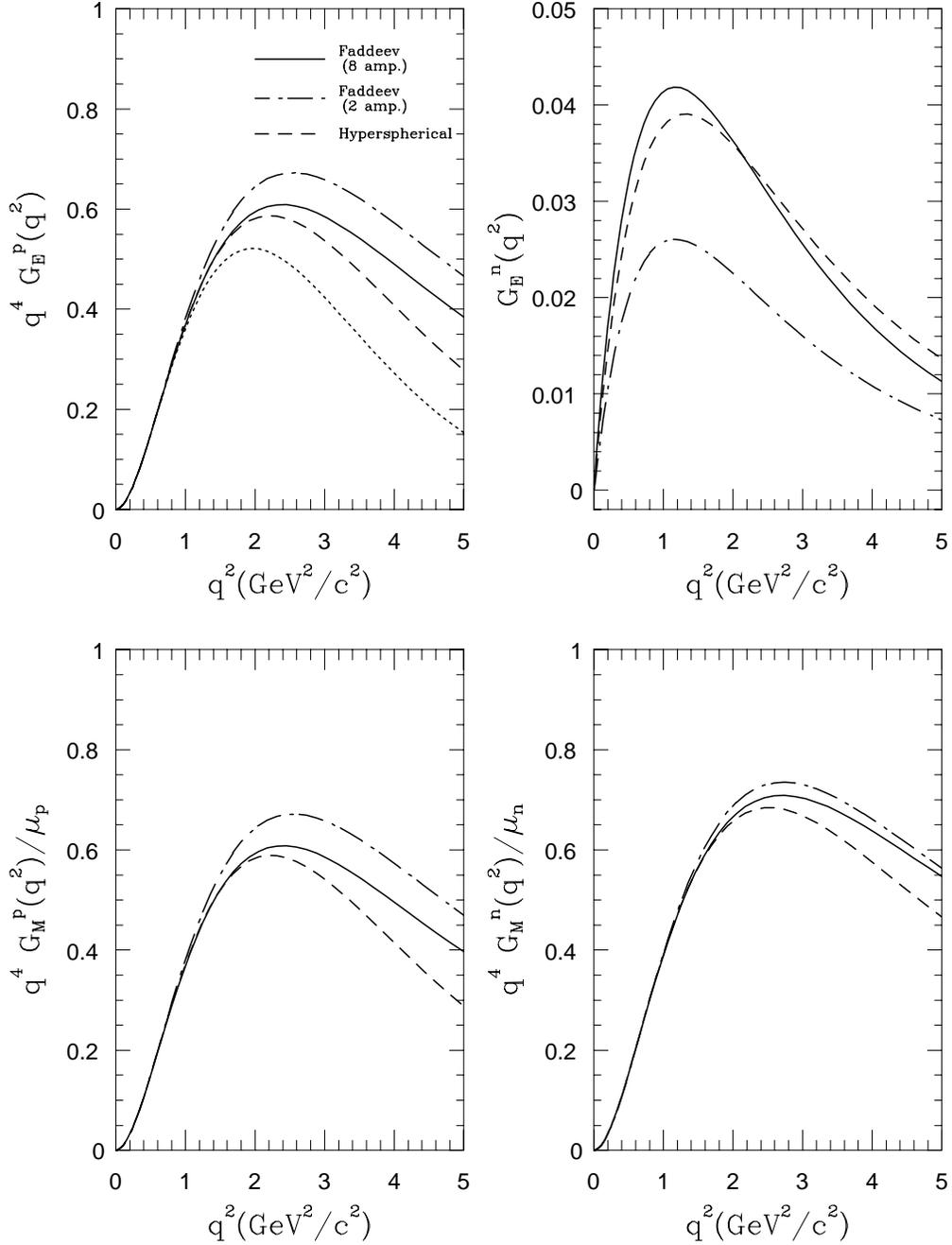, width=12.7cm} }
\end{center}
%\vspace{-3mm}
\caption{\small{Electric and magnetic form factors of the proton and the neutron 
up to 
$\vec{q}^2=5 \,({\rm GeV/c})^2$. Except for $G_E^n(\vec{q}^{2})$, where it is 
not 
necessary, they have been multiplied by a factor $q^4$ to emphasize the 
differences in the higher $q$ domain. Results are presented for wave functions 
calculated with the Faddeev equations (2 and 8 amplitudes) and the 
hyperspherical formalism (K=0 and 2). A comparison with results provided by a 
gaussian wave function can be made by looking at the upper-left figure (dotted 
line).}}
\end{figure}

\begin{figure}[htb!]
\vspace{4mm}
\begin{center}
\mbox{ \epsfig{file=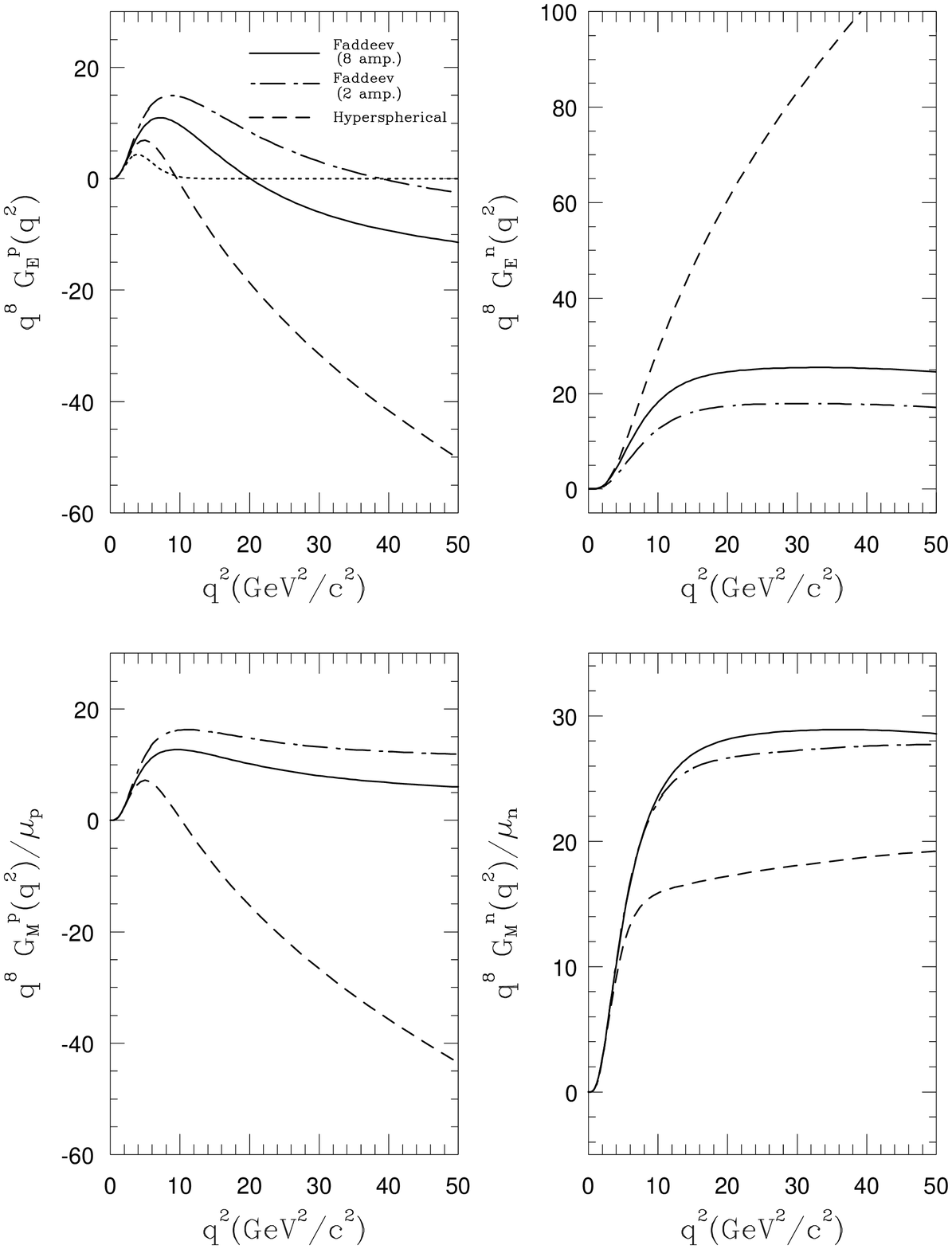, width=12.7cm}  }
\end{center}
%\vspace{-3mm}
\caption{\small{Same as in Fig. 5, but up to $\vec{q}^2=50 \,({\rm GeV/c})^2$. 
Results 
have 
been multiplied by $q^8$ in all cases to emphasize the asymptotic behavior.}}
\end{figure}  

The results for the form factors, $G_E^p(\vec{q}^{2}), \; G_E^n(\vec{q}^{2}), 
\; G_M^p(\vec{q}^{2})$ and $G_M^n(\vec{q}^{2})$ are given in Fig. 
5, for transferred momenta up to $\vec{q}^{2}=5 \, ({\rm GeV/c})^{2}$. In each 
case, the results are shown for the Faddeev calculations with two and eight 
amplitudes and  for the hyperspherical formalism (with the two waves $K=0$ 
and $K=2$). For $G_E^p(\vec{q}^{2})$, we also give results for 
a gaussian wave function corresponding to a matter radius close 
to that obtained in the other approaches. It is seen that the Faddeev 
calculations with two and eight amplitudes significantly differ from 
each other, especially for the charge neutron form factor. The eight amplitude 
results compare reasonably well with the results obtained in the hyperspherical 
formalism up to $\vec{q}^2=3 \, ({\rm GeV/c})^2$. Some discrepancy appears 
beyond this value, as seen in this figure. One also sees that the result 
using a gaussian wave function quickly differs from the above ones (a factor 
of about 2 at $\vec{q}^2=5 \, ({\rm GeV/c})^2$ ) and tends to $0$ more rapidly.
\clearpage  

In Fig. 6, we present results that concern the asymptotic behavior of the form 
factors. They have been divided by $q^{-8}$, which is the expected behavior at 
high $q$ (Sect. 2). These results, which should tend to a constant number, are 
shown up to $\vec{q}^{2} = 50 \, ({\rm GeV/c})^2$. Examination of the figures 
indeed indicates that, roughly, some asymptotic value seems to be reached. 
This demonstrates that constituent quark models, contrary to what is sometimes 
claimed, do lead to a power law behavior of form factors. This is achieved 
provided that they incorporate at least a minimal description of the short range 
correlations produced by gluon exchanges. Looking in more details, one may 
however notice that the results obtained with the hyperspherical formalism are 
somewhat higher and that their convergence to some asymptotic value is less 
clear than for the Faddeev approach. In this case, a slight variation with 
$\vec{q}^{2}$ is still visible, suggesting that the asymptotic behavior may not 
be reached yet. These different features have led us to make a more refined 
analysis.

As a side remark, let us mention that the smallness of the form factors at the 
highest values of $q^2$ under consideration (of the order of $10^{-5}-10^{-6}$) 
may cast doubt about their numerical accuracy. Tests for which an analytical 
result was available (see next section) has revealed that this accuracy was 
surprisingly good and could not be responsible for the above departures. 
In fact from a dimensional argument, the form factor given by Eq. (49) is 
expected to at least contain a factor, $\frac{1}{q^6}$ (see App. B), which 
already explains for a large part the smallness of the form factor. Some of the 
observed departures may however be due to the lack of accuracy in solving 
equations to obtain wave functions. 
%%%%%%%%%%%%%%%%%%66666666666666666666666666666666666666666666%%%%%%%%%%%%%%%%%
\section{ Discussion of the results}
Due to the difficulty to perform extensive calculations with the Faddeev 
approach, the discussion made below will essentially 
concerns the results obtained with the hyperspherical harmonic formalism. We 
will nevertheless attempt to establish some relationship between 
the two approaches. On the other hand, instead of working with the proton and 
neutron, charge and magnetic form factors, we 
will deal with matrix elements involving components of the wave function with a 
given symmetry, namely:
\begin{equation}
<S | O | S>,\;\; < S | O | MS>,\;\; < MS | O | MS> , \;\;< MA|  O|  MA>,
\end{equation}
whose expressions are given by Eq. (49). The reason to make this 
alternative presentation of the results is the possibility that 
they may not all have the same asymptotic behavior. This is expected 
from other three-body calculations where it was shown that, 
in momentum space, the component $\psi_{MS}$ was going to zero more rapidly 
than $\psi_{S}$ \cite{MOSK}. This result was however 
obtained with separable forces and, actually, for the Faddeev amplitude.
%%%%%%%%%%%%%%%%%%%%%%%%%%%%%%%6.1
\subsection{Calculations with the hyperspherical harmonics }
The advantage of the hyperspherical harmonic formalism is to provide 
a simple, but approximate expression of the wave function 
at short distances, which furthermore can be used for analytic 
calculations of the form factors. Thus, from the expressions 
$\psi_1(\rho)$ and $\psi_3(\rho)$ given by Eqs. (37) and (41), one 
gets for the  various matrix elements listed above:
\begin{eqnarray}
< S | O  |S > \;\;\; = \;\; 405 \sqrt{6} \;\frac{\alpha_1^2 
\beta_1}{(q^2+6\beta^2_1)^{7/2}} 
  \hspace*{23mm} \rightarrow \;\;\; \frac{7.88}{ q^7} ({\rm GeV/c})^7,  \\
< S | O | MS> \;\; = \; - 2835\sqrt{6} \frac{\alpha_1 \alpha_3 q^2}{ ( 
  q^2+6(\frac{\beta_1+\beta_3}{2})^{2} )^{9/2} } \hspace*{10mm}
  \rightarrow -\frac{13.5}{ q^7} ({\rm GeV/c})^7,  \\
< MS| O |MS> \; = \; 7290 \frac{\alpha^2_3}{q^8} \left( 64 
-\frac{175  \sqrt{6}  \beta_3 \,  q^{10}  + ..
    }{ (q^2+6\beta^2_3)^{11/2} }  \right) \;
  \rightarrow \;\;\; \frac{75.6}{q^8} ({\rm GeV/c})^8,  \\
<MA | O | MA > = - 2430 \frac{\alpha^2_3}{q^8} \left( 64   
-\frac{168  \sqrt{6} \beta_3   \, q^{10}  +.. 
   }{ (q^2+6\beta^2_3)^{11/2} }  \right)  
  \rightarrow -\frac{25.2}{q^8} ({\rm GeV/c})^8.\, 
\end{eqnarray}
For the two last matrix elements, only the terms dominant at high q and the 
first corrections to it have been retained here. A complete expression is 
given in App.~B. An expression identical to Eq. (53) was obtained in 
\cite{GIAN,BIJK}, but on a pure phenomenological basis, without relation to some 
microscopic dynamics. 

\begin{figure}[htb!]
\begin{center}
\mbox{ \epsfig{file=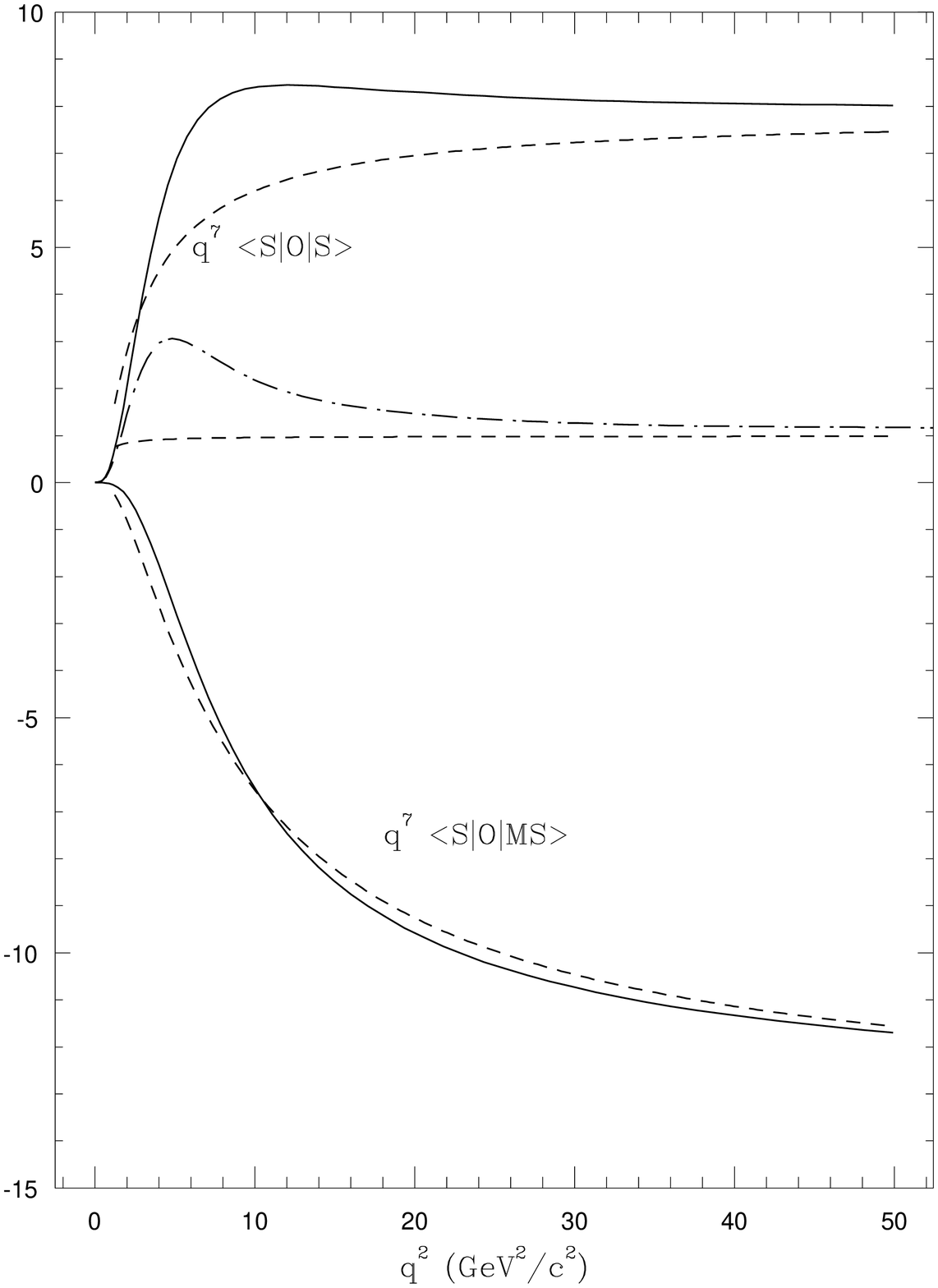, width=6cm} \hspace{8mm} 
 \epsfig{file=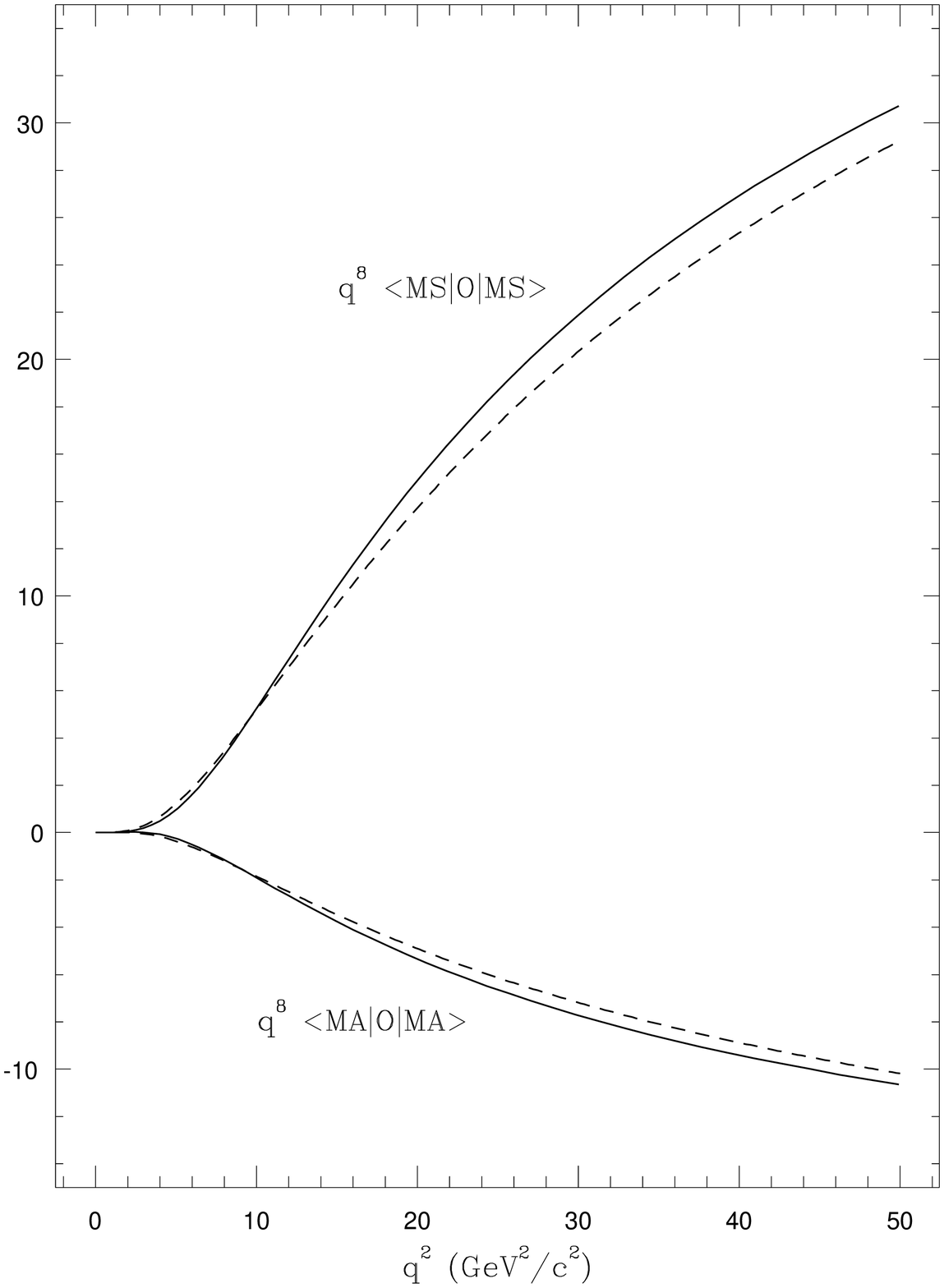, width=6cm} }
\mbox{\scriptsize \hspace*{3cm}  a) \hspace*{6.9cm}  b) \hspace*{2.6cm} } 
\end{center}
\caption{\small{Form factors calculated in the hyperspherical formalism up to 
$q^2=50 
\, ({\rm GeV/c})^2$. The form factors, $<S|O|S>$ and $<S|O|MS>$, have
been multiplied by $q^7$ (Fig. a) and the form factors, $<MS|O|MS>$ and 
$<MA|O|MA>$, by a factor $q^8$ (Fig. b). In each figure, we present 
the form factors obtained from the numerical wave functions, 
$\psi_1(\rho)$ and $\psi_3(\rho)$ (continuous line), as well as  
the analytic ones, Eqs. (53-56) (dashed line). The $<S|O|S>$ form 
factor in absence of spin-spin interaction is also shown in Fig. a (dash-dotted 
line for the numerical calculation and dashed line for the analytical one).}}
\end{figure}  

Being a good representation of the short range behavior of the calculated 
wave functions, the parametrizations of the components, $\psi_1(\rho)$ 
and $\psi_3(\rho)$, given by Eqs. (37) and (41) should provide accurate 
expectations for the form factor at high $q$ in the 
hyperspherical harmonic approach considered here. By comparing the form 
factors given by Eqs. (53-56) to those calculated with the actual 
wave functions, it is possible to determine the range of $q^2$ where the 
description of the wave functions by the parametrizated 
ones, Eqs. (37, 41), becomes relevant, as well as the role of non-perturbative 
effects they do not account for. The results are presented in Figs. 7a, b. 
They have been multiplied by a factor 
$q^7$ for $< S | O | S >$ and $<S | O | MS>$ and $q^8$ 
for $<MS | O | MS>$ and $<MA | O | MA>$. These factors are chosen in 
accordance with the asymptotic behavior expected from Eqs. (53-56). The results 
so obtained can be directly compared to the asymptotic numerical values given 
in Eqs. (53-56). Figure 7a also contains the result for the form factor, 
$<S|O|S>$, when the spin-spin part of the Bhaduri et al.'s force is removed.

Examination of the figures shows a good agreement between the form factors 
given by Eqs. (53-56) and those obtained with the calculated 
wave functions, $\psi_1(\rho)$ and $\psi_3(\rho)$, in the range, $q^2 
> 30 \, ({\rm GeV/c})^2$. With the help of the analytic expression, it 
is possible to have a discussion on the onset of the asymptotic behavior.

These results evidence two striking features. First, the form factors 
$<S | O | S >$ and $ <S | O | MS>$ at very high $q$ behave like $q^{-7}$, 
instead of $q^{-8}$ as it is expected in a complete calculation \cite{ALAB}. 
The $q^{-7}$ behavior can be understood using coun\-ting rules, similarly to 
the derivation of the $q^{-8}$ in the general case: the propagator of the 
set of three-quarks in the hyperspherical harmonic formalism introduces a 
factor $q^{-2}$, while the interaction, given by the Fourier transform of 
the $\frac{1}{\rho}$ hyperradial potential provides a factor $q^{-5}$.
Second, the form factors $<MS | O | MS>$ and $<MA |O |MA>$ behave 
like $q^{-8}$, but the asymptotic behavior is far to be reached in the range 
around $50 \, ({\rm GeV/c})^2$ (a factor 2-3 is missing).

The explanation for the wrong behavior of the form factors $ <S | O | S>$ and 
$<S | O | MS>$ is to be looked for in the approximation 
consisting in retaining the lowest $K$ values in the expansion of the wave 
function. Let us examine for example the spatial wave function for $K=0$, 
$\psi_1(\rho)$. Depending on the variable $\rho = \sqrt{\rho^2_{12}+\rho^2_3}$,  
at small $\rho_{12}$ ($\rho_{12} << \rho_3 $ ) it behaves like :
\begin{equation}
\psi_1(\rho)_{\rho_{12} \rightarrow 0} = \psi_1(\rho_3)+\frac{1}{2} 
\psi_1'(\rho_3) \frac{\rho_{12}^2}{\rho_3} +....
\end{equation}
The appearance of $\rho^2_{12}$ as the first 
non zero term in the expansion cannot account properly for the short range 
correlations of particles 1 and 2, the third particle being a spectator. As one 
obtains in the two-body case (see Sect. 3), a term linear in $\rho_{12}$ is 
expected. To get the right behavior, one should sum up an infinite set of 
contributions with $K$ ranging from 0 to $\infty$. Furthermore, a rapid 
convergence is not ensured at all. 

In the case without spin-spin force $(\psi_3(\rho)=0)$, we looked at some 
contributions of the next terms in the hyperspherical harmonic expansion 
corresponding to $K=4$ and $6$, Eq. (102)). The first one contributes 
to $<S | O | S>$ three times as much as the wave $K=0$ in the asymptotic 
domain, with the same sign. The contribution of the $K=6$ wave, which is 
antisymmetric in the $\rho_{12}$ and $\rho_3$ variables and, thus, introduces 
a richer structure in the wave function, could correct the result in the right 
direction. This is in agreement with the fact that reprodu\-cing the correct 
asymptotic behavior requires a somewhat fine description of the wave function 
at short distances, especially with respect to the $\phi$ 
hyperspherical angle.

Another related aspect is worthwhile to be mentioned. Up to a factor, the 
coefficients of the $q^{-7}$ term in Eq. (53)  and Eq. (54),   
$ \alpha_1.\alpha_1\beta_1$ and $\alpha_1.\alpha_3$ respectively, have a form 
similar to Eq. (12),     namely the product of the wave 
function at the origin, $\alpha_1$, which by itself contains non-perturbative 
effects, and its derivative at the same point, 
$\alpha_1\beta_1$ or $\alpha_3$. Factorizing out the wave function 
at the origin, $ \alpha_1 $, the last quantities, $\beta_1$ and 
$\frac{\alpha_3}{\alpha_1}$, for which an analytic expression can be obtained 
from Eqs. (38) and (42), are seen to be linearly dependent on the strength 
of the dominant part of the 
interaction at short distances, determined by  $\kappa$ and $\kappa'_{\sigma}$. 
This is not in accordance with the argument reminded in Sect. 2, which 
implies two gluon exchange and, therefore, a dependence on the square 
of these $\kappa$ factors.

On the other hand, from the smallness of the mixed symmetry state (1.5\%), it 
may be inferred that its contribution to the form factor is negligible. This is 
true at small momentum transfers, but not at high ones, as shown in Fig. 7a   or 
on the analytic expressions given by Eqs. (53, 54). Taking into account the 
expressions of $\alpha_3/\alpha_1$, Eq. (42) and $\beta_1$, Eq. (38), it is 
found that the matrix elements $ <S | O | MS >$ and $<S | O | S >$ at high $q$ 
are quite comparable. Indeed the ratio tends to a constant of the order of 
unity:
\begin{equation}
(\frac{<S | O | MS>}{<S | O | S >} )_{q \rightarrow \infty} = - 2\sqrt{2} \,
\frac{ \kappa'_{\sigma}  }{\kappa  +  \kappa'_{\sigma}  }   \simeq - 1.8
\end{equation}

The last aspect we want to discuss about the form factors, $ <S | O|  S> $ 
and \mbox{$ <S | O | MS>$}, is the role of non perturbative effects. In a pure 
Coulombian pro\-blem and in the approximation where only the $K=0$ wave is 
retained, the form factor approaches its asymptotic value from below. 
Examination of Fig. 7a   shows the opposite feature, as a consequence of having  
incorporated into the $\rho^2$ term of the description of the wave function at 
short distances a contribution depending on the energy of the system,  see Eq. 
(40), which indirectly involves the confining interaction. The above feature, 
which is strongly enhanced when the spin-spin force is neglected (see Fig. 7a),  
will be also seen in other calculations presented below. The observed structure 
at $q^2=4 \, ({\rm GeV/c})^2$ in the case without spin-spin force has to do with 
the change in the second derivative of the wave function around $\rho= 0.3 \, 
{\rm fm}$ that can be seen in Fig. 4. An immediate consequence is to make the 
onset of the asymptotic behavior to be reached quicker.

It is also interesting to look at the two other form factors, $<MS | O | MS >$ 
and $<MA | O | MA >$, although it is more difficult to rely on theoretical 
expectations, due to the lack of interpretation for the parameter $\beta_3$ 
entering the short-range description of the component $\psi_3(\rho)$, Eq. (41), 
on which they depend. We already noticed that these form factors behave like 
$q^{-8}$ at high $q$ (see analytical calculations, Eqs. (55, 56)). 
On this basis, one may think that their role should be negligible at high $q$, 
in comparison to $<S | O | S >$ or $<S | O | MS>$. This is hardly 
verified ($<MS  |O | MS > = 0.55$ \mbox{$< S | O | S >$} 
at $q^2=50 \, ({\rm GeV/c})^2$). 

Moreover, the values of these form factors multiplied by $q^8$, which should 
tend to a constant, are still increasing in the range $q^2 \simeq 50 \, ({\rm 
GeV/c})^2$ and far to have reached the asymptotic value (larger by a factor 2). 

In fact, assuming that the asymptotic behavior was reached in this range 
led us to the preliminary conclusion that this form factor was behaving like 
$q^{-7}$, like $<S | O | S >$ and $<S | O | MS>$. It is only from the 
examination of the analytical calculation, Eqs. (53-56), that we could determine 
that the asymptotic  behavior of the form factors, $ < MS | O | MS >$ 
and $<MA | O | MA>$, was $q^{-8}$ and understand, as well, the other features 
they evidence. In particular, as is seen from Eqs. (55, 56), the next to leading 
order contribution in $q^{-8}$ contains an extra factor $q^{-1}$, instead 
of $q^{-2}$ for the form factors, $<S | O | S >$ or $<S | O | MS>$. The 
contribution is about $60\%$ of the dominant term at $q=7\,({\rm GeV/c})$ and, 
since it is destructive, it provides an overall reduction of the form 
factor by a factor 2-3.  This feature strongly delays 
the onset of the asymptotic behavior for the form factors, $ < MS | O | MS >$ 
and $<MA | O | MA>$. Equations (55, 56) suggest that the asymptotic 
behavior is obtained within a few  \% at values of $q^2$ as large as 10000 
$({\rm GeV/c})^2$ (!).

So, we can conclude that calculating reliably the form factors at values of 
$q^2$ as high as those considered here supposes not only an accurate 
determination of the wave function at short distances, but also on the whole 
range. Calculations we did up to $q^2 = 200 \,({\rm GeV/c})^2$ have revealed 
moderate but regular oscillations around the asymptotic behavior of the form 
factor $<S | O | S >$, explaining a slight plateau that can hardly be seen 
in Fig. 7a around $q^2 = 40 \, ({\rm GeV/c})^2$, but shows up in the 
numerical values. From the checks we did, it turns out that these 
oscillations are in relation with the value of the hyperradius 
where the inner and outer Numerov solutions of the Schr\"odinger equations were 
matched. At this point, the third derivative of the numerical solution 
evidences a change in sign (giving rise to the oscillations) that does not seem 
to have any physical meaning.
%%%%%%%%%%%%%%%%%%%%%%%%%%%%%%%6.2
\subsection{Calculations in the Faddeev approach}
\begin{figure}[htb!]
\begin{center}
\mbox{ \epsfig{file=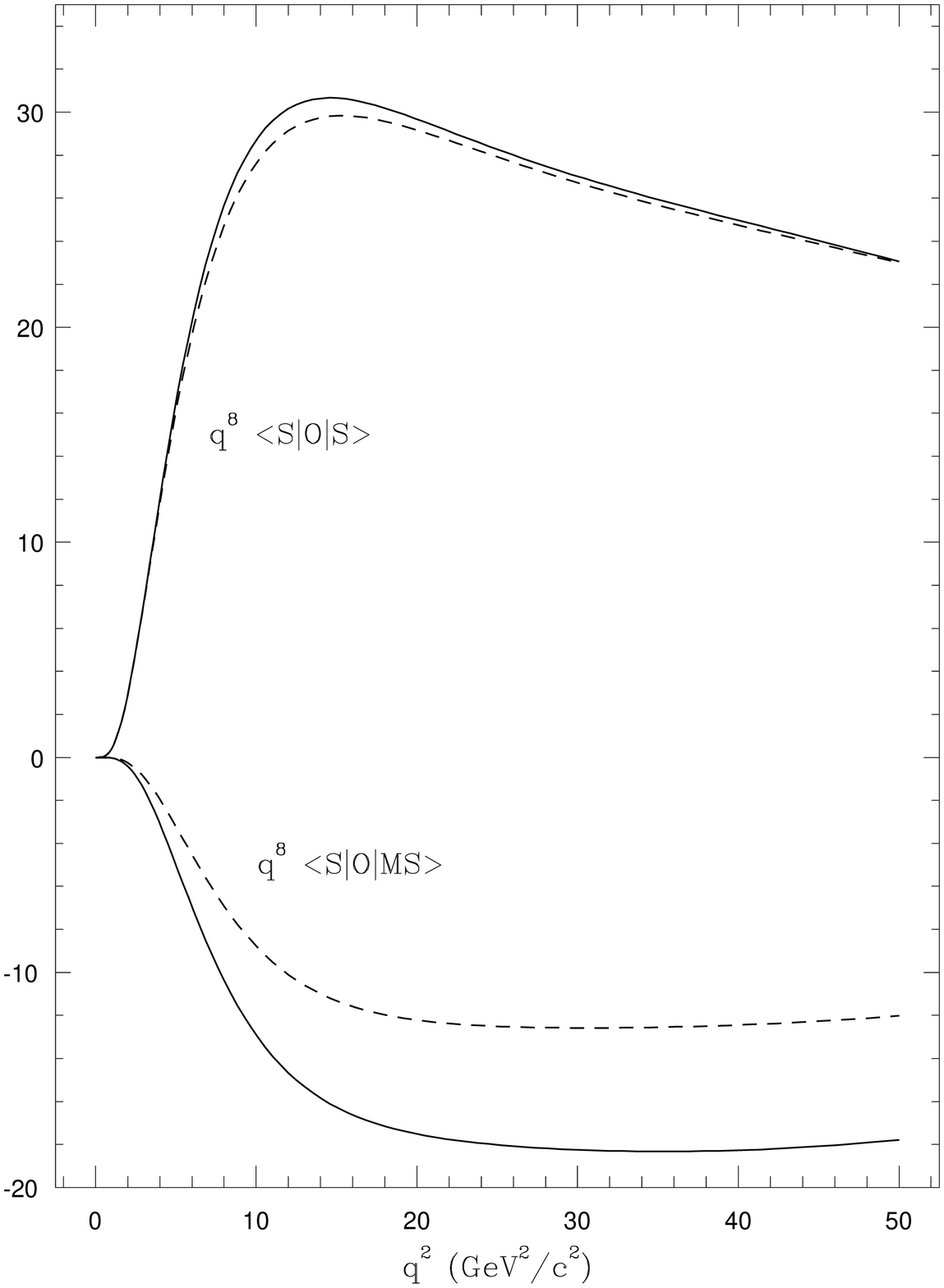, width=6cm} \hspace*{1cm}
\epsfig{file=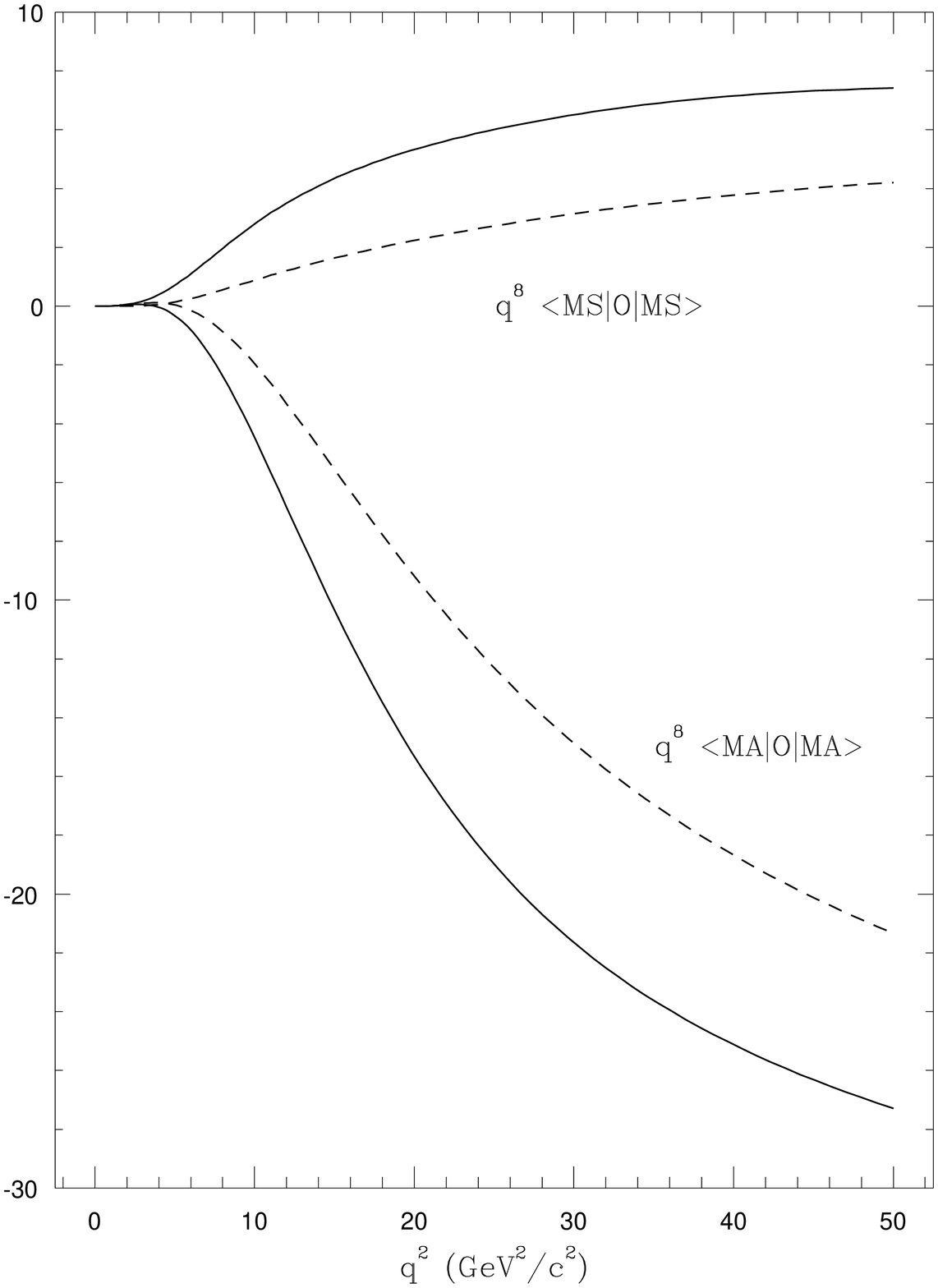, width=6cm} }
\end{center}
\vspace{-3mm}
\caption{\small{Form factors calculated in the Faddeev approach up 
to $q^2=50 \,({\rm GeV/c})^2$. The  dashed and continuous lines respectively 
represent the results for the 2 and 8 amplitude calculations.
The form factors, $<S|O|S>$, $<S|O|MS>$, $<MS|O|MS>$ 
and $<MA|O|MA>$, have been multiplied by a factor $q^8$.
The values, which these quantities are expected to converge to, are given in 
Eqs. (74).  }  }
\end{figure}  

One may wonder whether observations made about the calculations performed 
in the hyperspherical harmonic approach apply to those in the Faddeev approach, 
for which we have a priori no precise benchmark as far as the asymptotic 
beha\-vior of form factors is concerned. From examining the corresponding 
results presented in Fig. 8, we may notice that the convergence of the form 
factors $< S | O | S > $ and $ <S | O | MS>$ to some asymptotic behavior assumed 
to be $q^{-8}$ is the same as for the hyperspherical harmonic calculations with 
respect to $q^{-7}$: same rate and from above for the first one. As to the form 
factors \mbox{$< MS | O | MS > $} and $ < MA | O | MA > (+2 < A | O | A> )$, 
the absolute value of their product by $q^8$, especially for the second one,  
tends to steadily increase around $q^2 = 50 \,({\rm GeV/c})^2$, quite similarly 
to the results obtained with the hyperspherical harmonic approach. Due to the 
way calculations were performed, the contribution of the fully antisymme\-tric 
spin-isospin state to the last matrix element, $< A | O | A> $, could not be 
easily separated in the case of the 8 amplitude Faddeev approach where it is not 
zero. Its contribution is in any case quite small and should not affect the 
discussion relative to the matrix element, $ < MA | O | MA >$, to which it is 
admixed (what we reminded between parentheses).

Two more observations are in order. There is no evidence that, for $q 
\rightarrow \infty$, the form factors, $< MS | O | MS >$ or $< MA | O | MA >$, 
go to zero more rapidly than the form factors, $< S | O | S>  $ 
or  \mbox{$ <S | O | MS > $}, as is expected from some calculations (see Sect. 
7.1 below). On the other hand, on the basis of the same calculations, the form 
factors in the Faddeev approach are expected to go to zero more rapidly and to 
be smaller in absolute value at high $q$ than in the hyperspherical harmonic 
approach with the truncation space chosen. The comparison of the results 
indicates that this conclusion is supported in some part for the form 
factors, $ <S | O | S >$ and \mbox{ $< S | O | MS >$}, (a factor 0.4 at $q^2=50 
\,({\rm GeV/c})^2$ for $ <S | O | S >$ for instance), but not 
for \mbox{ $< MA | O | MA > (+ 2 < A | O | A>)$} (a factor 2.5). At this point, 
the slow convergence of this matrix element to an asymptotic value prevents one 
to make definite statements.

Another question of interest concerns the convergence of the form factors to 
their asymptotic behavior depending on the Faddeev calculation, two or eight 
amplitudes in the present case. Apart for a change in scale for the form factors 
involving the mixed symmetry state, the examination of our results does not 
evidence any significant change in the asymptotic behavior. The increase with 
$q$ of the quantity, $q^8 .( < MA | O | MA > (+2 < A | O | A>) \, )$,  is 
slightly less with the eight amplitude than with the two amplitude Faddeev 
calculation around $q^2 = 50 \,({\rm GeV/c})^2$. In any case, and contrary to 
the hyperspherical harmonic approach, there is no evidence at this point that 
the power law asymptotic behavior of form factors is affected by the truncation 
of the Faddeev type calculation which, in the simplest case (two amplitudes), 
already involves a more structured wave function. Only the overall size may 
depend on the truncation. In fact, an examination of the mixed symmetry 
components of the wave function at small distances for the 2 and 8 amplitude 
cases, around $\rho=0.05\,$fm, shows a discrepancy of a few percents only, 
without any relation to what is observed in the calculations performed for 
momentum transfers up to $q^2=50\,({\rm GeV/c})^2$. This is another 
indication that the asymptotic behavior is far to be reached for the matrix 
elements $<MS|O|MS>$ and $<MA|O|MA>$.
%%%%%%%%%%%%%%%%%%%%%%%%%%%%%%%%%%%%777777777777%%%%%%%%%%%%%%%%%%%%%%%%%%%%
\section{ Wave function approach to the asymptotic form factors}
Partly to reduce the gap between the two approaches we discussed, partly to 
answer some of the questions raised by the results obtained in the Faddeev 
approach, we here present further developments. The aim is to get an estimate of 
the asymptotic form factor as accurate as possible, starting from the 
observation that the wave function at the origin is rather well determined and 
independent on the approach (see Table 1). Such a program can be in principle 
performed because, similarly to the two-body case discussed in Sect. 3, the form 
factor at high $q$ is expected to only depend on the gluon exchange force, once 
the wave function at the origin is known. The expression for the form factors 
reads in momentum space:
\begin{equation}
<X | O | Y >  = \int \frac{d\vec{\kappa}_{12}}{(2\pi)^3} 
\frac{d \vec{\kappa}_{3} }{(2\pi)^3} \; 
\psi_{X} (\vec{\kappa}_{12}, \vec{\kappa}_3) \; \psi_{Y} (\vec{\kappa}_{12}, 
\vec{\kappa}_3 + \sqrt{ \frac{2}{3} } \; \vec{q} ) .
\end{equation}
Being essentially interested in the asymptotic behavior, some approximation can 
be done. 

We follow the qualitative lines developped by Alabiso and 
Schierholz \cite{ALAB},  whose approach essentially extends to the three-body 
system that one reminded for the two-body case in Sect. 3. Noting that the wave 
function is peaked at small values of the argument, two regions of the 
integration range over the variable $\vec{\kappa}_3$ are important: 
$\vec{\kappa}_3 = 0$ 
and $\vec{\kappa}_3 =- \sqrt{ \frac{2}{3} } \vec{q}$. One thus gets :
\begin{eqnarray}
< X | O | Y >_{q \rightarrow \infty}   =  \left( \int \frac{ d\vec{\kappa}_{12} 
}{(2\pi)^3} \frac{d\vec{\kappa}_3}{(2\pi)^3} \;
\psi_{X} (\vec{\kappa}_{12}, \vec{\kappa}_3) \right) \; \psi_{Y}( 
0,\sqrt{\frac{2}{3}} 
\; \vec{q} ) \;\;\;\;\;\; \nonumber \\
               +  \left( \int \frac{ d\vec{\kappa}_{12} }{(2\pi)^3} 
\frac{d\vec{\kappa}_3}{(2\pi)^3} \; \psi_{Y} (\vec{\kappa}_{12}, \vec{\kappa}_3) 
\right) \;  \psi_{X}( 0,-\sqrt{\frac{2}{3}} \; \vec{q} ).
\end{eqnarray}

Remembering that the integral is nothing but the value of the wave function in 
configuration space at the origin, and that 
only $\psi_{S}(\rho =0) \neq  0$, we come to the conclusion that two among the 
four matrix elements $<S | O | S >$,  $<S | O  |MS >$, \mbox{$<MS| O |MS>$} 
and $ < MA | O | MA >$ have a non zero contribution at the dominant order, 
which is given by $ \psi_Y (\vec{\kappa}_{12} =~0, \vec{\kappa}_3= 
\sqrt{\frac{2}{3}} \; \vec{q} )$ or $\psi_X ( \vec{\kappa}_{12}=0, 
\vec{\kappa}_3=-\sqrt{\frac{2}{3}} \; \vec{q} )$. They are:
\begin{eqnarray}
< S | O | S>_{q \rightarrow \infty}  =  2 \psi_S^{(r)} (\rho = 0) \;
\psi_{S}^{(k)}( \vec{\kappa}_{12}=0, \vec{\kappa}_3=\sqrt{\frac{2}{3}} \vec{q} 
),\\
< S | O | MS>_{q \rightarrow \infty} =   \psi_S^{(r)} (\rho = 0) \;
\psi_{MS}^{(k)}( \vec{\kappa}_{12}=0, \vec{\kappa}_3=\sqrt{\frac{2}{3}} \vec{q} 
).
\end{eqnarray}
The superscripts, $r$ and $k$, remind that the corresponding wave functions, 
which have been defined previously, refer to different spaces.
%%%%%%%%%%%%%%%%%%%%%%%%%%%%%%%%7.1
\subsection{Improving upon the hyperspherical harmonics results}
One can check that when Eqs. (61) and (62) are applied to the numerical 
solutions of the Bhaduri potential at short distances obtained in the 
hypersperical harmonic formalism, Eqs. (37) and (41), the asymptotic behavior of 
the form factors, Eqs. (53) and (54), is reproduced. Concerning  $<MS| O |MS>$ 
and $ < MA | O | MA >$, the next to dominant order approach confirms the 
asymptotic behavior obtained in Eqs. (55) and (56). 

In order to improve the calculation of the form factors in the 
hyperspherical approach, we can try to get a better determination of the wave 
function for high values of the momenta, what can be done iteratively from a 
zeroth order wave function. This one is chosen as the Fourier transform of a 
wave function of the form $ \psi_1(\rho) / \pi^{3/2} = \bar{\alpha} 
e^{-\bar{\beta} \rho} / \pi^{3/2} $ (it may correspond for instance to a 
Coulomb like potential or to the solution of the Bhaduri et al. potential at 
short distances Eq. (37), in which case, $\bar{\alpha}= \alpha_1$, 
$\bar{\beta}= \beta_1$). It reads:
\begin{eqnarray}
\psi^{(0)} ( \vec{\kappa}_{12},\vec{\kappa}_{3} )  & = & \int d\vec{\rho}_{12}\; 
d\vec{\rho}_3 \; e^{-i(\vec{\kappa}_{12}.\vec{\rho}_{12} 
+ \vec{\kappa}_{3} .\vec{\rho}_3) } \; \frac{\bar{\alpha}}{ \pi^{3/2} } \; e^{ 
-\bar{\beta} \sqrt{ \rho^2_{12} +\rho^2_3} } \nonumber \\
& = & \frac{ N^{1/2} }{ ( \kappa^2_{12} + \kappa^2_3 +\bar{\beta}^2 )^{7/2} },
\end{eqnarray}
where 
\begin{displaymath}
\vec{\kappa}_{12} = \frac{\vec{p}_1 -\vec{p}_2 }{\sqrt{2}} ,\; \;\;
\vec{\kappa}_3 = \sqrt{\frac{2}{3} } \; 
(\vec{p}_3-\frac{\vec{p}_1 +\vec{p}_2 }{2} ),\;\;\; 
N^{1/2}=120 \, \pi^{3/2} \, \bar{\alpha} \, \bar{\beta} . \;\;\;\;\; \nonumber
\end{displaymath}
The quantities, $\vec{p}_1, \vec{p}_2$ and $\vec{p}_3$, represent the 
momenta of the quarks 1, 2 and 3. Re\-pre\-senting only the high 
momentum behavior of the wave function, $\psi^{(0)} 
(\vec{\kappa}_{12},\vec{\kappa}_{3})$ and subsequent wave functions don't 
fulfill the usual normalization condition, which would imply a relation 
between $\bar{\alpha}$ and $\bar{\beta}$, see Eq. (98).

For an interaction dominated by a Coulomb force (or a Yukawa one) and for the 
wave function given by Eq. (63), it is possible 
to calculate analytically the wave function by an iteration procedure up to 
first order. Thus from the Schr\"odinger equation, 
\begin{equation}
|\psi>=\frac{1}{E-T} \, V |\psi>
 =\frac{1}{E-T} \, ( V_{12}+ V_{13}+V_{23}) |\psi>,
\end{equation} 
we can write (as it is done in the Faddeev formalism):
\begin{equation}
\psi=\psi_{12}+\psi_{13}+\psi_{23},
\end{equation}
so that 
\begin{equation}
\psi^{(1)}_{ij}=\frac{1}{E^{(0)}-T} V_{ij} \psi^{(0)}.
\end{equation}
The contribution due to the force 
between particles 1 and 2, which can be identified to the Faddeev amplitude 
$\psi_{12,3}$, Eq. (27), but in momentum space, reads: 
\begin{eqnarray}
\psi^{(1)}_{12} (\vec{\kappa}_{12},\vec{\kappa}_{3}) =  \frac{2 \sqrt{2} 
}{15\pi}   
(\kappa  +  \kappa'_{\sigma})  \, m_q \hspace*{6.5cm} \nonumber \\
\times \left(
\frac{3 \; N^{1/2}}{4\kappa_{12}(\kappa^2_{12}+\kappa^2_3+\bar{\beta}^2)^{7/2}}  
\, {\rm log}(\frac{ \sqrt{ \kappa^2_{12}+\kappa^2_3+\bar{\beta}^2}+\kappa_{12} }
{  \sqrt{ \kappa^2_{12}+\kappa^2_3+\bar{\beta}^2} - \kappa_{12} } ) 
\;\;\;\;\;\;\;\;\;\;\;\;\;\;\;\;\;\; \nonumber \right. 
\\  \left. 
+ \frac{3}{2} \, \frac{N^{1/2}}{ (\kappa^2_{12}+\kappa^2_3+\bar{\beta}^2)^{3} 
(\kappa^2_3+\bar{\beta}^2)} 
        +     \frac{N^{1/2}}{ (\kappa^2_{12}+\kappa^2_3+\bar{\beta}^2)^{2} 
(\kappa^2_3+\bar{\beta}^2)^2} \right).  
\end{eqnarray}

The total wave function is obtained by adding to (67) the contributions where 
the role of particles 1 and 3 or 2 and 3 are exchanged, corresponding to the 
interaction of the pairs 23 or 13. As a consistency check, it can be verified 
that avera\-ging the total wave function over the hyperspherical angle $\phi$ 
with the appropriate weight factor in the differential element of volume, 
${\rm sin}^2\phi \; {\rm cos}^2\phi$, allows one to recover the zeroth 
order wave function (63). This is achieved provided that the following 
condition is fulfilled for a Coulomb potential with strength, $\kappa  +  
\kappa'_{\sigma}$:
\begin{equation}
\bar{\beta} = \frac{8 \sqrt{2}}{5\pi} \, (\kappa  +  \kappa'_{\sigma} ) \; m_q.
\end{equation}

Expression (67) can also be used to calculate the mixed symmetry component. In 
such a case, the separate amplitudes with the pair of particles 12 having spin 0 
and 1 should be associated with the appropriate strength of the force, 
$\kappa + 3 \kappa'_{\sigma}$ and $\kappa-\kappa'_{\sigma}$. The Yukawa nature 
of the spin-spin force may also be accounted for. In our opinion, the wave 
function so obtained may be usefully compared to the two amplitude Faddeev 
calculation presented in Sect. 4.1. It certainly has not all the physics 
contained in this one, especially that related to the confinement but it 
probably has a large part of the physics relative to the description of the form 
factor at high momentum transfers. In this respect, it is noticed that the 
amplitude (67) behaves like $\kappa^{-8}_3$ when $ \kappa_3$ goes to $\infty$, 
while the zeroth order amplitude (63) behaves like $\kappa_3^{-7}$. As it will 
be shown below, this property determines the behavior of the form factor at 
high $q$ for the dominant contributions.

The expression of the total wave function (28) together with that of the 
individual contributions given by (67) for one of them can be used to improve 
the form factors calculated in the hyperspherical formalism. In comparison with
the most complete 8 amplitude Faddeev calculation, the form factors, 
$< S | O | S >$ and $< S | O | MS>$, obtained on the basis of Eqs. (61, 62),
are found to be smaller by a factor 2-3 in the asymptotic domain around $50 \, 
({\rm GeV/c})^2$. As to the ratio, $\frac{<S|O|MS>}{<S|O|S>}$, an expectation 
can be obtained in the limit where one retains the dominant 
contribution (due to that part of the total wave function involving the terms 
obtained from Eq. (67) 
by exchanging the role of particles 1 and 3 or 2 and 3):
\begin{equation}
(\frac{<S | O | MS>}{<S | O | S >} )_{q \rightarrow \infty} = - 
\frac{1}{\sqrt{2}} \,
\frac{ \kappa'_{\sigma}  }{\kappa  +  \kappa'_{\sigma}  }   \simeq - 0.45.
\end{equation}
It is found too small by a factor 1.5. Notice that the above factor in Eq. (69) 
is smaller than in Eq. (58) by a factor 4, showing once more the failure of the 
hyperspherical harmonic formalism to make accurate predictions in the present 
domain when it is restricted to the lower K values. As to the other form 
factors, $<MS | O | MS >$ and $ < MA | O | MA >$, a prediction made in the 
spirit  of Eqs. (61, 62), but accounting for the minimal dependence of wave 
functions on $\vec{\kappa}_{12}$ or $\vec{\kappa}_3$, leads to a $q^{-10} \; 
{\rm log}\,q$ dependence of these quantities at high $q$. This implies that they 
should tend to zero with $q$ more rapidly than the matrix elements, 
$< S | O | S >$ and $< S | O | MS>$, while the full calculations rather indicate 
an opposite situation around $q^2=50\,({\rm GeV/c})^2$. The slow convergence to 
the asymptotic behavior we do find in some cases, especially for the form 
factors, $<MS | O | MS >$ and $ < MA | O | MA >$, calculated in the 
hyperspherical harmonic formalism, invites to some caution in concluding from 
this discrepancy.
 
In order to understand these results and their difference with the most complete 
ones, let us note that to get a $q^{-8}$ asymptotic dependence of the 
form factors, one has to pick up terms proportional to $\rho^2$ in the 
expansion of the product of the initial and  final wave functions at small 
distances. This follows from dimensional arguments (the elementary volume 
involves six powers of $\rho$). To get such terms, one can combine  $\rho^2$ 
terms coming from the wave function in the initial (final) state with the zeroth 
order wave function (the wave function at the origin) in the final (initial) 
state. The contribution of these ones to the form factors are likely to be 
accounted for by Eqs. (61, 62). One can also combine terms of the first 
order in $\rho$ in each state. These ones have obviously no contribution  
to the asymptotic expressions of the form factors given by Eqs. (61, 62). 

The failure of Eqs. (61, 62) to account for the asymptotic behavior can also be 
analyzed by considering the structure of the wave functions in momentum space. 
With a wave function as simple as (63), the dominance of the low momenta 
components in the matrix element (59), which is essential to get Eqs. (61, 62), 
necessarily applies to either wave function of the initial or the final state. 
With the wave function (67), which results from a first iteration over the 
previous one, but contains different factors, other possibilities are offered. 
The dominance of the contribution of low momentum components may involve one 
factor in the initial state wave function with respect to some integration 
variable and a factor in the final state wave function with respect to another 
integration variable. This is illustrated by the following contribution 
of the dominant part of a matrix element (Eq. (59)) involving the Faddeev 
amplitudes 
$\psi_{13,2}\; \psi_{23,1}$: 
\begin{eqnarray}
 \left[ \int  d\vec{\kappa}_{12} \; d\vec{\kappa}_3 \; 
 \frac{1}{ (\kappa^2_{12}+\kappa^2_3+\bar{\beta}^2)^2 \;\; ( \, (-\frac{1}{2} 
\vec{\kappa}_3 - \frac{\sqrt{3}}{2} \vec{\kappa}_{12} )^2 + \bar{\beta}^2  )^2  
} 
    \;\;\;\;\;\;\;\;\;\;\;\;\;\;\;\;\;\;\;\; \right. \nonumber \\  \left.
   \frac{1}{   (\,(-\frac{1}{2}(\vec{\kappa}_3 + \sqrt{\frac{2}{3}} \vec{q} ) + 
\frac{\sqrt{3}}{2} \vec{\kappa}_{12} )^2 + \bar{\beta}^2)^2 
     \;\; (  \kappa^2_{12} + ( \vec{\kappa}_3 + \sqrt{\frac{2}{3}} \vec{q} )^2  
+  \bar{\beta}^2  )^2} \right] \simeq 
   \;\;\;\;\;\;\;\;\; \nonumber \\ 
 \left[ \int   \frac{1}{ (\kappa '^2_{12} + \kappa '^2_3 + \bar{\beta}^2 )^2 \; 
(\kappa '^2_3 + \bar{\beta}^2 )^2 
 ( \, (-\frac{1}{2} (\vec{\kappa}'_3 + \sqrt{\frac{2}{3}} \vec{q} ) - 
\frac{\sqrt{3}}{2} \vec{\kappa}'_{12} )^2+\bar{\beta}^2  )^2  }  
 \;\;\;\;\; \right.  \nonumber \\ 
   \left. \frac{1}{ ( \,(\frac{1}{2} \vec{\kappa}'_{12} + \frac{\sqrt{3}}{2} 
\vec{\kappa}'_3 )^2   + 
     (  -\frac{1}{2} \vec{\kappa}'_3 + \frac{\sqrt{3}}{2} \vec{\kappa}'_{12} + 
\sqrt{\frac{2}{3}} \vec{q} )^2 + \bar{\beta}^2   )^2 \; } 
   d\vec{\kappa }'_{12} \; d\vec{\kappa }'_3    \right]_{\kappa '_{12},\kappa 
'_3 \sim \bar{\beta}}  \nonumber \\ 
 + \left[ \int  \frac{1}{ (\kappa ''^2_{12}+\kappa ''^2_3+\bar{\beta}^2)^{2} \; 
(\kappa ''^2_3 + \bar{\beta}^2 )^2 
  \; ( \,(-\frac{1}{2}(\vec{\kappa}^{''}_3 - \sqrt{\frac{2}{3}} \vec{q} ) - 
\frac{\sqrt{3}}{2} \vec{\kappa}^{''}_{12} )^2+\bar{\beta}^2  )^2 } 
  \right. \nonumber \\ 
  \left. \frac{1}{ ( \,(\frac{1}{2} \vec{\kappa}^{''}_{12} + \frac{\sqrt{3}}{2} 
\vec{\kappa}^{''}_{3} )^2   + 
        (  -\frac{1}{2} \vec{\kappa}^{''}_3 + \frac{\sqrt{3}}{2} 
\vec{\kappa}^{''}_{12} -\sqrt{\frac{2}{3}} \vec{q} )^2+\bar{\beta}^2   )^2 \;  
}      d\vec{\kappa }^{''}_{12} \; d\vec{\kappa }^{''}_3    \right]_{\kappa 
''_{12},\kappa ''_3 \sim \bar{\beta}}    \nonumber \\  
 + (\frac{2}{\sqrt{3}})^{3} \left[ \int  
 \frac{1}{ ( \,(\frac{1}{3}(\vec{\kappa}'_3 + 2 \vec{\kappa}^{''}_3 + 
\sqrt{\frac{2}{3}} \vec{q})^2 + \kappa '^2_3 + \bar{\beta}^2 )^2 \; 
                   (\kappa '^2_3 + \bar{\beta}^2 )^2 \; (\kappa ''^2_3 + 
\bar{\beta}^2 )^2 }                    \right.          \nonumber \\  \left.
 \frac{1}{ ( \, (\frac{1}{3}(\vec{\kappa}^{''}_3 + 2 \vec{\kappa}'_3 - 
\sqrt{\frac{2}{3}} \vec{q})^2 + \kappa ''^2_3 + \bar{\beta}^2 )^2 } 
  d\vec{\kappa}'_3 \; d\vec{\kappa}^{''}_3   \right]_{\kappa '_3,\kappa ''_3 
\sim \bar{\beta} } .     
\end{eqnarray}

The two first terms on the r.h.s. of (70) are nothing but a contribution 
similar to that given in Eq. (60). They correspond to the processes represented 
in Fig. 9a. The third term involves the short range radial dependence of 
wave functions of both the initial and final states. It corresponds to the 
contribution of a diagram like that drawn in Fig. 9d. 

The integration in the above equation can be performed in the limit of large q, 
where the $q^{-8}$ dependence factorizes out. The result is: 
\begin{equation}
 \left( \frac{128 }{3 \pi} + 81(\frac{2}{\sqrt{3}})^{3} \right) 
 \frac{\pi^4}{q^8 \bar{\beta}^2},
\end{equation}
where the two terms respectively correspond to the two first terms and the 
third one on the r.h.s. of Eq. (70).
It is worthwhile to make a comparison with the contribution of a ``diagonal 
term", $\psi_{13,2}\; \psi_{13,2}$,
or $\psi_{23,1}\; \psi_{23,1}$, given by:
\begin{equation} 
 \frac{16}{3\pi}  \frac{\pi^4}{q^8 \bar{\beta}^2}.
\end{equation}
\noindent
It is seen that the contribution from a non-diagonal term, Eq. (71), is more 
efficient to contribute at high $q$, roughly a factor 50 larger. This 
is in relation with the expectation that the form factor at high $q$ requires a 
momentum transfer from the quark which is stroken 
by the photon to the two other quarks. It is likely that a more complete 
calculation will produce for 
the ``diagonal term'' an extra $q^{-1}$ factor at least, which the above 
suppression factor partly accounts for (see Sect. 7.3 below).

\begin{figure}[htb!]
\begin{center}
\mbox{ \epsfig{ file=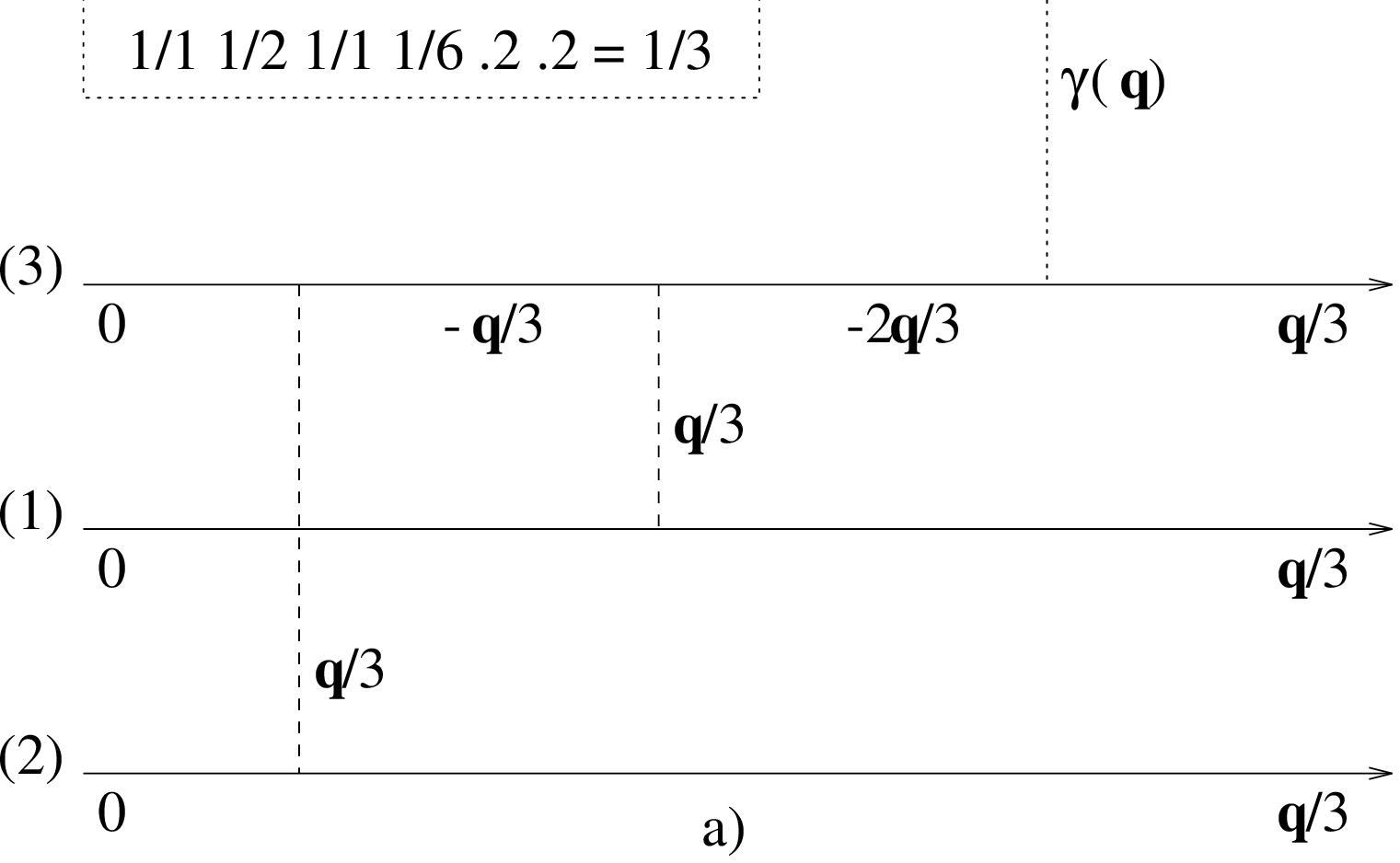, angle=270, width=8cm}} 
\vspace*{2mm}
\end{center}
\begin{center}
\mbox{ \epsfig{ file=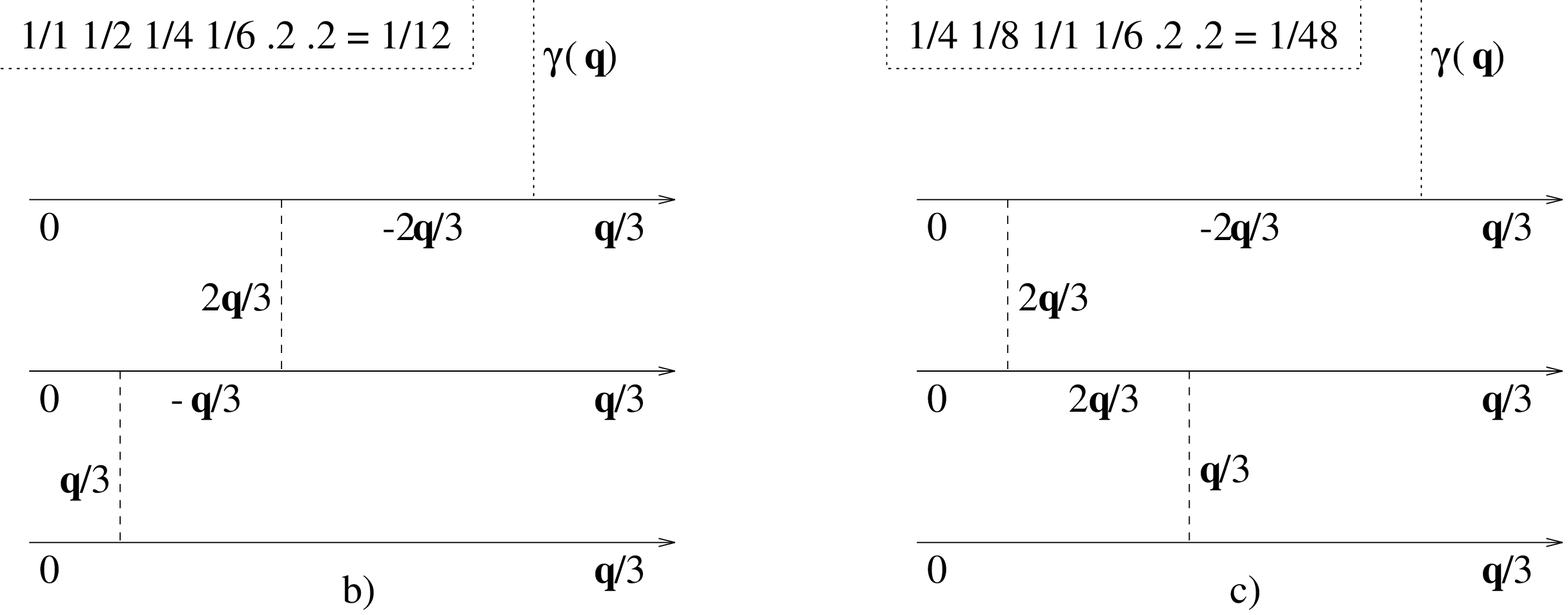, angle=270, width=13cm}}
\vspace*{2mm} 
\end{center}
\begin{center}
\mbox{ \epsfig{ file=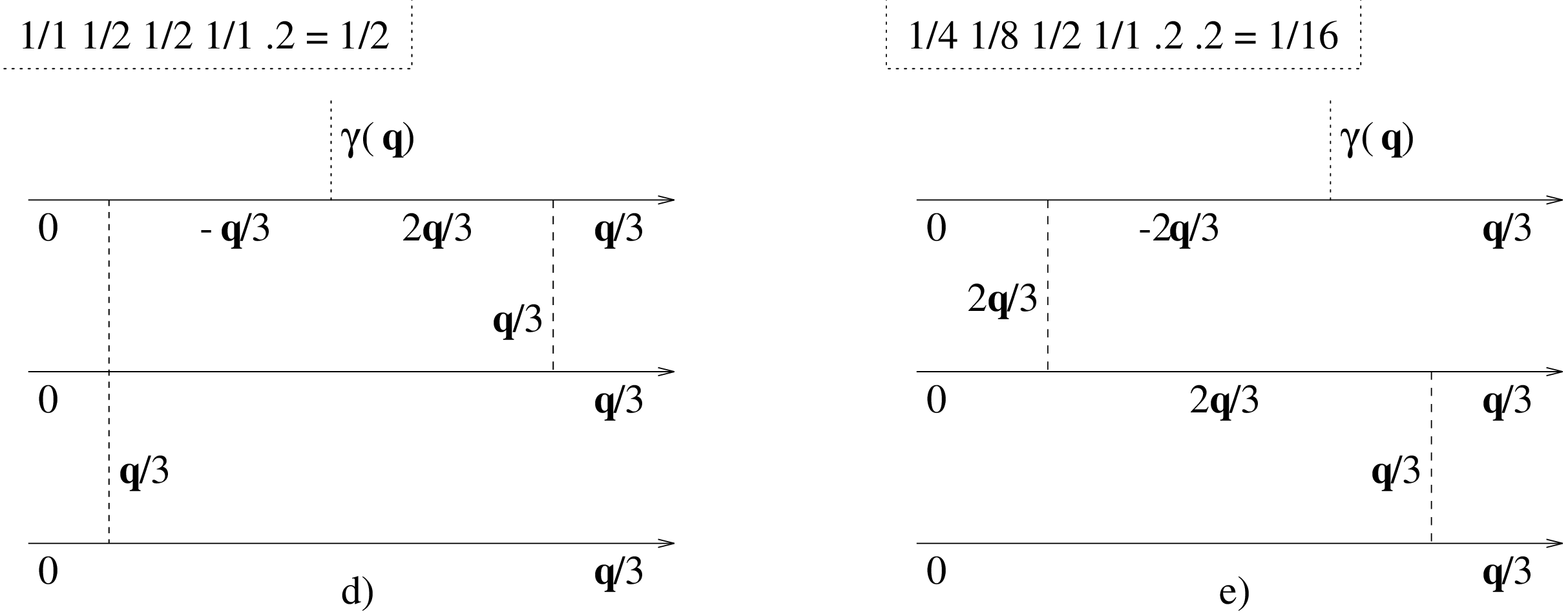, angle=270, width=13cm}} 
\vspace*{-2mm} 
\end{center}
\caption{\small{Representation of diagrams contributing to the asymptotic form 
factor 
with the indication of various factors entering the gluon or quark propagators. 
These ones, which can be written as $1/( n\,(\frac{\vec{q}}{3})^2)$, are 
represented in the figure by $1/n$, the factor $(\frac{\vec{q}}{3})^2$ being 
factored out. Reading diagram a) from left to right for instance, one explicitly 
gets: $n=1$ for the first gluon propagator, $n=1+1$ for the first quark 
propagator, $n=1$ for the second gluon propagator and $n=4+1+1$ for the second 
quark propagator.  The weight accounting for undisplayed diagrams 
is also indicated: a factor 2 for similar ones where the role of 
the final and initial states is exchanged and, in most cases, a factor 2 for 
similar diagrams where the role of particles 1 and 2 are exchanged. Factors 
accounting for the different spin-isospin components of the wave function are 
not incorporated here (rather see Table 2).}}
\end{figure}
\clearpage

%%%%%%%%%%%%%%%%%%%%%%%%7.2  
\subsection{Further development}
From Eq. (59), and the wave functions provided by Eq. (67), one can now make a 
more complete calculation of the dominant term at high $q$, including some 
contribution due to the spin-spin force. The different form factors then read:
\begin{eqnarray}
<S | O |S> \;\; =   \frac{729 \sqrt{3}}{\pi} \frac{(\bar{\alpha} \, 
m_q)^2}{q^8} \;\;\; \hspace*{6.2cm} 
  \nonumber \\
 \;\;\;\;\;\;\;\;\; [ \frac{512}{15\pi} ( \frac{1}{\sqrt{3}}  + \frac{1}{48} 
 {\rm log} \frac{2+\sqrt{3}}{2-\sqrt{3}} ) \; + \; 9 \; ] 
 (\kappa+\kappa'_{\sigma})^2, \nonumber \\
<S | O |MS> \;   =  - \frac{729 \sqrt{6}}{\pi} \frac{(\bar{\alpha}\,m_q)^2}{q^8} 
\hspace*{6.2cm} \nonumber \\
 \;\;\;\;\;\;\;\;\; [ \frac{32}{5\pi} ( \frac{1}{\sqrt{3}} + \frac{1}{36} 
 {\rm log} \frac{2+\sqrt{3}}{2-\sqrt{3}} ) \; + \; \frac{15}{4}] 
 (\kappa+\kappa'_{\sigma}) \, 2\kappa'_{\sigma},  \nonumber \\
<MS | O |MS>  \;  =  \; \frac{729 \sqrt{3}}{\pi} 
\frac{(\bar{\alpha}\,m_q)^2}{q^8}  
 \; [\; 0 \; + \; 3 \;] (2 \kappa'_{\sigma})^2, \hspace*{3.2cm} \;\;\;
  \nonumber \\
<MA | O |MA>  =  \frac{729 \sqrt{3}}{\pi} \frac{(\bar{\alpha}\,m_q)^2}{q^8}  
 \; [\; 0 \; - \; 12 \;] (2 \kappa'_{\sigma})^2, \hspace*{3.4cm}\; 
\end{eqnarray} 
The first term in the squared brackets corresponds to the improvement 
essentialy based on Eqs. (61, 62) discussed in Sect. 7.1. The second 
one corresponds to a further improvement, which goes beyond these equations 
and was discussed at the end of the same section. The two terms have the same 
origin as the two first terms in the r.h.s. of Eq. (70) on the one hand and the 
third one on the other. They can be shown to have a close relation with the 
diagrams of Fig. 9 (see below). 

A slightly different but more complete expression, as far as the asymptotic 
behavior and the effect of the spin-spin force are concerned, can be obtained. 
We start from a zeroth order, Eq. (115), strongly peaked at zero momenta, and  
proceed to two iterations obtaining a wave function, Eq. (120),  which is more 
appropriate than  Eq. (67) as far as the short range correlation is concerned 
(no average over the $\phi$ angle), but is not normalizable. Then, the 
asymptotic behavior is given by :
\begin{eqnarray}
<S |O |S> & = & \frac{729\sqrt{3}}{\pi} \frac{(\bar{\alpha}\,m_q)^2}{q^8} \; 
      [7 ( (\kappa + \kappa'_{\sigma})^2 - 2 \kappa '^2_{\sigma} ) \, C + 9 
(\kappa+\kappa'_{\sigma})^2] \nonumber \\
& = & (3.1C+17.9) \, ({\rm GeV/c})^8 \, q^{-8} \nonumber \\
& = & (19.8,\, 21.0) \, ({\rm GeV/c})^8 \, q^{-8} ,\nonumber \\
<S |O |MS> & = & -\frac{729\sqrt{6}}{\pi} \frac{(\bar{\alpha}\,m_q)^2}{q^8} \; 
      [\frac{9}{4} \, C +\frac{15}{4}] 
         (\kappa+\kappa'_{\sigma}) \, 2\kappa'_{\sigma} \;\;\;\; \nonumber \\
    & = &  -(7.9C+13.2) \, ({\rm GeV/c})^8 \, q^{-8}   \nonumber \\
    & = & - (17.9,\, 21.1) \, ({\rm GeV/c})^8 \, q^{-8}  ,   \nonumber \\
<MS |O |MS> & = & \frac{729\sqrt{3}}{\pi} \frac{(\bar{\alpha}\,m_q)^2}{q^8} 
 \; [\; 0 \; + \; 3 \;] (2 \kappa'_{\sigma})^2 \nonumber \\
& = & 9.3 \, ({\rm GeV/c})^8 \, q^{-8}  ,\nonumber \\
<MA |O |MA> & = & \frac{729\sqrt{3}}{\pi} \frac{(\bar{\alpha}\,m_q)^2}{q^8} 
 \; [\; 0 \; - \; 12 \;] (2 \kappa'_{\sigma})^2 , \nonumber \\
 & = & -37.2 \, ({\rm GeV/c})^8 \, q^{-8} .
\end{eqnarray} 
The factor $C$ accounts for extra non-perturbative effects, not included 
in the value of the wave function at the origin, $\bar{\alpha}$ (see appendix 
C). Although there is no close relationship, differences between (73) and (74) 
can give some insight on the effect of the truncation of the Faddeev 
amplitude (remember that the wave function from (67) can be assimilated to the 
two amplitude Faddeev calculation). It is noticed that they only affect the 
first term of each bracket. The main difference is represented by the 
contribution of the term $ -2 \kappa '^2_{\sigma}$ to $<S |O |S>$. This one 
could not occur in results presented in Eq. (73), which were obtained using the 
fully spatially symmetric component of the wave function, Eq. (63), as a zeroth 
order. The absence of difference 
for the second term in the bracket is due to the fact that, once the wave 
function at the origin, $\bar{\alpha}$, is determined, the calculation involves 
the first order terms of an expansion of the wave function around the 
origin. These terms, see Eqs. (113, 114), are completely determined 
by the $ \frac{1}{r}$ behavior of the potential at short distances. The 
numerical values have been calculated using $\bar{\alpha} = 19.64 \, {\rm 
fm}^{-3}$, $ m_q=337 \, {\rm MeV} $, $\kappa=102.67 \,{\rm MeV} \, {\rm fm} $ 
and $\kappa'_{\sigma}= 1.66 \, \kappa$, the range given in the brackets being 
determined by the values  $C=0.6$ and $C=1.0$. This range can be directly 
compared to the  values that form factors  shown in Fig. 8 are taking in the  
asymptotic domain.

\begin{table}[htb!]
\caption{\small{Detail of the coefficients entering Eq. (74) with indication of 
the two interacting quark pairs and the diagram of Fig. 9 they 
correspond to.}}
\begin{center}
\begin{tabular}{lcccc}
\hline
     &            &                             &             &           \\ 
          & \small{$<S|O|S>$}  & $\frac{<S|O|MS>}{\sqrt{2}}$ & 
     \small{$<MS|O|MS>$} & \small{$<MA|O|MA>$} \\
[2.ex]\hline
      &            &                             &             &           \\ 
       $V_{23}V_{13}O$    & $\frac{16}{3}$  & $-\frac{8}{3}$      & 0 &  0 \\ 
  a                &                 &                     &   &    \\ [2.ex]
$V_{12}(V_{13}+V_{23})O$  & $\frac{4}{3}$ &  $\frac{1}{3}$ & 0 &  0 \\ 
  b                &                 &                     &   &    \\ [2.ex]
$(V_{13}+V_{23})V_{12}O$  & $\frac{1}{3}$ & $\frac{1}{12}$ & 0 &  0 \\ 
  c                &                 &                     &   &    \\ [2.ex]
${\rm sum}$               &     7        & -$\frac{9}{4}$  & 0 &  0 \\ [3.ex]
$V_{23}OV_{13}$           &     8        &   -4            & 4 & -12 \\ 
  d                       &          &                     &   &    \\ [2.ex]
$(V_{13}+V_{23})OV_{12}$  &    1    &  $\frac{1}{4}$       & -1 & 0 \\ 
  e                       &         &                      &   &    \\ [2.ex]
${\rm sum}$               &    9    & -$\frac{15}{4}$      & 3 & -12 \\ [3.ex] 
\hline
\end{tabular}
\end{center}
\end{table}

The different factors appearing in Eq. (74) can also be directly obtained, 
up to an overall factor, from considering the five types of diagrams shown 
in Fig. 9. The detail is given in Table 2. Three of them (a, b, c) involve 
two gluon exchanges in the initial or the final state wave functions. The 
two other ones (d, e)  involve one-gluon exchange in both the initial 
and the final state wave functions. These two type contributions 
correspond to the first and second terms in the squared bracket appearing in 
Eqs. (74). They involve the wave function at the origin multiplied by its second 
derivative, $\psi^{(r)}(0) \, \psi^{(r)''}(0)$, for the first one and the 
product of the derivatives at the origin, $\psi^{(r)'}(0) \, \psi^{(r)'}(0)$, 
for the second one. These expressions are those that are relevant for 
determining the asymptotic behavior of the form factor for the three-body case, 
while the expression, $\psi^{(r)}(0) \, \psi^{(r)'}(0)$, is appropriate for the 
two-body case (incidentally, this expression  works for the hyperspherical 
harmonic calculation limited to the wave $K=0$ (see Eq. (53) together with Eqs. 
(37, 38)). We finally remark that, at high $q$, the combination of the various 
form factors, which corresponds to a scalar-isoscalar quark density, has the 
peculiarity to factorize into a simple form: 
\begin{eqnarray}
<S |O |S> +\frac{1}{2} (<MS|O|MS>+<MA|O|MA>) = 
\;\;\;\;\;\;\;\;\;\;\;\;\;\;\;\;\;\;\; \nonumber \\ 
\frac{729\sqrt{3}}{\pi} \, \frac{(\bar{\alpha}\,m_q)^2}{q^8} 
\; [\; 7 \, C \;+ \; 9 \;]\;
( \, (\kappa + \kappa'_{\sigma})^2 - 2 \kappa'^2_{\sigma}\, ).
\end{eqnarray}
%%%%%%%%%%%%%%%%%%%%%%%%%%%%%%%
\subsection{Remaining discrepancies} 
The comparison of the Faddeev results in the asymptotic domain with the 
expectation given by the matrix element $<S |O |S>$ of Eq. (74) 
improves over that made previously in this section (7.1). On the other hand, 
the results of the eight amplitude Faddeev calculations 
are definitively closer to what is expected from (74) than the two 
amplitude Faddeev calculations. The remaining discrepancies may have their 
origin either in the truncation of the total Faddeev amplitude or in the fact 
that the asymptotic behavior has not been reached yet, unless both are related. 
At present, it is not clear whether the discrepancies should 
correspond to corrections of the order, $\frac{1}{q}$, as in the case of 
hyperspherical harmonic results obtained previously, see Eqs. (55) and (56), 
or of the order,  $\frac{1}{q^2}$ (up to ${\rm log} \, q $ terms). Numerically, 
either one is possible in the higher range of momentum transfers we explored.  

From examining the general expression for the form factor, Eq. (59), or more 
particularly the expression for the wave function obtained from Eq. (67), it is 
reasonable to consider that the theoretical corrections to the leading term 
should be of the order $\frac{1}{q^2}$ (up to ${\rm log} \, q $ terms) in a 
complete calculation. A rigorous demonstration is difficult however. Any 
attempt to show it from an expansion of the form factor in powers 
of $\frac{1}{q}$ indicates the existence of terms that linearly diverge 
in the limit $q \rightarrow \infty$, as it can be checked from the expression 
of a particular contribution, see third term in the r.h.s. of Eq. (70) 
(discarding the limitation $\kappa '_3,\kappa ''_3 \sim \bar{\beta}$ ). In 
practice, they are cut off by a factor $q$, hence corrections of relative 
order $\frac{1}{q}$. It is possible that the total sum of these 
terms cancels out, leaving a correction of relative 
order $\frac{1}{q^2}$, while in an incomplete calculation they would partly 
survive. An example, though referred to the two-body case, is 
detailed in the appendix D. It also provides some hint for making rigorous 
statements as to corrections of the relative order $\frac{1}{q^2}$ to 
the leading order contributions. 

On the other hand, from $QCD$ based 
calculations, the opinion is that corrections should be of the 
order $\frac{1}{q^2}$  \cite{MULL}, but this estimate incorporates the full 
quark propagator and, especially, accounts for Z-type diagrams neglected here. 
One cannot infer from this calculation that the present one should 
evidence similar features. Notice that $QCD$ based arguments hold up 
to ${\rm log} \, q $ terms. These ones partly originate from a relativistic 
treatment and should not be confused with those mentioned above, which are 
due to the three-body nature of the problem under consideration. 
The absence of convergence of some form factors, $<MA|O|MA>$ in both the 
hyperspherical and Faddeev aproaches and $<MS|O|MS>$ in the hyperspherical 
approach, makes the situation more confusing. As they are sensitive 
to the spin-spin part, one may expect them to depend on its short range 
behavior, which deviates from a Coulomb type potential we considered in our 
theo\-retical analysis. Assuming that the effect of the exponential factor in 
the force can be approximated by exp(-$\rho /r_0$) with $\rho = 3/q$, part 
of the large dependence of the above form factors could be explained. No 
similar dependence is seen in the other form factors however. Furthermore, 
one should not forget that we are dealing with a coupled channel problem, 
which can change the behavior of wave functions at short distances with 
respect to expectations based on perturbative arguments. In fact, this is a 
likely explanation for the slow convergence of the form factors, $<MA|O|MA>$  
and $<MS|O|MS>$, to their asymptotic behavior.
%%%%%%%%%%%%%%%%%%%%%%%%%%%%%%%%%
\subsection{Physical form factors}
Throughout this section and the previous one, we discussed particular 
contributions to the physical form 
factors with the idea to facilitate the analysis of their asymptotic behavior. 
Using Eqs. (74), the expressions  of the physical form factors in the 
asymptotic 
regime are: 
\begin{eqnarray}     
G_E^p(\vec{q}^{2}) =  \frac{729\sqrt{3}}{\pi} 
\frac{(\bar{\alpha}\,m_q)^2}{q^8}
  \hspace*{7.9cm} \nonumber \\  
  \left( (7C+9)(\kappa + \kappa'_{\sigma})^2 -(9C+15) (\kappa+\kappa'_{\sigma}) 
  \, \kappa'_{\sigma} -(14C+18)\kappa'^2_{\sigma} \right) \nonumber \\ 
  =(-8.1C-14.7) \, ({\rm GeV/c})^8 \, q^{-8} 
  = -(19.6, \, 22.8) \,({\rm GeV/c})^8 \, q^{-8}, 
  \;\;\;\;\;\;\nonumber \\
G_E^n(\vec{q}^{2}) =  \frac{729\sqrt{3}}{\pi} 
\frac{(\bar{\alpha}\,m_q)^2}{q^8}
  (9C+15) (\kappa+\kappa'_{\sigma}) \, \kappa'_{\sigma} 
  \hspace*{4.3cm}  \nonumber \\
  =(11.2C+18.6) \, ({\rm GeV/c})^8 \, q^{-8}
  = (25.3, \, 29.8) \, ({\rm GeV/c})^8 \, q^{-8} ,\;\;\;\;\;\;\;\;\;\; 
  \nonumber \\ 
G_M^p(\vec{q}^{2}) / \mu_p =  \frac{729\sqrt{3}}{\pi} 
  \frac{(\bar{\alpha}\,m_q)^2}{q^8} \, \frac{m_N}{m_q \, \mu_p} 
  \hspace*{6.1cm} \nonumber \\ 
  \left( (7C+9)(\kappa + \kappa'_{\sigma})^2 -(9C+15) (\kappa+\kappa'_{\sigma}) 
  \, \kappa'_{\sigma} -(14C-14)\kappa'^2_{\sigma} \right) \nonumber \\ 
  =(-8.2C+10.2) \, ({\rm GeV/c})^8 \, q^{-8}
  = (5.3, \, 2.0) \, ({\rm GeV/c})^8 \, q^{-8} , 
\;\;\;\;\;\;\;\;\;\;\;\; \nonumber \\ 
G_M^n(\vec{q}^{2}) / \mu_n = \frac{729\sqrt{3}}{\pi} 
\frac{(\bar{\alpha}\,m_q)^2}{q^8} \, \frac{m_N}{m_q \, \mu_n}  
\hspace*{6cm} \nonumber \\
\left( -(\frac{14}{3}C+6)(\kappa + \kappa'_{\sigma})^2 +(3C+5) 
(\kappa+\kappa'_{\sigma}) \, \kappa'_{\sigma} 
+(\frac{28}{3}C-20)\kappa'^2_{\sigma} \right) \nonumber \\ 
=(-2.6C+32.3) \, ({\rm GeV/c})^8 \, q^{-8}
= (30.7, \, 29.7) \, ({\rm GeV/c})^8 \, q^{-8}. 
\;\;\;\;\;\;\;\;\;\;
\end{eqnarray}
The negative sign for $G_E^p(\vec{q}^{2})$, which implies a change in sign 
for this form factor between $0$ and $\infty$, is in agreement with numerical 
calculations presented in Fig. 6. This feature, which contradicts what is 
generally expected from QCD, has its origin in the unrealistic value 
of $\kappa'_{\sigma}$ in the Bhaduri et al's quark-quark force 
(see next section). As expected from Eq. (76) above, the proton magnetic 
form factor is much smaller than the neutron one in the asymptotic regime. 
Again, this feature is strongly sensitive to the value of $\kappa'_{\sigma}$.

Comparison of the different calculations of the form factors considered 
along the text with Eqs. (76) shows that the 8 amplitude 
Faddeev calculation is the most convergent to the asymptotic behavior. 
It cannot be said however that it has definitively been reached 
in the range $30-50 \, ({\rm GeV/c})^2$, 
confirming what can be inferred from examination of Fig. 8. 
%%%%%%%%%%%%%8888888888888888888888%%%%%%%%%%%%%%%%%%%%%%%%%%%%%%%%%%%%%%%%%%%
\section{Making the calculations of the form factors more realistic}
In the previous sections, we mainly discussed the reliability of predictions 
with respect to technical aspects of the calculations regarding the description 
of the nucleon in terms of quarks. Any comparison with measurements was eluded. 
The reason for that is the existence of other effects to be accounted for. They 
are successively discussed in the next subsections.
\subsection{Quark form factors}
\begin{figure}[htb!]
\begin{center}
\mbox{ \epsfig{ file=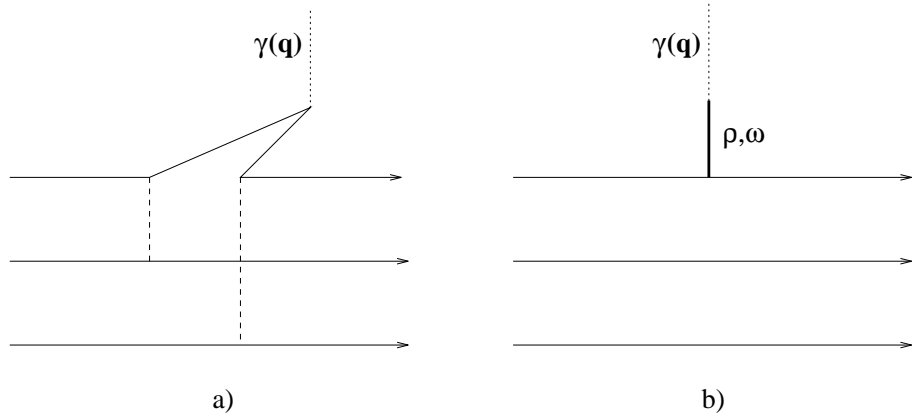, angle=270, width=12cm}} 
\vspace*{-3mm} 
\end{center}
\caption{\small{Extra contributions to form factors that have to be included, 
due to a coupling to a quark-antiquark pair (a) or a vector meson (b).}}
\end{figure}  

In a constituent quark model like the one we used, it is quite legitimate to 
account for some electromagnetic form factor of the quarks. Contributions due to 
the coupling of the photon to a quark-antiquark pair, shown in Fig. 10a, or to a 
vector meson, $\omega$ or $\rho$, shown in Fig. 10b, cannot be generated by the 
non-relativistic constituent quark model and, consequently, their effect has to 
be added by hand \cite{VALE,CANO2}. Processes represented in Fig. 10b lead to a 
quite successful phenomenology at the nucleon (and meson) level, which 
is known as vector meson dominance. This can be transposed at the quark level 
without any change. The same statement partly holds for the contribution 
of the pion cloud to the proton and neutron form factors, which may be 
accounted for in some part in the $\rho$ contribution. Some contribution 
may also come from the Darwin-Foldy term (for the charge density), but this one 
may be partly incorporated in the $\omega$ or $\rho$ contribution. 

For the proton charge squared radius, the latter can contribute an amount of 
$0.4 \, {\rm fm}^2$ to be added to the matter squared radius, $0.25 \, {\rm 
fm}^2$. The pion cloud, for that part not included in the $\rho$ meson can 
contribute an extra $0.10-0.15 \, {\rm fm}^2$, making the proton and neutron 
charge squared radius close to the measured ones. At non-zero momentum 
transfers, the same mechanism tends to make the prediction for the form factors 
closer to measurements in the range, $q^2=0-1 \; ({\rm GeV/c})^2$ \cite{CANO2}. 
At higher momentum transfers, the hypothesis of the vector meson dominance 
doesn't lead to any improvement. On the contrary, assuming the full validity 
of this hypothesis, it adds an extra factor, $q^{-2}$, to the discrepancy by a 
factor $q^{-4}$ between the result of non-relativistic calculations and the 
$QCD$ expectations. 
%%%%%%%%%%%%%%%%%%%%%%%%%%%%%%%%%%%%%%%%%%%%%%8.2
\subsection{Improved description of the nucleon spectroscopy}
An other aspect concerns the force between quarks. The Bhaduri et al.'s force is 
known to miss the position of the Roper resonance by a large amount. To remedy 
this situation, one proposal based essentially on a phenomenological basis, has 
been to add a three-quark force \cite{DESP1}. This one has not led to any 
significant discrepancy in the low energy phenomenology of pionic baryon decays. 
On the contrary, it improves the situation in many cases \cite{VALE,CANO1}. From 
the expression of this force:
\begin{equation}
V^{(3)}_{II} = \frac{1}{2} \sum_{i \neq j \neq k} \frac{V_0}{m_i m_j m_k}
\frac{e^{-m_0 r_{ij}}}{m_0 r_{ij}} \frac{e^{-m_0 r_{ik}}}{m_0 r_{ik}},
\end{equation}
which involves two Yukawa factors, it is expected that the form factor 
should have a $q^{-6}$ asymptotic behavior, instead of $q^{-8}$. This might 
contribute to make the form factors closer to the $q^{-4}$ behavior expected 
from $QCD$. While the above three-body force may account in a phenomenological 
way for omitted relativistic effects, we nevertheless believe that the 
effect it provides has no relationship to the boost effects that could 
provide the appropriate behavior and should naturally appear in a full 
relativistic calculation (see Sect. 8.4 below). 

Calculations using the nucleon wave function obtained from solving the 
corresponding Schr\"odinger equation and the above three-body force 
have been performed in the hyperspherical harmonic formalism with $K=0$ 
and $K=2$. Due to a strong short range attraction (the force 
leads to a $\frac{1}{\rho^2}$ term, similar to the centrifugal barrier), 
the solution behaves approximately like $\frac{1}{\rho^{\frac{3}{2}}}$ at 
short distances, instead of 1 for the Bhaduri et al.'s potential \cite{BHAD}. 
It results that the dominant contribution to the form factor, $<S|O|S>$, 
scales like $q^{-3}$ at large momentum transfers, differing from the 
expectation by three powers of $q$. The discrepancy has its source in 
the inability of the hyperspherical harmonic formalism with a limited 
number of terms to correctly reproduce the short range correlations between 
quarks produced by the three-body force given by Eq. (77). The problem is 
similar to that encountered for the Coulomb part of the force in the same 
approach, with the difference that it is more severe here. The calculated 
form factor is shown in Fig. 11 (dash-dotted line).

\begin{figure}[htb!]
\begin{center}
\mbox{ \epsfig{file=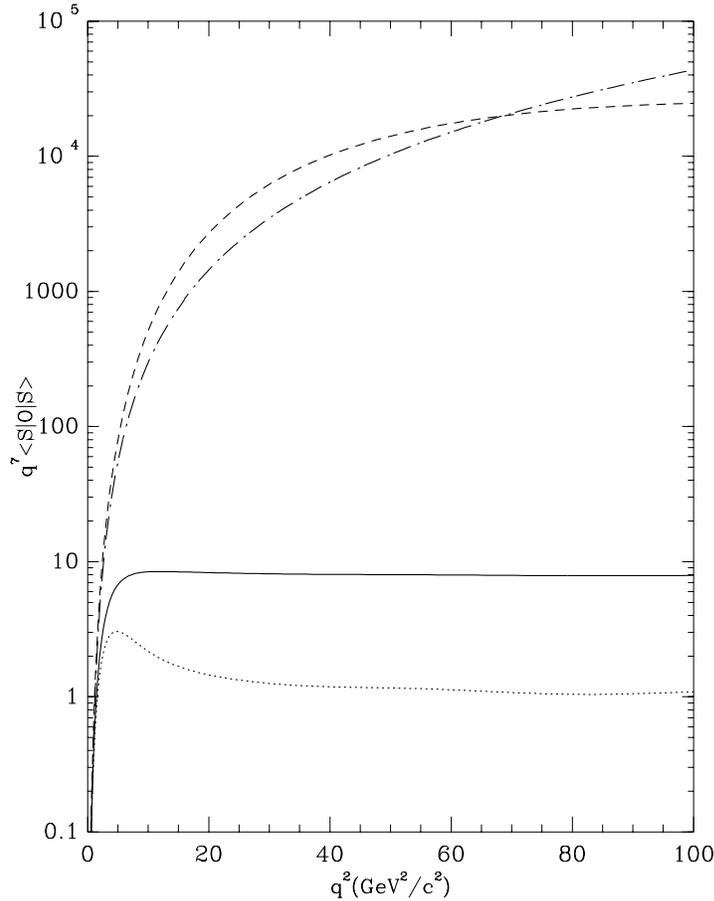, width=9cm} }
\end{center}
\caption{\small{Form factors calculated with three-body forces and without. 
Those 
presented here correspond to the $<S|O|S>$ transition matrix element. 
They have been multiplied by a factor $q^7$ to evidence the convergence 
to the asymptotic value expected for two of them (potential of Bhaduri et al. 
and potential including a gaussian three-body force, Eq. (78), continuous and 
dashed lines respectively). The other curves represent results for the 
three-body force given by Eq. (77) (dash-dotted line) and for the Bhaduri's 
model in absence of spin-spin force (dotted line).}}
\end{figure}  

Due to the proximity of some collapse in calculating the nucleon wave function 
with the above mentioned three-quark force, calculations were also 
performed with a Gaussian type force:
\begin{equation}
V^{(3)}_{III} = V_0 \, {\rm ex}p(- \sum_{i \leq j} \frac{r^2_{ij}}{\lambda^2}).
\end{equation}
This force is much less singular at short distances than $V^{(3)}_{II}$ 
and tends to zero more quickly for high momentum transfers, 
like ${\rm exp}( - \frac{\lambda^2}{4} (\kappa^2_{12} + \kappa^2_{3}) )$. The 
resulting nucleon spectroscopy is equally good \cite{CANO1}. The form factors so 
obtained do not differ much from the previous ones and, like them, overshoot 
those obtained with the Bhaduri et al.'s force by a large factor in the range 
around $50\,({\rm GeV/c})^2$ (see Fig. 11, dashed line). Contrary to the other 
calculation, the present one roughly evidences a behavior $q^{-7}$. This result 
is expected in the hyperspherical harmonic formalism with a restriction to $K=0$ 
and $K=2$ when the Coulomb part of the potential takes over the gaussian 
part. Surprisingly, it does not seem to be strongly affected by the large 
strength of the gaussian three-body force. While the asymptotic power law is 
identical to that obtained with the Bhaduri et al.'s force (continuous line in 
Fig. 11), the overall size is changed by three orders of magnitude. 

The comparison of the various results 
obtained from forces without and with three-body forces  tends therefore
to show that extra contributions to the description of the force between 
quarks, especially at short distances, can considerably modify the 
onset of the asymptotic behavior of the form factors, making it to occur at 
higher momentum transfers or with a different pattern. Thus, the form factors 
in the model involving the three-body force given by Eq. (77) reach their
asymptotic behavior from below. As to the form factor calculated with the 
other three-body force, Eq. (78), the apparent onset of the asymptotic behavior  
that is seen in Fig. 11 is misleading. The product of the form factor 
multiplied by $q^7$ª represented in the figure passes through a maximum 
and then should decrease to reach a value which is of the order 
of $0.7 \, 10^4 \, ({\rm GeV/c})^7$. Except for a difference in scale, 
the pattern exhibited by this result is quite similar to that one 
obtained for the Bhaduri et al.'s force without spin-spin force (dotted line in 
the same figure).
      
We would also like to notice that the difference by three orders of magnitude
for the form factors at high q produced when three-body forces are included 
is not a distinctive feature. A similar effect \cite{THEU} has been 
obtained with a two-body quark force resulting from meson exchange \cite{GRAZ}.
The common feature has perhaps some relationship to the small mean squared 
radius of the nucleon that these models predict, of the order 
of $0.10 \, {\rm fm}^2$.
%%%%%%%%%%%%%%%%%%%%%%%%%%%%%%%%%%%%%%%%%8.3
\subsection{Description of the spin-spin force}
It is known that the spin-spin part of the quark-quark force we used, 
essentially fitted on the difference of the $\pi$- and $\rho$-meson masses, 
is too large with respect to a genuine gluon exchange force by a 
factor 6 \cite{CANO1}. 
Correcting for this factor would decrease the contribution of the mixed 
symmetry states to the form factors and, indirectly, would accelerate the 
convergence to their asymptotic value since these particular contributions 
have a slower convergence. It would also probably remove some of the sign 
changes that are observed for some calculated form factors, especially 
$G_E^p(\vec{q}^{2})$ and $G_M^p(\vec{q}^{2})$. These changes are not 
expected to occur and, furthermore, are not seen in the measurements.

The correction by a factor 6 is not the main problem however.
Apart for the fact that it should be compensated by something else, the 
radial part of the force, which is given by a $\delta$ function 
in the non-relativistic limit, has been replaced by a less singular function, 
namely a Yukawa type potential. The advantage of this one 
is that it can be incorporated into a Schr\"odinger equation while avoiding a 
collapse of the solution in some cases \cite{BHAD3}. It is noticed that the 
spin independent part of the force also contains a $\delta$ type force, but 
this one is generally neglected. 

In a perturbative calculation of the asymptotic form factors, there would not 
be any difficulty to use a $\delta$ type force. The propagator 
of the gluons in Figs. 2 would be replaced by a constant, removing for each of 
them a $q^{-2}$ factor in predicting the behavior of 
the asymptotic form factor. The result so obtained should be closer to the 
asymptotic $QCD$ prediction. This feature however supposes 
approximations that are not supported by a more rigorous derivation of the 
force. The $\delta$ type force is expected to be made less 
singular by the presence of relativistic normalization factors, 
$\frac{m_q}{E^f} \; \frac{m_q}{E^i}$. The resulting non-locality is difficult 
to account for, and in absence of a better solution, employing a Yukawa 
potential instead of a $\delta$ type avoids to 
make unrealistic predictions.
%%%%%%%%%%%%%%%%%%%%%%%%%%%%%%%%%%8.4 
\subsection{Relativistic corrections}
A complete relativistic calculation of the form factor is rather tedious. Some 
recipe to account for part of these effects has been used in the past. 
One of them consists in changing the argument of the form factor as 
follows \cite{FRIA,BUCH,KUPE}:
\begin{equation}
F^{r.}(q^2)=F^{n.r.}( \frac{q^2}{1+\frac{q^2}{4m^2_N}} ).
\end{equation}
This expression, which has its origin in a naive estimate of the effect due to 
the Lorentz contraction, leads to a constant at large q, in contradiction 
with the $QCD$ expectation. It turns out to be valid only at small q 
\cite{AMGH,FRIA2}. A modified expression, motivated by the $QCD$ result, has 
also been used \cite{MITR,STAN}. 

The discussion presented in Sect. 2 indicates that the non-relativistic 
calculation misses an important factor due to a boost effect which mainly 
affects particles interacting by exchanging spin 1 bosons. This factor may be 
roughly of the order of:
\begin{equation}
(\frac{m_q+E_{\bar{q}}}{2m_q})^4
\end{equation}
where $E_{\bar{q}}$ is an  energy corresponding to an average momentum, 
$\bar{q}$, carried by the quarks in the Breit frame, see Fig. 3. At large q, 
this energy varies like a fraction of q and the above factor, Eq. (80), provides 
the $q^4$ factor discrepancy between the non-relativistic and relativistic 
calculations, while allowing one to recover the non-relativistic prediction in 
the limit of small momentum transfers. There is no point to insist on the 
approximate character of this factor. It results from an analysis of what a 
non-relativistic calculation partly misses and the  determination of the 
coefficient of the $q^4$ factor in Eq. (80) is likely to be uncertain. 
Furthermore, corrections of the same order are expected from Z-type diagrams, 
which involve the negative energy part of the quark propagator. In absence of a 
more refined analysis, we will simply mention an estimate made by incorporating 
in our calculation the factor given by Eq. (80)  together with a value of 
$\bar{q}$ between $\frac{q}{6}$ (momentum of initial and final quarks in the 
Breit frame) and $\frac{q}{2}$ (momentum of quarks in the intermediate state of 
Fig. 9.d). This amounts to a factor $\frac{ q^4}{2304 \, m_q^4}$ at high q. 
Disregarding the effect of the mixed symmetry component due to its uncertain 
character (see above), we considered the contribution of the matrix element 
$<S|O|S>$ to the quantities 
$q^4 \, G_M^p(\vec{q}^{2}) / \mu_p$ and $q^4 \, G_M^n(\vec{q}^{2}) / \mu_n$.
Results so obtained for the Bhaduri et al.'s quark interaction model 
have the size that experiment suggests, $0.4 \, ({\rm GeV/c})^4$ \cite{SLAC}, 
but are off by three orders of magnitude for the other models incorporating 
three-quark forces. 
%%%%%%%%%%%%%%%%%%%%%888888888888888888888888888888%%%%%%%%%%%%%%%%%%%%%%%%%%%%
\section{Conclusion}
We have studied the nucleon form factors using a non-relativistic constituent 
quark model. The emphasis has been put more on general features pertinent to 
their calculation than on a fine description of observables. A special attention 
has been given to the asymptotic behavior of the form factors at high momentum 
transfer. 

Contrary to what is sometimes mentioned, it is found that the 
non-relati\-vis\-tic quark model does lead to power law form factors at high 
$q$. This power law is related to the force between quarks in momentum space 
and thus can be predicted if it is well defined. It is our belief 
that this approach, possibly corrected for exchange currents and relativity, is 
completely equivalent to the one used in the $QCD$ asymptotic regime where 
uncorrelated wave functions are used together with an effective operator 
accounting for three-quarks absorbing a virtual photon while exchanging two 
gluons. In the present approach however, the calculation of the form factors 
is performed in a unique model of the nucleon and tends to cover the 
full range of momentum transfers from 0 to $\infty$. 

The exact power law behavior, which is expected to be $q^{-8}$ in the 
non-relativistic framework used in the present study, has been found to depend 
on the approximations made in treating the three-body system. While the 
hyperspherical harmonic formalism with a restriction to the lowest values of the 
grand orbital, $K$, is generally a good approximation for determining the 
binding energy, it has been found to systematically lead here to the wrong power 
law, $q^{-7}$ instead of $q^{-8}$. To remedy this situation, the full set of 
$K$ values is required. This is expected from the behavior of the wave function 
of two quarks at a short distance, on which it depends linearly. As far as we 
can see, the Faddeev approach is unsensitive to this difficulty, but we did not 
get the certitude that the magnitude of the asymptotic form factor is 
independent on the truncation of the Faddeev amplitude although it could be. In 
any case, this dependence seems to be small. At low momentum transfers, the form 
factors calculated with the 2 amplitude Faddeev approach or with the 
hyperspherical formalim are in good agreement with the more complete 8 amplitude 
Faddeev calculation up to $q^2=2\,({\rm GeV/c})^2$ for the former and 
$q^2=3\,({\rm GeV/c})^2$ for the latter.

The convergence of the form factors to their asymptotic value is another 
feature we looked at carefully. By only considering form factors in the range 
around $q^2=50\,({\rm GeV/c})^2$, it has often been difficult to guess their 
power law behavior. This has led us to examine the wave functions at short 
distances and, from them, to determine  the asymptotic behavior. In a few cases, 
we had to conclude that form factors at $q^2=50\,({\rm GeV/c})^2$ were off their 
asymptotic value by a factor 2-3. This occured especially for form factors 
involving the transition between mixed symmetry components, $MS \leftrightarrow 
MS$ and $MA \leftrightarrow MA$ and is due in these cases to sizeable  next to 
leading order contributions, which seem to be of the relative order $q^{-1}$ but 
may also be of relative order $q^{-2}\; {\rm log} \,q$ as generally expected 
theoretically. 

Most often, the behavior of the form factor at high $q$ is illustrated by a 
process where a quark, which has been stroken by a virtual photon, shares the 
momentum it received with the other two quarks by the successive exchange 
of two gluons. In practice, calculations involve several processes with 
different time orderings of the photon absorption and gluons exchanges. In the 
present work, the dominant contribution rather comes from a process 
where the same quark that is stroken by the photon has previously 
and subsequently exchanged a gluon with each of the other quarks. It is 
likely that this should also hold in $QCD$, but this has to be checked. In 
any case, we believe it is worthwhile to emphasize the point as some 
arguments and calculations have been developped without regard to 
this  type of process. Its contribution is essentially determined by the 
square of the derivative of the wave function at the origin. This 
differs from the two-body case or the simplest calculation with the 
harmonics hyperspherical formalism, which only involve the first derivative.

While we have not been able to fully answer some of the questions raised in the 
introduction, we showed that some caution is required in dealing with the 
asymptotic behavior of form factors. Their onset may not occur as quickly as 
usually expected. Most important, it supposes an accurate determination of the 
three-quark wave function at short distances. Present calculations provide 
enough evidence that the high momentum transfer behavior in a non-relativistic 
approach is given by $q^{-8}$. We believe that the difference with the power 
law, $q^{-4}$, expected from $QCD$ is most likely due to relativistic (boost) 
effects that have not been considered in the present work. Consequently, it 
is inappropriate to use the $q^{-4}$ behavior to get constraints on the 
non-relativistic three-quark wave function. Contrary to a belief that probably 
originates from a non-relativistic approach or from studies with spinless 
particles, a $q^{-4}$ behavior of the form factor does not imply a similar 
behavior of the wave function at high momenta. Along the same lines, it is 
inapproriate to use a truncated hyperspherical harmonic approach to analyze form 
factors at high momentum transfers. In either case \cite{GIAN,STRO,GAVI,BIJK}, a 
bias is systematically introduced, which prevents one to make relevant 
conclusions.

For the future, it would be worthwhile to study more carefully the next to 
leading order corrections to the form factors, both mathematically and 
nume\-rically. The questions to be answered are the relative order of these 
corrections, $\frac{1}{q}$, or $\frac{1}{q^2}$ (up to ${\rm log} \, q $ terms), 
and the sensitivity to the truncation generally performed in solving the Faddeev 
equation. Obviously, a full relativistic calculation would be desirable. It is 
likely that the questions raised here will appear there too. We hope that the 
answers we got will be of some relevance for this more difficult problem.  

{\bf Acknoledgement}
One of us (BSB) acknowledges financial support from the CICYT and is grateful 
to hospitality of Valencia University, where a part of this work was performed.
This work has been partially supported by DGESIC under grant PB97-1401-C02-01 
and by EC-TMR network under contract ERB TMR X-CT96-0008.

%%%%%%%%%%%%%%%%%%%%%%%%%%%%%appendices%%%%%%%%%%%%%%%%%%%%%%%%%%%%%%%%%%%%%%%%

\appendix
\section{Wave functions in the hyperspherical harmonic forma\-lism} 
\subsection{Jabobi coordinates}
For a wide range of physical potentials (mainly two-body
potentials depending on the relative coordinates), the three-body
problem is conveniently treated by  using Jacobi coordinates. For
the equal mass ($m$) case, the Jacobi coordinates can be defined as
corresponding to the vectors:
\begin{eqnarray}
 \vec{\rho}_{12} & = & \frac{\vec{r}_1 - \vec{r}_2}{\sqrt{2}}, 
 \nonumber\\
 \vec{\rho}_{3} \; & = & \sqrt{2/3} \, (\vec{r}_3 -\frac{\vec{r}_1 
+\vec{r}_2}{2}).
\end{eqnarray}
Altogether with the center of mass $(\vec{R})$ coordinate, they
determine completely the position of the system.

The kinetic energy operator can be separated as
\begin{equation}
T = \sum_i \frac{\vec{p}_i\,^2}{2m} = \frac{\vec{P}\,^2}{6m} 
+ \frac{\vec{p}\,_{\rho_{12}}^2}{2m}
+ \frac{\vec{p}\,_{\rho_{3}}^2}{2m} \equiv T_{c.m.} + T_{int}, 
\end{equation}
containing the center of mass kinetic energy, $\frac{\vec{P}\,^2}{6m} 
\equiv T_{c.m.}$, plus the internal kinetic energy, $T_{int}$, in terms of
the Jacobi momenta. Then for potentials depending only on the Jacobi
coordinates,  the center of mass
motion can be factorized out.
\subsection{Hyperspherical coordinates and harmonics}
From $ \vec{\rho}_{12} $ and $ \vec{\rho}_3 $, one can define hyperspherical
coordinates $\rho \in  [0, \infty), \phi \in [0, \pi/2]$ in the form:
\begin{eqnarray}
\rho_{12} & = & \rho \, \sin \phi, \nonumber\\
\rho_3 \, & = & \rho \, \cos \phi,
\end{eqnarray}
the differential volume element $d^3 \vec{\rho}_{12} \; d^3 \vec{\rho}_3$ being 
then written as:
\begin{equation}
\rho^5 \; d \rho \; d \Omega \equiv \rho^5 \; d \rho \; d \phi \; \sin^2
\phi \; \cos^2 \phi \; d \hat{\rho}_{12} \; d \hat{\rho}_3.
\end{equation}

The internal kinetic energy is then expressed in terms of $\rho$
and $\Omega \equiv \{ \theta_{12}, \varphi_{12}, \theta_3, \varphi_3,
\phi \}$, (where ($\theta, \varphi)$ are the spherical angles of 
$\vec{\rho}_{12}$ and $\vec{\rho}_3$) in the form:
\begin{equation}
T_{int} = - \frac{1}{2m} \left( \frac{\partial^2}{\partial \rho^2}
+ \frac{5}{\rho} \frac{\partial}{\partial \rho} + \frac{K^2 (\Omega)}{\rho^2}
\right),
\end{equation}
where $K^2 (\Omega)$ is an angular operator whose eigenfunctions are called
the hyperspherical harmo\-nics (HH), $Y_{[K]} (\Omega)$:
\begin{equation}
K^2 (\Omega) \, Y_{[K]} (\Omega) = - K (K + 4) \, Y_{[K]} (\Omega),
\end{equation}
$K$, the so called grand orbital number, defines the parity of the
HH as $ (-)^K $. One can choose appropriate combinations of  hyperspherical
harmonics  with definite values of the total orbital angular
momentum $L (\vec{L} = \vec{l}_{12} + \vec{l}_3)$ and its third
projection $M_L$ and with definite spatial symmetry:
$$ Y_{[(K, sym)]}^{(L, M_L)}  \, (\Omega). $$

In  terms of these, the wave function of a system of total momentum
$\vec{P}=0$ and spin and parity, $J^P$, reads
\begin{equation}
|\Psi (\rho, \Omega) > = < \rho \Omega| \Psi > = \sum_{\begin{array}{l} K, L\\
sym \end{array}} \psi_{K,L} (\rho)  \left[ Y_{[K, sym]}^{(L, M_L)}(\Omega)
|sym> \right]_{J^P} \equiv \sum_{j} \psi_{j} (\rho) Y_j (\Omega),
\end{equation}
where $|sym>$ stands for a spin-isospin wave function of definite symmetry  and 
the square bracket implies the coupling to total angular momentum $J$ with the 
required overall symmetry. The subindex $j$ of $\psi_j (\rho)$ and $Y_j(\Omega)$ 
corresponds to a simplified notation, $j$ comprising $K$, angular momentum and 
spin-isospin variables.

The state $|\Psi (\rho, \Omega) >$ satisfies the Schr\"{o}dinger equation:
\begin{equation}
 \left[   - \frac{1}{2m} \left( \frac{\partial^2}{\partial \rho^2} + 
\frac{5}{\rho} \frac{\partial}{\partial \rho} +
\frac{K^2 (\Omega)}{\rho^2} \right) + V + 3m - E  \right]
| \Psi (\rho, \Omega) > = 0.
\end{equation}
By substituting $| \Psi (\rho, \Omega) >$ by its expansion, Eq. (87), and 
projecting it, one gets a system of coupled equations (its number depending on 
the number of $Y_j (\Omega)$ terms kept in the expansion):
\begin{eqnarray}
\left[ - \frac{1}{2m} \left( \frac{1}{\rho^{5/2}} 
\frac{d^2}{d \rho^2} \rho^{5/2} -
\frac{(K + 2)^2 - 1/4}{\rho^2} \right) + 3m - E  \right]
\psi_j (\rho)  \nonumber \\
+ \sum_{j' } V_{jj'} (\rho) \; \psi_{j'} (\rho) = 0,
\end{eqnarray}
where the matrix elements of the potential are given by
\begin{equation}
V_{j, j'} (\rho) = \int d \Omega \;  Y_{j'}^* (\Omega)
\; V (\rho, \Omega, s, \tau) \; Y_j (\Omega),
\end{equation}
In this equation, $s, \tau$ indicate the possible dependences of the potential
on spin and isospin. The norma\-lization condition  is:
\begin{equation}
\sum_j \int d \rho \; \rho^5 \, | \psi_j (\rho) |^2 = 1 \, .
\end{equation}
\subsection{Details about wave functions}
For the moment, we are going to truncate the expansion (87) by
considering only $K = 0, 2$ terms. Furthermore, if we consider a
restricted potential as the one given by (17) but without
spin-spin interaction, the ground state does not contain $K = 2$
component. Then we can write:
\begin{equation}
| \Psi (\rho, \Omega) >_{\frac{1}{2}^+}  = \frac{1}{\pi \sqrt{\pi}}
\psi_1 (\rho) | S >,
\end{equation}
satisfying
\begin{equation}
\left[\frac{1}{2m_q} (- \frac{1}{\rho^{5/2}} \frac{d^2}{d \rho^2} \rho^{5/2}
+ \frac{15}{4 \rho^2}) - \frac{4 \sqrt{2}}{\pi} \frac{\kappa}{\rho}
+ \frac{16 \sqrt{2}}{5 \pi} \frac{\rho}{a^2} - \frac{3 D}{2} + 3m_q - E 
\right] \psi_1 (\rho) = 0 \, . 
\end{equation}
The terms $- \frac{4 \sqrt{2}}{\pi} \frac{\kappa}{\rho}$ and $\frac{16 
\sqrt{2}}{
5 \pi} \frac{\rho}{a^2}$ represent respectively the monopole components of
the Coulomb and linear parts of the quark-quark interaction:
\begin{equation}
- \frac{4 \sqrt{2}}{\pi} \frac{\kappa}{\rho} = - 3 \int d \, \Omega 
\left( \frac{1}{ \pi \sqrt{\pi}} \right)^2 \frac{\kappa}{2 \sqrt{2} \rho \sin
\phi}, 
\end{equation}
\begin{equation}
 \frac{16 \sqrt{2}}{5 \pi} \frac{\rho}{a^2} =  3 \int d \, \Omega 
\left( \frac{1}{ \pi \sqrt{\pi}} \right)^2 \frac{\rho \sqrt{2}
\sin \phi}{2 a^2}. 
\end{equation}
The relative distance $r_{12}$ appearing in the potential (17) has been 
replaced by its expression in terms of the hyperspherical 
coordinates, $\sqrt{2} \rho \,{\rm sin} \phi $, and the factor 3 accounts 
for the interaction between the three pairs 12, 23 and 13 which contribute 
equally to (94, 95). The solution of Eq. (93) at short distance can be 
expanded in terms of the powers of $\rho$. It reads:
\begin{eqnarray}
\psi_1(\rho)_{\rho \rightarrow 0} = \alpha'_1 \left[ 1 - \frac{8 \sqrt{2} }{5 
\pi 
} \kappa \; m_q \; \rho \hspace*{7cm} \right. \nonumber \\         
  \left. - \frac{1}{12} \left( 2m_q (E -3m_q + \frac{3D}{2} ) 
  - 5(\frac{8\sqrt{2} }{5 \pi }  \kappa m_q )^2  \right) \rho^2  
  \nonumber \hspace*{2cm} \right. \\
\left.  + \frac{1}{21} \left( \frac{32\sqrt{2}}{5\pi}\frac{m_q}{a^2} + 
\frac{68\sqrt{2}}{15\pi} \, \kappa (E -3m_q + \frac{3D}{2} ) m_q^2
 - \frac{25}{12} (\frac{8\sqrt{2} }{5 \pi }  \kappa m_q)^3 \right) \; 
\rho^3+..\right],
\end{eqnarray}
where $\alpha'_1$ is a constant (the value of the wave function at the origin).

The coefficient of the term linear in $\rho$ is determined by the Coulomb part 
of the force, while the energy of the system appears 
in the second term in $\rho^2$ of the expansion. The intensity of the 
confining potential affects the third term in $\rho^3$
of the expansion. It however implicitly appears in the energy $E$. 

A particular case of Eq. (96) is that one where the confining potential 
is neglected. Equation (93) then reduces to that of a Coulombian type 
problem. The solution is given by:
\begin{equation}
\psi(\rho) = \alpha e^{-\beta\rho},
\end{equation}
with $ \beta = \frac{8\sqrt{2} }{5 \pi } \kappa m_q $
and $E-3m_q + \frac{3D}{2} = -\frac{\beta^2}{2m_q} = 
-\frac{64}{25\pi^2}\kappa^2 m_q$.\\
From the  normalization condition of the radial wave function  given by:
\begin{equation}
\int d\rho \; \rho^5 \; \psi^2(\rho) = 1,
\end{equation}
one gets $\alpha = \frac{(2 \beta)^3}{\sqrt{120}}$.

In the case where the spin-spin interaction is turned on, one has to solve a 
set of coupled equations. For our purpose here, it is enough to keep 
the most singular part of the spin-spin 
force $(\frac{\kappa'_{\sigma}}{r_{ij}} \, \vec{\sigma}_i .\vec{\sigma}_j)$
(the complete expression has been used for the numerical calculations
detailed in the main text):
\begin{eqnarray}
\left[ \frac{1}{2m_q} \left( -\frac{1}{\rho^{5/2}} \frac{d^2}{d\rho^2} 
\rho^{5/2} + \frac{15}{4} \frac{1}{\rho^2} \right) 
- \frac{4\sqrt{2}}{\pi } ( \kappa + \kappa'_{\sigma} ) \frac{1}{\rho}+ \frac{16 
\sqrt{2} }{5 \pi a^2}\; \rho \hspace*{2cm} \nonumber \right.  \\
 \left. - \frac{3D}{2} + 3m_q - E \right] \psi_1(\rho) =  \frac{8}{5\pi}  
\kappa'_{\sigma} \frac{1}{\rho} \, \psi_3 (\rho),\\
\left[ \frac{1}{2m_q} \left( -\frac{1}{\rho^{5/2}} \frac{d^2}{d\rho^2} 
\rho^{5/2} + \frac{63}{4} \frac{1}{\rho^2}\right) 
- \frac{4\sqrt{2}}{\pi } (\frac{38}{35} \kappa + \frac{10}{7}\kappa'_{\sigma} ) 
\frac{1}{\rho} + \frac{16 \sqrt{2} }{5 \pi a^2}\; \frac{62}{63} \,
\rho \hspace*{2cm} \nonumber \right. \\ \left. - \frac{3D}{2} + 3m_q - E 
\right] 
\psi_3(\rho)= \frac{16}{5\pi}  \kappa'_{\sigma} \frac{1}{\rho} \, \psi_1(\rho),
\end{eqnarray}
where $\psi_3 (\rho) \equiv \psi_{K=2, L= 0}(\rho)$. In the limit 
$\rho \rightarrow 0$, it is easy to show that to remove the $\frac{1}{\rho^2}$ 
singularity on the left hand side of the second equation, $\psi_3(\rho)$ has to 
go to zero as :
\begin{equation}
\psi_3(\rho)_{\rho \rightarrow 0} =  \frac{32 }{35 \pi} \kappa'_{\sigma} m_q \; 
\rho \; \psi_1(0).
\end{equation}

Similarly, other components in the limit $\rho \rightarrow 0$ may be determined. 
An expression of the wave function beyond $K = 0,2$, i.e. including $K = 4$ and 
$K = 6$ terms, and limited to those terms linears in the variable $\rho$ is the 
following: 
\begin{eqnarray}
(\Psi_{HH})_{linear}  \simeq \frac{1}{\pi \sqrt{\pi}} 
\left[ 1-\frac{8\sqrt{2} }{5 \pi } ( \kappa + \kappa'_{\sigma} ) 
\; m_q \; \rho \right. \nonumber 
\hspace*{4cm} \\
\left. . \left( 1 - \frac{2}{21} \left( \;
({\rm cos}^2\phi-{\rm sin}^2\phi)^2+4\;{\rm cos}^2\phi \; {\rm sin}^2\phi \; 
(\hat{\rho}_{12}.\hat{\rho}_3)^2-\frac{1}{2} \right) \hspace*{2cm}
\right. \right. \nonumber \\ \left. \left. - 
\frac{8}{231}({\rm cos}^2\phi-{\rm sin}^2\phi) 
\left( 
({\rm cos}^2\phi-{\rm sin}^2\phi)^2 -12 \;{\rm cos}^2\phi \; {\rm sin}^2\phi\; 
(\hat{\rho}_{12}.\hat{\rho}_3)^2 \right) +..\right) |S> \right. \nonumber 
\\ \left. + \frac{32\sqrt{2}}{5\pi}\ \kappa'_{\sigma} m_q \, \rho \left( 
({\rm cos}^2\phi-{\rm sin}^2\phi)|MS> - 
2\; {\rm sin}\phi \; {\rm cos}\phi \; \hat{\rho}_{12}.\hat{\rho}_3|MA>   
\right)  +... \right].
\end{eqnarray}
%%%%%%%%%%%%%%%%%%%%%%%%%%%%%%%%%%%%%%%%BBBB
\section{Details about calculations of the  form factors}
The complete expressions of the form factors calculated from the analytic 
expressions of the wave functions given by Eqs. (37) and (41) are as follows: 
\begin{eqnarray}
< S | O  |S > = 405 \sqrt{6} \, \frac{\alpha_1^2 
\beta_1}{(q^2+6\beta^2_1)^{7/2}}, 
\;\;\;\;\;\;\;\;\;\;\;\;\;\;\;\;\;\;
\;\;\;\;\;\;\;\;\;\;\;\;\;\;\;\;\;\;\;\;\;\;\;\;\;\;\;\;\;\;\;\;\;\; \\
< S | O | MS> = - 2835\sqrt{6} \, \frac{\alpha_1 \alpha_3 q^2}{ ( 
q^2+6(\frac{\beta_1+\beta_3}{2})^{2} )^{9/2} },\;\;\;\;\;\;\;\;\;\;\;
\;\;\;\;\;\;\;\;\;\;\;\;\;\;\;\;\;\;\;\;\;\;\;\; \\
< MS| O |MS> = 7290 \frac{\alpha^2_3}{q^8} \left( 64 -\frac{ \beta_3 \sqrt{6} 
}{ (q^2+6\beta^2_3)^{11/2} } \right. 
\;\;\;\;\;\;\;\;\;\;\;\;\;\;\;\;\;\;\;\;\;\;\;\;\;\;\;\;\;\;\; \\
\left. \left[ 175 q^{10} + 574 q^{8} \tilde{\beta}^2_3 + 
924 q^6 \tilde{\beta}^4_3 + 792 q^4 \tilde{\beta}^6_3 
 + 352 q^2 \tilde{\beta}^8_3 + 64 \tilde{\beta}^{10}_3 \right] \right), 
\nonumber \\
<MA | O | MA > = - 2430 \frac{\alpha^2_3}{q^8} \left( 64   -\frac{ \beta_3 
\sqrt{6} }{ (q^2+6\beta^2_3)^{11/2} } \right. 
\;\;\;\;\;\;\;\;\;\;\;\;\;\;\;\;\;\;\;\;\;\;\;\;\;\;\;\; \\
\left. \left[ 168 q^{10} + 588 q^8 \tilde{\beta}^2_3 + 924 q^6 \tilde{\beta}^4_3 
+ 792 q^4 \tilde{\beta}^6_3 + 352 q^2 \tilde{\beta}^8_3 
+ 64 \tilde{\beta}^{10}_3 \right] \right) \nonumber ,
\end{eqnarray}
where $\tilde{\beta}_3 = \sqrt{6} \, \beta_3$. 
While close expressions of form factors can be obtained in a few cases, as 
above, quite generally and especially for more realistic 
wave functions incorporating terms depending linearly on the variables, 
$r_{12}, \; r_{13}$ and $r_{23}$, this is not possible. 

One may however be interested to get some prediction for the asymptotic form 
factor. Starting from the expression of the matrix element, $<X|O|Y>$, given by 
Eq. (49), and after integration over the orientations of the  $\rho_{12}$ and 
$\rho_3$ variables, which can be performed in most cases, one is left with a 
quantity of the following form: 
\begin{equation}
I(q) = \frac{16}{\pi} \sqrt{\frac{3}{2}} \, \frac{1}{q} \, \int d\phi \, 
{\rm sin}^2\phi \, {\rm cos}^2\phi \;\; d\rho \, \rho^4 \; 
{\rm sin}(\sqrt{\frac{2}{3}} \, q \rho \, {\rm cos} \, \phi ) \; H(\rho,\phi),
\end{equation} 
where $H(\rho,\phi)$ involves the nucleon radial wave functions. The overall 
factor, $\frac{16}{\pi}$, arises from a factor $(4\pi)^2$ due to the integration 
over the angles relative to the vectors, $\vec{\rho}_{12}$ and $\vec{\rho}_3$, 
and another one, $(\frac{1}{\pi \sqrt{\pi}})^2$, appearing in the normalization 
of the wave function, Eq. (102). Assuming that $H(\rho,\phi)$ can be expanded 
at small $\rho$ as: 
\begin{equation}
H(\rho,\phi)=F_0 + \rho \, F_1(\phi) +...+ \rho^n \, F_n(\phi) +.. ,
\end{equation}
it is clear, from dimensional arguments, that the $\rho^n$ term will produce a 
contribution to the form factor proportional to $q^{-(6+n)}$, besides 
$\delta(\vec{q})$ functions or its derivatives which are irrelevant at high q. 
To get the corresponding coefficient, $a_n$, one has to perform the integration 
over the $\rho$ and $\phi$ variables: 
\begin{equation}
a_n = \frac{16}{\pi} \; \sqrt[6+n]{\frac{3}{2}} \; {\rm lim}_{\epsilon 
\rightarrow 0} 
\int d\phi \, {\rm sin}^2\phi \, {\rm cos}^2\phi \; dx \, x^{4+n} \; 
{\rm sin}(x \, {\rm cos}\,\phi) \, F_n(\phi) \, e^{-\epsilon x}.
\end{equation}
As in Eq. (13), the factor $e^{-\epsilon x}$ allows one to get rid of the part 
involving the $\delta(\vec{q})$ function or its derivatives. After integration 
over $x$, Eq. (109) becomes: 
\begin{equation}
a_n = \frac{16}{\pi} \; \sqrt[6+n]{\frac{3}{2}} \; {\rm lim}_{\epsilon 
\rightarrow 0} (-\frac{d\;}{d\epsilon})^{4+n} \int d\phi \, {\rm sin}^2\phi \, 
{\rm cos}^2\phi \, \frac{F_n(\phi)}{{\rm cos}^2\phi+\epsilon^2} .
\end{equation}

As a check, one can show from expression (37) that
\begin{equation}
F_n(\phi)= \alpha_1^2 \,   \frac{ (-2 \, \beta_1)^n }{n!}.
\end{equation}
Putting it in (110) and using the relation:
\begin{equation}
\int d\phi 
\frac{{\rm sin}^2\phi \, {\rm cos}^2\phi}{{\rm cos}^2\phi + \epsilon^2} = 
\frac{\pi}{2} \, \left(\frac{1}{2} + \epsilon^2 - \epsilon \sqrt{1+ 
\epsilon^2}\right),
\end{equation}
the different terms of the expansion of the form factor (103) in terms of the 
powers of $\frac{1}{q}$, $\frac{1}{q^{6+n}}\;(n=0,1,..)$, are recovered.
%%%%%%%%%%%%%%%%%%%%%%%%%%%%%%%%%%%%%%%%%%%%%%%%%CCCCC
\section{Short distance behavior of the wave function in
the Faddeev formalism and asymptotic form factors}
\subsection{Short distance wave functions}
In the Faddeev approach, and for the considered interaction given by
(17), the short range behavior of the wave function (28) can be obtained
from Eq. (26). So, for the amplitudes $\psi^{0,1}_{12,3} (
\vec{r}\,_1, \vec{r}\,_2, \vec{r}\,_3) $:
\begin{eqnarray}
\psi^{0}_{12,3}(\vec{r}_1,\vec{r}_2,\vec{r}_3) \propto 1 - \frac{3}{4} \; 
(\kappa 
+ 3\kappa'_{\sigma} ) \; m_q \; r_{12}  +... , \nonumber \\
\psi^{1}_{12,3}(\vec{r}_1,\vec{r}_2,\vec{r}_3) \propto 1 - \frac{3}{4} \; 
(\kappa 
-  \kappa'_{\sigma} ) \; m_q \; r_{12} +....\;\; ,
\end{eqnarray}
and for the full wave function  and up to an undetermined factor:
\begin{eqnarray}
\Psi_{F, 0}(\vec{r}_1,\vec{r}_2,\vec{r}_3) \propto \left( 1- 
\frac{1}{4}(\kappa + \kappa'_{\sigma}) \; m_q \, 
(r_{12}+r_{13}+r_{23})+...\right) |S> 
 \nonumber \;\;\;\;\;\;\;\;\;\;\;\;\;\; \\  - \frac{1}{4 }\kappa'_{\sigma} m_q 
 \left( \left( ( 2r_{12} - r_{13} - r_{23} )+..\right) |MS> + \sqrt{3} \left( 
(r_{23} - r_{13})+..\right)  |MA> \right).
\end{eqnarray}

It is worth to notice that from this wave function one can recover the
first terms of the expansion, Eq. (102), by making the projection on
appropriate hyperspherical harmonics. It is also noticed that this wave
function, after averaging over all directions of $\vec{r}_{12}$, has
a term linear in its modulus, $r_{12}$, or equivalently in
$\sin \phi$, while the truncated wave function given by Eq. (102) depends
on the square of $\sin \phi$. As this linear term in $r_{12}$ corresponds
to a first order in the interaction, the wave function in momentum space
given by Eq. (67) should have some track of it. In fact by
taking the Fourier transform of Eq. (67) it is possible to recover
(113). The proportionality factor is then found to be $\frac{1}{3} 
\frac{\bar{\alpha}}{\pi \sqrt{\pi}}$.

An expression for the momentum space wave function reproducing the short
range behavior of the r-space wave function (114), may by useful.
Starting from the following zeroth order completely symmetrical wave
function (see permutation relations at the end of this appendix),
\begin{eqnarray}
\Psi_{12,3}^{(0)} + \Psi_{32,1}^{(0)} + \Psi_{13,2}^{(0)} = \frac{1}{3} 
\frac{\bar{\alpha}}{\pi\sqrt{\pi}} (2\pi)^6 
\left( \delta(\vec{\kappa}_3) \; \delta(\vec{\kappa}_{12}) \right. 
\;\;\;\;\;\;\;\;\;\;\;\;\;\;\;\;\;\;\;\;\nonumber \\   \left.
+ \delta(-\frac{1}{2}\vec{\kappa}_3 + \frac{\sqrt{3}}{2} \vec{\kappa}_{12}) 
\; \delta(\frac{1}{2}\vec{\kappa}_{12} + \frac{\sqrt{3}}{2} \vec{\kappa}_3) 
\right. \;\;\;\;\;\;\;\;\;\;\;\; \nonumber \\ \left.
+ \delta(-\frac{1}{2}\vec{\kappa}_3 - \frac{\sqrt{3}}{2} \vec{\kappa}_{12}) 
\; \delta(\frac{1}{2}\vec{\kappa}_{12} - \frac{\sqrt{3}}{2} \vec{\kappa}_3) 
  \right) \; |S>,
\end{eqnarray}
one can iteratively obtain a first order wave function from the 
momentum space Faddeev equation. For one of the amplitudes, 
$\Psi^{(1)}_{12,3} (\kappa_{12}, \kappa_{3})$ for instance:
\begin{eqnarray}
\Psi^{(1)}_{12,3}(\kappa_{12}, \kappa_3)= - \frac{1}{\sqrt{2}} 
\frac{m_q}{\kappa_{12}^2+ \kappa_3^2 + \beta^{(1)^2}} \int 
\frac{ \; d\vec{\kappa}'_{12} \; d\vec{\kappa}'_3 }{(2 \pi)^3} \; 
V(\kappa_{12},\kappa'_{12}) \delta(\vec{\kappa}_3 - \vec{\kappa}'_3)
 \hspace*{1.5cm} \nonumber \\ 
 \left( \Psi^{(0)}_{12,3} (\kappa'_{12}, \kappa'_3)+ 
(1 \leftrightarrow 3 )+(2 \leftrightarrow 3 ) \right),
\end{eqnarray}
where the potential, here given for definiteness for its Coulomb part and in 
accordance with Eq. (17), is defined as 
\begin{displaymath}
V(\kappa_{12},\kappa'_{12})= - 
\frac{4\pi \,\kappa}{(\vec{\kappa}_{12} - \vec{\kappa}'_{12})^2} \, .
\end{displaymath}
Then
\begin{eqnarray}
\Psi_{12,3}^{(1)} + \Psi_{32,1}^{(1)} + \Psi_{13,2}^{(1)} = \frac{
\bar{\alpha}}{\pi 
\sqrt{\pi}}\frac{4\pi}{\sqrt{2}} m_q \, C^{(1)} 
\left( I_3 \; [( \kappa + \kappa'_{\sigma} ) |S> + 2 \kappa'_{\sigma} |MS> ]  
\hspace*{1.5cm}  \right. \nonumber \\
 + \left.  I_1
 \; [( \kappa + \kappa'_{\sigma} ) |S> + 2 \kappa'_{\sigma} ( -\frac{1}{2}|MS> 
+\frac{\sqrt{3}}{2} |MA>)]   \right. \;\;\; \nonumber \\  
 + \left.  I_2
 \; [( \kappa + \kappa'_{\sigma} ) |S> + 2 \kappa'_{\sigma} ( -\frac{1}{2}|MS> 
-\frac{\sqrt{3}}{2} |MA>)]     \right), 
 \end{eqnarray}
where: 
 \begin{eqnarray}
 I_3=\frac{(2\pi)^3 \; \delta(\vec{\kappa_3}) }{(\kappa_{12}^2 + \beta^{(1)^2}) 
\kappa_{12}^2}, \;\;\;\;\;\;\;\;\;\;\;\;\;\;\;\;\;\;\; \nonumber \\
 I_1=   \frac{(2\pi)^3 \; \delta(-\frac{1}{2}\vec{\kappa}_3 + 
\frac{\sqrt{3}}{2} \vec{\kappa}_{12}) }{
 ( (\frac{1}{2} \vec{\kappa}_{12} + \frac{\sqrt{3}}{2} \vec{\kappa}_3 )^2 + 
\beta^{(1)^2}) (\frac{1}{2} \vec{\kappa}_{12} + \frac{\sqrt{3}}{2} 
\vec{\kappa}_3 )^2}, \nonumber \\
 I_2= \frac{(2\pi)^3 \; \delta(-\frac{1}{2}\vec{\kappa}_3 - \frac{\sqrt{3}}{2} 
\vec{\kappa}_{12}) }{
 ( (\frac{1}{2} \vec{\kappa}_{12} - \frac{\sqrt{3}}{2} \vec{\kappa}_3 )^2 + 
\beta^{(1)^2}) (\frac{1}{2} \vec{\kappa}_{12} - \frac{\sqrt{3}}{2} 
\vec{\kappa}_3 )^2} , 
\end{eqnarray}
and $C^{(1)}$ is a normalization constant. The quantities $I_1$ and $I_2$
are obtained from $I_3$ by performing a circular permutation and re-expressing 
the variables in terms of $\vec{\kappa}_{12}$ and $\vec{\kappa}_3$.
The $\frac{1}{\kappa_{12}^4}$ behavior at high values of $\kappa_{12}$
is essential to reproduce the linear dependence in $\vec{r}_{12}$ in Eq. (114). 

From Eq. (116), multiplying both members by the factor $\kappa_{12}^4$,
taking the limit $\kappa_{12} \rightarrow \infty $ and integrating over the 
variable $\vec{\kappa}_3$, one gets for the Coulomb type potential:
\begin{eqnarray}
  \lim_{\kappa_{12} \rightarrow \infty} \kappa_{12}^4 \int d\vec{\kappa}'_3 
\Psi_{12,3}(\kappa_{12}, \kappa'_3)  
\;\;\;\;\;\;\;\;\;\;\;\;\;\;\;\;\;\;\;\;\;\;\;\;\;\;\;\;\;\;\;\;\;\;\;\;\;\; 
\nonumber \\
 = \frac{4\pi}{\sqrt{2}} m_q    (\kappa + \kappa'_{\sigma}) \int 
  \frac{ \; d\vec{\kappa}'_{12} \; d\vec{\kappa}'_3 }{(2 
\pi)^3} \left( \Psi_{12,3}(\kappa'_{12}, \kappa'_3) +  (1 \leftrightarrow 3 
)+(2 \leftrightarrow 3 ) \right).
\end{eqnarray}
Up to a common factor, the l.h.s. represents the coefficient of the $r_{12}$ 
term  (see the first line on the r.h.s. of (114)) while the r.h.s. represents 
the configuration space wave function at the origin, $\Psi(0)$. Now, by using 
the wave function given by (117) and under the requirement of recovering the 
coefficient of the linear term in $r$ in(114) one gets $\beta^{(1)} = 
\frac{3}{\sqrt{2}} m_q (\kappa + \kappa '_\sigma)$. In this way, the 
corresponding wave function can be used in calculations where this property may 
be required as in the calculation of the form factor in the asymptotic regime. 
This value also ensures that the wave function at the origin is unchanged at 
this order, with the result $C^{(1)} = 1$.
 
It is also possible to iterate over the solution given by Eq. (117) ((114) in 
configuration space). The amplitude, $\Psi_{12,3}^{(2)}$, so obtained, reads:
\begin{eqnarray}
\Psi_{12,3}^{(2)} = \frac{\bar{\alpha}}{\pi \sqrt{\pi}} (\frac{4\pi}{\sqrt{2}} 
m_q )^2 \; C^{(2)} \;\;\;\;\;\;\;\;\;\;\;\;\;\;\;\;\;\;\;\;\;\;\;\;\;\;\;\;\;\; 
\nonumber \\ \left( [ \; (\kappa + \kappa'_{\sigma})^2 (J_3 + J_1 + J_2) 
 +(2 \kappa'_{\sigma})^2 (J_3 - \frac{1}{2} J_1 -\frac{1}{2} J_2) ] \; |S> 
\;\;\;\;\; \right. \nonumber \\
 \left. + (\kappa + \kappa'_{\sigma}) (2 \kappa'_{\sigma}) [(J_3 + J_1 + J_2) + 
 (J_3 - \frac{1}{2} J_1 -\frac{1}{2} J_2) ] \; |MS> \;\; \right. \nonumber \\ 
 \left. +\frac{\sqrt{3}}{2} (\kappa + \kappa'_{\sigma}) (2 \kappa'_{\sigma}) 
(J_1 - J_2 )\; |MA> 
 \;\;\;\;\;\;\;\;\;\;\;\;\;\;\;\;\;\;\;\;\;\;\; \right. \nonumber \\
 \left. + \frac{\sqrt{3}}{2} (2 \kappa'_{\sigma})^2  (J_1 - J_2 )\;|A> \right), 
 \;\;\;\;\;\;\;\;\;\;\;\;\;\;\;\;\;\;\;\;\;\;\; 
\end{eqnarray}
where $J_{i}$ is related to $I_{i}$ as follows:
\begin{equation}
J_{i} = \frac{1}{\kappa_{12}^2 + \kappa_3^2 + \beta^{(2)^2}} \; 
\int \frac{ \; d\vec{\kappa}'_{12} \; d\vec{\kappa}'_3 }{(2 \pi)^3} \; 
\frac{\delta(\vec{\kappa}_3 - \vec{\kappa}'_3)}{ (\vec{\kappa}_{12} - 
\vec{\kappa}'_{12})^2} \; I'_{i} \, ,
\end{equation}
with the result:
\begin{eqnarray}
J_3=\frac{1}{\kappa_{12}^2+ \beta^{(2)^2}} \;   \frac{1}{\kappa_{12}\beta^{(1)}} 
 {\rm arctg}(\frac{\beta^{(1)}}{\kappa_{12}}) \;
  \frac{(2\pi)^3 \delta(\vec{\kappa}_3)}{4 \pi \beta^{(1)}} , \nonumber \\
J_1=  \frac{1}{\kappa_{12}^2 + \kappa_3^2 + \beta^{(2)^2}} 
 \;  \frac{1}{(\vec{\kappa}_{12}+\frac{1}{\sqrt{3}} \vec{\kappa}_3)^2} \;
 (\frac{2}{\sqrt{3}})^3 \,
  \frac{1}{(\frac{4}{3} \kappa_3^2 + \beta^{(1)^2}) \frac{4}{3} \kappa_3^2},  
\nonumber \\
J_2=  \frac{1}{\kappa_{12}^2 + \kappa_3^2 + \beta^{(2)^2}} 
 \;  \frac{1}{(\vec{\kappa}_{12}-\frac{1}{\sqrt{3}} \vec{\kappa}_3)^2} \; 
 (\frac{2}{\sqrt{3}})^3 \,
  \frac{1}{(\frac{4}{3} \kappa_3^2 + \beta^{(1)^2}) \frac{4}{3} \kappa_3^2}.  
\end{eqnarray}
For simplicity, we omitted the dependence of the functions, $I_{i}$ on the 
arguments, $\vec{\kappa}'_{12}$ and $\vec{\kappa}'_3$,  what is reminded by 
a `` $'$ '' at I. 

In the limit where $\beta^{(1)}$ can be considered as an infinitesimally small 
quantity, the last factors in $J_1$ and $J_2$ tend to the function 
$\delta(\vec{\kappa}_3)$, making  the quantities $J_1$, $J_2$ and $J_3$ 
equal to each other. Equation (120) simplifies and identifies to that part of 
Eq. (117) corresponding to the same Faddeev amplitude, $\Psi_{12,3}$, 
proportional to $I_3$. On the other hand, the presence of the 
$\delta(\vec{\kappa}_3)$ in Eq. (122) has no physical foundation. It results 
from the iteractive character of the calculation and should disappear in a 
complete one, as partly realized when going from the first iteration, Eq. (117), 
to the second one, Eq. (120).

The parameter, $\beta^{(2)}$, may be determined by requiring that Eq. (119) is 
satisfied when the expression of the wave function, Eq. (120), is used. 
This can be done once $\beta^{(1)}$ has been determined from the previous 
iteration ($\beta^{(1)}= 
\frac{3}{ \sqrt{2}} m_q (\kappa + \kappa'_{\sigma})$). Disregarding this value, 
one may also assume $\beta^{(1)} = \beta^{(2)}$ and determine the corresponding 
common value of these parameters by fulfilling Eq. (119). The value of $C^{(2)}$ 
is determined so that the configuration space wave function at the origin, 
$\Psi(0)$, is equal to that one in the previous iteration or its value from a 
full 
calculation:
\begin{equation}
  \Psi(0) = \int 
  \frac{ \; d\vec{\kappa}'_{12} \; d\vec{\kappa}'_3 }{(2 \pi)^6} 
  \left( \Psi_{12,3}(\kappa'_{12}, \kappa'_3) +  (1 \leftrightarrow 3 )+(2 
\leftrightarrow 3 ) \right) .
\end{equation} 

For a pure Coulombian problem and depending on the approach, values 
$C^{(2)}=1.0$ and $C^{(2)}=0.66$, corresponding respectively to $\beta^{(2)} = 
1.105 \, \kappa m_q$ and $\beta^{(2)}=1.403 \,\kappa m_q $, are obtained, 
providing some uncertainty range. Ultimately, after an infinite series of 
iterations, the value of the parameter corresponding to $\beta^{(2)}$ should 
converge to that related to the binding energy. This one is expected to be equal 
to $\beta=\frac{8\sqrt{2} }{5 \pi } \kappa m_q $ in the hypercentral 
approximation ($K=0$), Eq. (97), and $\beta= 1.016\,\frac{8\sqrt{2} }{5 \pi } 
\kappa m_q = 0.732 \, \kappa m_q$ in a more complete calculation 
\cite{MART,KOK}.
%%%%%%%%%%%%%%%%%%%%%%%C.2     
\subsection{Asymptotic form factors}
Concerning the asymptotic behavior of the form factors, it is interesting
to apply Eqs. (107-109) to the first order term of the wave function given
by (114) which involves terms $r_{12}$ and $r_{13} + r_{23}$. 
After integration over the various 
angles and up to some factor, 
one is left with the following expression of $F_1(\phi)$: 
\begin{eqnarray}
F_1(\phi)  & = &  {\rm sin} \, \phi \;\;\;\;\;\;\;\;\; 
({\rm for}\; r_{12}), \nonumber \\
        & = &  \frac{|{\rm sin} \, \phi + \sqrt{3 }\,{\rm cos} \, \phi |^3 
        - |{\rm sin} \, \phi - \sqrt{3} \, {\rm cos} \, \phi |^3 
        }{ 6\sqrt{3} \, {\rm sin} \, \phi \; {\rm cos} \, \phi } 
\;\;\;\; ({\rm for} \;r_{13}+r_{23}).
\end{eqnarray}
The first term in Eq. (124) can be dealt with easily. Using the relation:
\begin{equation}
\int d\phi \; \frac{{\rm sin}^2\phi \; {\rm cos}^2\phi
}{{\rm cos}^2\phi + \epsilon^2} \, {\rm sin}\,\phi 
= \frac{2}{3} - \epsilon \, (1+\epsilon^2)\,(\frac{\pi}{2}-{\rm 
arctg}\,\epsilon),
\end{equation}
it is seen that, in absence of $\epsilon^5$ terms in the expansion, there is no 
contribution to the coefficient of the $q^{-7}$ term, $a_1$. 
The second term in Eq. (124) can be written as a term similar to the previous 
one:
\begin{equation}
\int d\phi \; \frac{{\rm sin}^2\phi \; {\rm cos}^2\phi}{{\rm cos}^2\phi 
+ \epsilon^2} \, \frac{1}{{\rm sin}\,\phi} = 1 - \epsilon \, (\frac{\pi}{2}
-{\rm arctg}\,\epsilon),
\end{equation}
plus another one integrated over the range, $\frac{1}{2}<{\rm cos}\,\phi < 1$. 
As this one safely converges when $\epsilon \rightarrow 0$, it is immediate that 
its expansion in terms of powers of $\epsilon$ will exhibit only even powers 
and, therefore, will not contribute to $a_1$. Thus, the linear term in $\rho$ 
appearing in the expansion of the correct wave function around the origin does 
not contribute to the form factor at the order $q^{-7}$, as expected. This is 
not a trivial result however. A linear dependence of the wave function on the 
variable $\rho_3$, for instance, which has no physical ground, but turns out to 
be undistinguishable from a  contribution of a $\rho_{12}$ dependence in the 
approximated calculation retaining the $K=0$ wave, leads to a  $q^{-4}$ 
asymptotic behavior of the form factor. This indicates that the $\rho_3$ 
dependence of the wave function has to be carefully determined, otherwise it 
would lead to a bias in calculating form factors in the asymptotic domain.

Expression (120), which involves the effect of two gluon exchanges without any 
restriction, in contrast to Eq. (67), can be used to determine 
the asymptotic behavior of form factors. When it is averaged over the various 
angles, including $\phi$, it allows one to recover the high  momentum behavior 
of the wave function (63) obtained in the hyperspherical formalism, with the 
same front factor, independently of the value taken by $C^{(2)}$.     

Throughout this paper, and especially to get the expressions given in this 
appendix, we employed the following permutation relations: \\
exchange of particles 1 and 3:  
\begin{eqnarray}
\vec{\rho}_3 \rightarrow -\frac{1}{2}\vec{\rho}_3 +
\frac{\sqrt{3}}{2} \vec{\rho}_{12},\;\;\;\;\;\;\;\;\;\;\;\;\;\;\;\;\;\;\;\;\;
\vec{\rho}_{12} \rightarrow \frac{1}{2}\vec{\rho}_{12} +
\frac{\sqrt{3}}{2} \vec{\rho}_3,\;\;\;\;\;\;\;\;\;\;\; \nonumber \\
\vec{\kappa}_3 \rightarrow -\frac{1}{2}\vec{\kappa}_3 +
\frac{\sqrt{3}}{2} \vec{\kappa}_{12},\;\;\;\;\;\;\;\;\;\;\;\;\;\;\;\;\;\;\;\;
\vec{\kappa}_{12} \rightarrow \frac{1}{2}\vec{\kappa}_{12} + 
\frac{\sqrt{3}}{2} \vec{\kappa}_3,\;\;\;\;\;\;\;\;\;\;\; \nonumber \\
|MS> \rightarrow -\frac{1}{2}|MS> +\frac{\sqrt{3}}{2} |MA>,\;\;\;|MA> 
\rightarrow \frac{1}{2}|MA> +\frac{\sqrt{3}}{2} |MS>,
\end{eqnarray}
exchange of particles 2 and 3: 
\begin{eqnarray} 
\vec{\rho}_3 \rightarrow -\frac{1}{2}\vec{\rho}_3 -
\frac{\sqrt{3}}{2} \vec{\rho}_{12}, \;\;\;\;\;\;\;\;\;\;\;\;\;\;\;\;\;\;\;\;\;
\vec{\rho}_{12} \rightarrow \frac{1}{2}\vec{\rho}_{12} -
\frac{\sqrt{3}}{2} \vec{\rho}_3, \;\;\;\;\;\;\;\;\;\;\;\; \nonumber 
\\ \vec{\kappa}_3 \rightarrow -\frac{1}{2}\vec{\kappa}_3 -
\frac{\sqrt{3}}{2} \vec{\kappa}_{12}, \;\;\;\;\;\;\;\;\;\;\;\;\;\;\;\;\;\;\;\;
\vec{\kappa}_{12} \rightarrow \frac{1}{2}\vec{\kappa}_{12} -
\frac{\sqrt{3}}{2} \vec{\kappa}_3, \;\;\;\;\;\;\;\;\;\;\;\; \nonumber \\
|MS> \rightarrow -\frac{1}{2}|MS> -\frac{\sqrt{3}}{2} |MA>,\;\;\;|MA> 
\rightarrow \frac{1}{2}|MA> -\frac{\sqrt{3}}{2} |MS>.
\end{eqnarray}
%%%%%%%%%%%%%%%%%%%%%%%%%%%%%%%%%%%%%%%%%%%%%% 
\section{Details about the $q^{-8}$ asymptotic behavior of the form 
factors and terms beyond}
We first consider in this appendix the effect of a truncation of the total wave 
function calculated in the Faddeev approach, which has been limited to the ten 
partial waves with the lowest values of the angular momentum of the pair of 
particles 1 and 2: $l=0,\, 1,\, 2,\, 3$ and $4$ for each spin state of the same 
pair, 0 and 1. We especially look at the contribution of the term which 
arises from the wave function given by Eq. (114) and involves the 
quantity $r_{13} \, r_{23}$. Its contribution to form factors, which is the 
dominant one, involves $l$ values ranging from $0$ to $\infty$, while the 
contribution of the other term $r_{12}(r_{13}+r_{23})$, which is less important, 
only implies $l=0$. The first contribution may thus be sensitive to some 
truncation, while the second one cannot. Let's remind that the truncation under 
consideration is different from that on the Faddeev amplitude, for which two 
cases with 2 and 8 partial waves have been considered in this work.

Using an approach similar to the one presented in the appendix B, we calculated 
the contributions to the coefficient of the $q^{-8}$ factor due to different 
waves. We separated those for even and odd values of $l$, corresponding 
respectively to the matrix elements $<S|O|S>, <S|O|MS> $ and $<MS|O|MS>$ on the 
one hand, and $<MA|O|MA>$ on the other. Assuming that the total sum is 
normalized to one, we found that the contribution for a given $l$ is given by:
\begin{equation}
\frac{ 900(-1)^{(l+1)} }{(2l-5)(2l-3)(2l-1)(2l+3)(2l+5)(2l+7)}.
\end{equation}
The sum thus reads for even values of $l$:
\begin{equation}
1=0.5714(=\frac{4}{7}) +0.4329(=\frac{100}{231})-0.0040(=\frac{4}{1001})-...
\end{equation}
and for odd values:
\begin{equation}
1=0.952(=\frac{20}{21})+0.046(=\frac{20}{429})+....
\end{equation}

The above result shows that no serious discrepancy is introduced by the 
truncation of the total wave function we made and that the origin of the 
discrepancy of the calculated form factors and their expectations in the 
asymptotic regime has to be looked for elsewhere. On the contrary, 
calculations using phenomenological wave functions limited to $l=0$ would 
in any case miss a sizeable contribution.

We mentioned in the text the possibility that the correction to the dominant 
$q^{-8}$ term be of the relative order $q^{-1}$, with the result of delaying 
the convergence of the form factors to their asymptotic value. We here give 
details about an example dealing with a two-body system. This is not directly 
relevant to the three-body case of interest in this work, but we nevertheless 
believe it could cast some light on this one. 

The expression of the form factor for a system of two equal mass particles is 
given by:
\begin{equation}
F(\vec{q}^{2}) = \int \frac{d\vec{k}}{(2 \pi)^3} \; \varphi(\vec{k}) 
\; \varphi( \vec{k}+ \frac{1}{2} \vec{q} ),
\end{equation}
where $\varphi(\vec{k})$ represents the wave function of the system under 
consideration. For the lowest state of the hydrogenic system discussed in Sect. 
3, $\varphi(\vec{k}) =\sqrt{4 \pi} \frac{4(\kappa)^{5/2}}{ (k^2+\kappa^2)^2}.$ 
Hence:
\begin{equation}
F(\vec{q}^{2})  =  \int \frac{d\vec{k}}{(2 \pi)^3} \;
 \frac{64 \pi \, \kappa^5}{(k^2+\kappa^2)^2 ((\vec{k}+\frac{1}{2}\vec{q})^2 + 
\kappa^2)^2}  = \frac{\kappa^4}{(\frac{1}{16}q^2+\kappa^2)^2}.
\end{equation}
In the limit of large $q^2$, the form factor reads:
\begin{equation}
F(\vec{q}^{2}) = \frac{256 \, \kappa^4}{q^4}(1- 32 \frac{\kappa^2}{q^2}+...),
\end{equation}
evidencing a $q^{-2}$ correction to the leading order term. It is interesting 
to compare this result with that of a slightly different expression: 
\begin{equation}
\tilde{F}(\vec{q}^{2}) = \int \frac{d\vec{k}}{(2 \pi)^3} \;
\frac{64 \pi \, \kappa^5}{(k^2+\kappa^2)^2 (k^2+\frac{1}{4}q^2 + \kappa^2)^2},
\end{equation}
corresponding to the first term of an expansion of the integrand in Eq. (133) in 
terms of the quantity $\frac{\vec{k}.\vec{q}}{k^2+\frac{q^2}{4}+\kappa^2}$. 
Its expression is given by:
\begin{equation}
\tilde{F}(\vec{q}^{2}) = 
\frac{128 \, \kappa^4}{q^4} 
\left(1+\frac{\kappa}{\sqrt{\kappa^2+\frac{1}{4}q^2}}-
\frac{4\kappa}{\kappa + \sqrt{\kappa^2+\frac{1}{4}q^2} } \right).
\end{equation}
In the limit of large $q$, it becomes:
\begin{equation}
\tilde{F}(\vec{q}^{2}) = \frac{128 \, \kappa^4}{q^4}(1-6\frac{\kappa}{q}+...).
\end{equation}

The dominant term at high $q$ is a half of that one in the former calculation, 
Eq. (134), which is in relation with the fact that only the low value of k in 
Eq. (135) contributes to it, whereas in Eq. (132), the values of $\vec{k}$ 
around $-\frac{1}{2}\vec{q}$ also contribute for an equal amount. More important 
here, the correction to the dominant term is of relative order $q^{-1}$ instead 
of $q^{-2}$. This is in agreement with an expansion in terms of powers of 
$q^{-2}$ of the integrand in Eq. (135). Beyond the $q^{-4}$ term which 
converges, the next one in $q^{-6}$ diverges linearly 
and should be effectively cut-off by a factor $q$, hence the correction of 
relative order $q^{-1}$. The same argument should apply to the full expression 
of the form factor, Eq. (132), but it turns out that the different $q^{-1}$ 
corrections cancel out in this case, indicating that the way the variable 
$\vec{q}$ appears in this expression is of special relevance for the existence 
of $q^{-2}$ corrections instead of $q^{-1}$, as expected at first sight. In any 
case, the above discussion shows the difficulty to make statements in the more 
complicated three-body case.

The difference as to the $q^{-1}$ corrections has probably to do with the 
difference in the mathematical properties of $F(\vec{q}^{2})$ and 
$\tilde{F}(\vec{q}^{2})$ when the variable $q$ is made complex, 
$q \rightarrow |q| e^{i\theta}$ with $\theta$ varying from $0$ to $\pi$. An 
absence of change is the sign that the function under consideration is a 
function of $q^2$. When the above transformation is made on 
$\tilde{F}(\vec{q}^{2})$, a singularity occurs for $\theta=\frac{1}{2}\pi$ and 
the integration variable taking the value $k=\sqrt{\frac{|q|^2}{4} - \kappa^2}$. 
This does not occur for $F(\vec{q}^{2})$, Eq. (133), due to the presence at the 
denominator of the term $\vec{k}.\vec{q}$ besides the quantity 
$k^2+\frac{q^2}{4}+\kappa^2$. The weight of the singularity is reduced by the 
condition that $\vec{k}$ should be orthogonal to $\vec{q}$.

One can imagine that the above results be extended for some part to the 
three-body case. The absence of an exact analytic expression for the three-body 
wave function prevents one to make definite statements however. We nevertheless 
expect that corrections of relative order $q^{-2}$ should be associated to 
${\rm log}\,q$ factors, indicating that the results for the two-body system 
cannot be transposed as such to the three-body system.

\end{document}